\documentclass[aps,pra,notitlepage,reprint,floatfix,amsmath,amssymb]{revtex4-1}

\usepackage{siunitx}
\usepackage{xcolor}
\usepackage{graphicx}
\graphicspath{{figures/}}         

\usepackage[caption=false]{subfig} 
\usepackage{hyperref}

\newcommand{\ppol}{p}
\newcommand{\spol}{s}
\newcommand{\etal}{\textit{et al.}~}
\newcommand{\ve}{\epsilon}

\newcommand{\zxp}{\zeta({\textbf x}_{\|})}
\newcommand{\bxp}{{\mathbf x}_{\|}}

\renewcommand{\vec}[1]{\mathbf{{#1}}}
\newcommand{\pvec}[1]{\mathbf{{#1}}_\parallel }
\newcommand{\vecUnit}[1]{\hat{\mathbf{#1}}}

\newcommand{\nn}{\nonumber}

\newcommand{\ofq}{(\vec{q})}
\newcommand{\ofp}{(\vec{p})}
\newcommand{\ofpq}{(\vec{p}\,|\,\vec{q})}

\newcommand{\dtwox}{{\mathrm{d}^2 x_\parallel}}
\newcommand{\dtwoq}{{\frac{\mathrm{d}^2q}{(2 \pi)^2}}}

\DeclareMathOperator{\klb}{\mathnormal{\mathbf{k}_l^b}}
\DeclareMathOperator{\kma}{\mathnormal{\mathbf{k}_m^a}}

\newcommand{\Vie}[3]{\mathop{\mathbf{#1}_{#2}^{#3}}}
\newcommand{\Cie}[3]{\mathop{\mathcal{#1}_{#2}^{#3}}}

\newcommand{\BCie}[3]{\mathop{\boldsymbol{ \mathcal{#1}}_{#2}^{#3}}}
\newcommand{\ofuipi}[4]{\mathop{(\mathbf{#1}_{#2}\, |\,\mathbf{#3}_{#4})}}
\newcommand{\dtwopi}[2]{\mathop{\frac{\mathrm{d}^2{#1}_{#2}}{(2 \pi)^2}}}

\newcommand{\of}[1]{\mathop{(\mathbf{#1})}}

\DeclareMathOperator{\st}{ | }

\begin{document}


\title{Physics of polarized light scattering from weakly rough dielectric surfaces: Yoneda and Brewster scattering phenomena}
\author{J.-P. Banon$^1$}
\author{\O. S. Hetland$^1$}
\author{I. Simonsen$^{1,2}$}
\affiliation{$^1$Department of Physics, NTNU -- Norwegian University of Science and Technology, NO-7491 Trondheim, Norway}
\affiliation{$^2$Surface du Verre et Interfaces, UMR 125 CNRS/Saint-Gobain, F-93303 Aubervilliers, France}

\date{\today}

\begin{abstract}
The optical Yoneda and Brewster scattering phenomena are studied theoretically based on perturbative solutions of the reduced Rayleigh equations. The Yoneda phenomenon is characterized as an enhancement of the intensity of the diffuse light scattered by a randomly rough interface between two dielectric media when the light is observed in the optically denser medium.
The intensity enhancement occurs above a critical angle of scattering which is independent of the angle of incidence of the excitation. The Brewster scattering phenomenon is characterized by a zero scattered intensity either in the reflected or transmitted light for an angle of scattering which depends on the angle of incidence. We also describe a generalization of the Brewster scattering phenomenon for outgoing evanescent waves and circularly-polarized waves.
The physical mechanisms responsible for these phenomena are described in terms of simple notions such as scalar waves, oscillating and rotating dipoles and geometrical arguments, and are valid in a regime of weakly rough interfaces.
\end{abstract}

\maketitle 

\section{Introduction}
When light is scattered in either reflection or transmission from or through a weakly rough interface, two phenomena of interest can be observed in the scattered intensity distributions. These are the \emph{Yoneda phenomenon}, relatable to the idea of total internal reflection, and the \emph{Brewster scattering phenomenon}, relatable to the polarizing angle.

The Yoneda phenomenon is characterized as an enhancement of the intensity of the light scattered diffusely by a randomly rough interface between two dielectric media when the light is observed in the optically denser medium.
The intensity enhancement occurs above a critical angle of scattering which is independent of the angle of incidence of the excitation. This critical angle is always the polar angle, in the denser medium, for which the wavenumber of a plane wave turns non-propagating in the less dense medium \cite{Hetland2016a,Hetland2017}.
Although well known in the scattering of x-rays from both metallic~\cite{Yoneda1963,Vineyard1982,Sinha1988,Gorodnichev1988,Leskova1997,Renaud2009} and non-metallic~\cite{Dosch1987,Stepanov2000,Kitahara2002,Gasse2016} surfaces, a paper by Kawanishi~\etal \cite{Kawanishi1997} marks their first explicit appearance in optics\footnote{In an earlier numerical investigation of light scattering from one-dimensional dielectric rough surfaces, Nieto-Vesperinas and S\'{a}nchez-Gil~\cite{Nieto-Vesperinas1992} observed ``sidelobes'' in the angular intensity distributions. They did, however, not associate this observation with the Yoneda phenomenon.}.
Kawanishi \textit{et al.}, by the use of the stochastic functional approach, studied the case where a two-dimensional randomly rough interface between two dielectric media is illuminated by \textit{p}- or \textit{s}-polarized light from either medium.
They obtained several interesting properties of the reflected and transmitted light that are associated with the phenomenon of total internal reflection when the medium of observation is the optically denser medium.
These include the appearance of Yoneda peaks, which were described by the authors as ``quasi-anomalous scattering peaks.''
As an interpretation of their results, the authors suggested that the Yoneda peaks may be associated with the presence of lateral waves~\cite{Book:Tamir1982} propagating along the interface in the optically less dense medium.
Although the mathematical origin of the Yoneda effect has been shown through various perturbative approaches based on the reduced Rayleigh equations (RRE), a physical interpretation of the effect is still under discussion; a summary of which can be found in Ref.~\citenum{Hetland2016a}.
Optical Yoneda peaks were recently observed experimentally for a configuration of reflection from a randomly rough dielectric interface, when the medium of incidence was the optically denser medium \cite{Gonzalez-Alcalde2016}.

\medskip
The \emph{Brewster angle} is maybe the best known planar surface reflection effect where the polarization of light plays a major role.
Proposed as a \emph{polarizing angle} by Sir David Brewster in 1812 \cite{Brewster1815}, its exact definition has been a slight matter of debate in modern times \cite{Lakhtakia1989}. For isotropic dielectric non-magnetic materials, however, it may be defined to be the angle of incidence, onto a planar dielectric surface, for which the reflection amplitude for \textit{p}-polarized light (light polarized in the plane given by the incident light and the surface normal) is zero.

A complete physical understanding of the Brewster phenomenon is, at best, non trivial. The most common explanation for the gradual disappearance of the reflection amplitude is based on the radiation pattern of dipoles induced in the scattering substrate \cite{jackson,book:hecht2002}. This idea is not new, and can be traced back to investigations by e.g. Sommerfeld \cite{Sommerfeld1923}.
Modelling the scattering from a rough surface as a layer of polarizable spheres led Greffet and Sentenac \cite{Greffet1991} to the same conclusion. In a later collaboration with Calvo-Perez, this point of view was reinforced through the development of, and the results given by, the Mean Field Theory (MFT) \cite{Sentenac1998,Calvo-Perez1999}.
However, amongst others Lekner \cite{Lekner1987} argues that even if the dipole argument holds great explanatory power for a wide range of scattering systems, he challenges the argument for the case of the Brewster angle for dielectric media. His main issue with the argument is that the accelerated electrons cannot oscillate as dipoles in the transmitted medium in the case of the wave approaching an interface with vacuum on the opposing side, since the argument goes that the dipoles are oriented according to the field in the refracted wave. Also, there is an analog to Brewster's angle for longitudinal acoustic waves called Green's angle, and in this case the radiation from each scatterer does not have dipole character~\cite{Lekner1987}.

These and other conceptual issues in the explanatory model for the Brewster angle are attempted to be reconciled by Doyle \cite{Doyle1985} in his work with a factored form of the Fresnel equations.
Inspired by the work of Sein \cite{Sein1970}, and Pattanayak and Wolf \cite{Pattanayak1972} on the interpretation and generality of the extinction theorem, Doyle claims that the proper understanding of the Brewster phenomenon has been hampered by the attention given to surface sources through a slightly misunderstood interpretation of the Ewald-Oseen extinction theorem~\cite{Ewald1916}.
Doyle emphasizes the participation of the entire media in the creation of the reflected wave, and makes use of Ewald's original concept of ``wave triads''.
Doyle's factored form of the Fresnel equations separately expresses the scattering pattern from individual dipoles and the coherent scattering function of the dipole array, and manages in this way to explain the polarizing angles for any combination of transparent media.


Kawanishi~\etal~\cite{Kawanishi1997} observed angles of zero scattering intensity to first order in their approach in the distributions of the intensity of the incoherently scattered light when the incident light was p-polarized. Due to their resemblance to the Brewster angle in the reflectivity from a flat interface, they dubbed these angles the ``Brewster scattering angles''. These angles were observed in both reflection \emph{and} transmission, for light incident from either medium.

Both the Brewster scattering angles and Yoneda peaks were recently observed and discussed in numerical simulations of scattering in both reflection and transmission from weakly rough surfaces \cite{Hetland2016a,Hetland2017}, and also in a film geometry \cite{Banon2018a} where it was claimed that the phase shifts associated with these phenomena impact the angular positions of interference rings of diffusely scattered light, known as Sel\'{e}nyi rings.

In this paper we seek to further illuminate the phenomena of Brewster scattering angles and Yoneda peaks and more generally identify the fundamental mechanisms at play in the scattering of polarized light by a weakly rough surface.
After describing the statistical properties of the interface in  Sec.~\ref{sec:scatt:sys}, we derive, in Sec.~\ref{sec:theory}, a set of reduced Rayleigh equations (RREs) for the case of electromagnetic scattering inspired by the work of Soubret \etal \cite{Soubret2001a} and give the corresponding RRE for scalar waves subjected to the continuity of the scalar field and its normal derivatives with respect to the interface.
Furthermore, we give an approximate solution of the RREs to first order in the surface profile function in a series expansion of the reflection and transmission amplitudes. The first order perturbative solution will be our main tool of investigation in Sec.~\ref{sec:results}. Section~\ref{sec:pheno} is devoted to summarizing some phenomenological observations which have been obtained in the literature before embarking in Sec.~\ref{sec:interpretation} into a more in-depth analysis of the reflection and transmission amplitudes with special care given to their physical interpretation.
In particular, we show how the response can be factorized as a product of a term reminiscent of a scalar wave response and a term encoding the component of the response specific to polarization. Such a factorization is a clear signature of two aspects of scattering by arrays of dipoles; the radiated power is controlled both by the interference of the spherical-like waves emitted by each atomic source and their individual characteristic dipolar radiation.
Once the general physical interpretation of the equations is clarified, we explain in detail the origin of the Yoneda phenomenon in Sec.~\ref{sec:yoneda}, and show that it is fundamentally a single scattering, scalar wave phenomenon.
The Brewster scattering phenomenon, and more generally all polarization induced effects, are then discussed thoroughly in Secs.~\ref{sec:brewster_1storder}-\ref{sec:full_distribution}.
We first restrict the analysis of the Brewster scattering phenomenon to scattering in the plane of incidence and derive a one-line criterion for predicting the Brewster scattering angle which allows for a simple geometrical interpretation.
A detour via the analysis of the polarization properties of the radiation of oscillating and rotating dipoles in free space is made in Sec.~\ref{sec:dipole} in order to facilitate the intuitive understanding of the full angular distribution of scattering by a rough surface discussed in Sec.~\ref{sec:full_distribution}.
Finally, Sec.~\ref{sec:conclusion} summarizes the conclusions we have drawn from this study and suggests experimental setups to test some interesting predictions made by the theory.

\section{Scattering systems}\label{sec:scatt:sys}

\begin{figure}[ht]
\centering
\includegraphics[width=0.95\linewidth , trim= 6.cm 4.cm 5.cm 2.cm,clip]{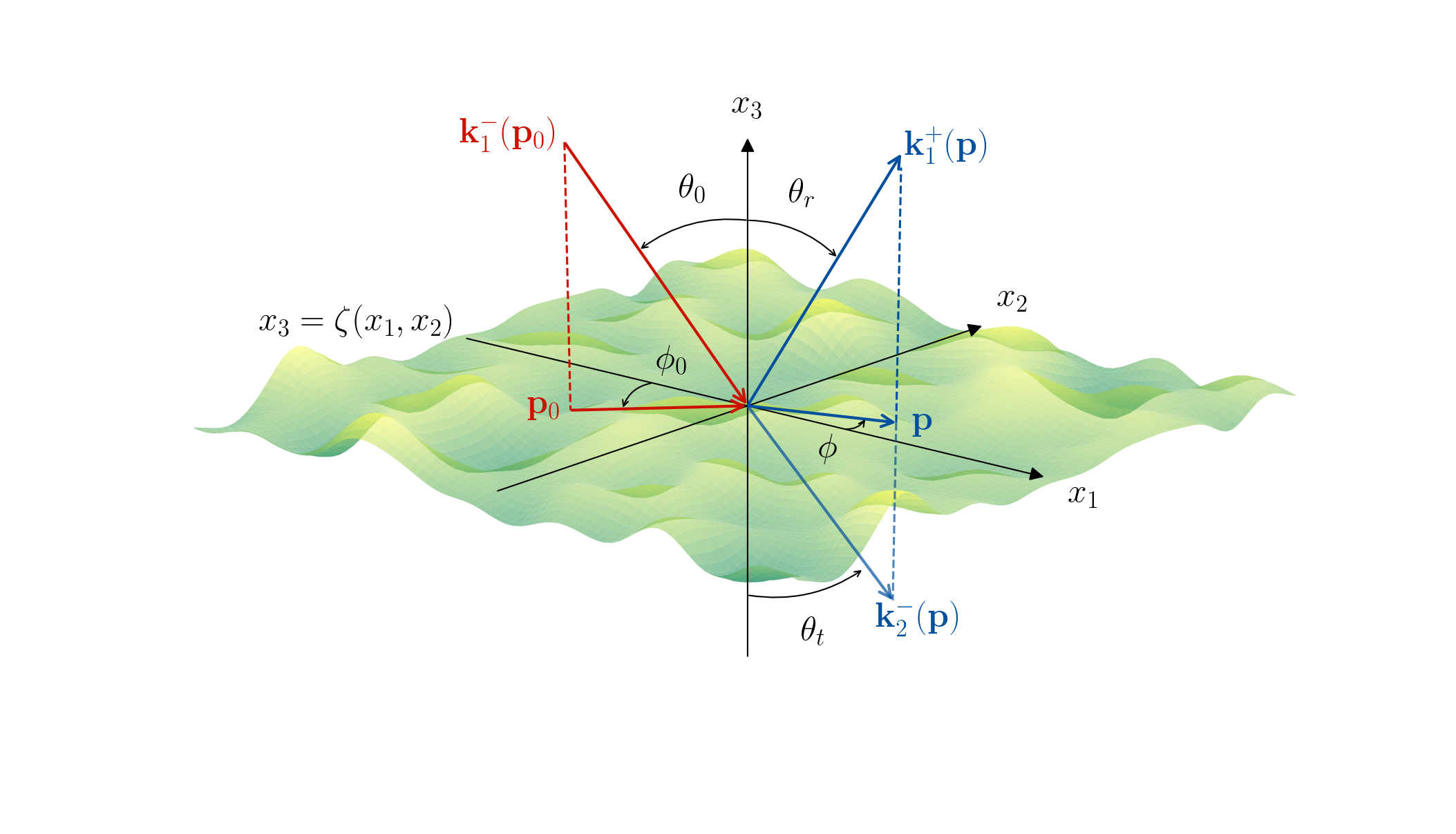}
\caption{Definitions of the angles of incidence and scattering, together with the relevant wave vectors.
}
\label{fig:system}
\end{figure}

The system we study in this work consists of a non-magnetic dielectric medium (medium 1), whose dielectric constant is $\ve_1 > 0$ (refractive index $n_1 = \sqrt{\epsilon_1}$), in the region $x_3 > \zxp$, and a non-magnetic dielectric medium (medium 2), whose dielectric constant is $\ve_2 > 0$ (refractive index $n_2 = \sqrt{\epsilon_2}$), in the region $x_3 < \zxp$ (Fig. 1).
The definition of the geometry is set in the three-dimensional space endowed with a Cartesian coordinate system $(O,\hat{\mathbf{e}}_1,\hat{\mathbf{e}}_2,\hat{\mathbf{e}}_3)$, with the vector plane $(\hat{\mathbf{e}}_1,\hat{\mathbf{e}}_2)$ parallel to the mean plane of the interface.
The origin, $O$, of the coordinate system can be arbitrarily chosen, only affecting the complex reflection and transmission amplitudes by an overall phase factor which plays no role in the intensity of the scattered light.
A point is then represented as $\mathbf{x} = \sum_{i = 1}^3 x_i \hat{\mathbf{e}}_i = \bxp + x_3 \, \hat{\mathbf{e}}_3$.
An overview of a typical system geometry is provided in Fig.~\ref{fig:system}.
The surface profile function $\zeta$ will be assumed to be a realization of a continuous, differentiable, single-valued, stationary, isotropic, Gaussian random process with zero mean and given auto-correlation. More specifically, the surface profile function is assumed to satisfy the following properties:
\begin{subequations}
\begin{align}
\left\langle \zeta(\mathbf{x}_\parallel) \right\rangle &= 0 \: , \\
\left\langle \zeta(\mathbf{x}_\parallel) \zeta(\mathbf{x}_\parallel') \right\rangle &= \sigma^2 \: W(\mathbf{x}_\parallel - \mathbf{x}_\parallel') \: .\label{eq:auto_covariance}
\end{align}
\end{subequations}
Here and in the following, the angle brackets denote an average over an ensemble of realizations of the stochastic process, $\sigma$ denotes the rms roughness, and $W$ is the height auto-correlation function normalized so that $W(\mathbf{0}) = 1$. In particular, we will deal with the special case of a Gaussian auto-correlation function defined by
\begin{equation}
W(\mathbf{x}_\parallel) = \exp \left( - \frac{|\mathbf{x}_\parallel|^2}{a^2} \right) \: ,
\end{equation}
where $a$ is the correlation length.
The corresponding power spectrum (defined as the Fourier transform of $W$) is then
\begin{equation}
g(\Vie{p}{}{}) = \pi a^2 \exp \left( - \frac{|\Vie{p}{}{}|^2 a^2}{4} \right) \: ,
\label{eq:power_spectrum}
\end{equation}
with $\Vie{p}{}{} =  p_1 \: \hat{\mathbf{e}}_1 + p_2 \: \hat{\mathbf{e}}_2$.
%


\section{Theory}\label{sec:theory}

The theoretical approach used in this work to study the scattering of light from the systems of interest is based on the so-called reduced Rayleigh equations. A reduced Rayleigh equation is an inhomogeneous integral equation in which the integral kernel encodes the materials and geometry of the scattering system, and the unknowns are the reflection or transmission amplitudes for each polarization.
First derived by Brown \etal \cite{Brown1984}, the reduced Rayleigh equation is obtained from the Rayleigh solution to the electromagnetic boundary problem. Using inspiration drawn from the extinction theorem it is possible to ``reduce'' the full Rayleigh equations through the elimination of either the reflected or transmitted field.
In the following, in order to establish the notation and highlight the main assumptions of the method, we will briefly recall the key ideas of the derivation of the reduced Rayleigh equations for a system composed of two media separated by a rough interface.
We will use, to our knowledge, the most general form of the reduced Rayleigh equations for a single interface derived by Soubret \etal in Ref.~\citenum{Soubret2001a} and used by these authors in Refs.~\citenum{Soubret2001a} and \citenum{Soubret2001} in the case of a single interface system and a film geometry.

\subsection{The reduced Rayleigh equations}
In this work we assume the electromagnetic response of the media to be modeled by non-magnetic, homogeneous, isotropic, linear constitutive relations in the frequency domain.
We consider the presence of an electromagnetic field $(\vec{E},\vec{H})$ in the whole space, where their restriction will be denoted by a subscript $j$ depending on the medium in which they are evaluated. As an example, the electric field evaluated at a point $\vec{x}$ in medium 1 at time $t$ is denoted $\vec{E}_1 (\vec{x},t)$.
The source free Maxwell equations, together with homogeneous, linear and isotropic constitutive relations in the frequency domain, result in the fact that the electric and magnetic fields satisfy the Helmholtz equation in each region. Namely, for $j \in \{1,2\}$,
\begin{equation}
	\nabla^2 \mathbf{E}_j (\Vie{x}{}{},\omega) + \epsilon_j (\omega) \: \left(\frac{\omega}{c} \right)^2 \, \mathbf{E}_j (\Vie{x}{}{},\omega) = \mathbf{0} \: \mathrm{,}
	\label{helmoholtz}
\end{equation}
and similarly for $\mathbf{H}$.
Here, $\nabla^2$ denotes the vector Laplace operator and $c$ is the speed of light in vacuum. Here onward, we will drop the time, or frequency, dependence, since we assume a stationary regime at a fixed frequency where time contributes only by an overall phase factor $\exp(- i \omega t)$.
It is well known that a solution to the Helmholtz equation can be written as a linear combination of plane waves, thus the electric field in each region can be represented as
\begin{align}
	\mathbf{E}_j (\mathbf{x}) = \sum_{a=\pm} \: \int_{\mathbb{R}^2} \: &\left[ \Cie{E}{j,p}{a} \ofq \,
  \hat{\mathbf{e}}_{p,j}^{a} \ofq  + \Cie{E}{j,s}{a} \ofq \, \hat{\mathbf{e}}_s \ofq  \right] \,
  \nn\\
  &\times\exp \left( i \, \Vie{k}{j}{a} \ofq \cdot \;\mathbf{x} \right) \dtwopi{q}{} \: \mathrm{,}
	\label{fieldexpansion}
\end{align}
where we have defined
\begin{subequations}
\begin{align}
	\Vie{k}{j}{\pm} \ofq &= \mathbf{q} \pm \alpha_j \ofq \, \hat{\mathbf{e}}_3	\: , \\
	\alpha_j \ofq &= \sqrt{k_j^2 - \mathbf{q}^2}, \quad \mathrm{Re} \: (\alpha_j), \mathrm{Im} \: (\alpha_j) \geq 0 \: ,\\
\hat{\mathbf{e}}_s \ofq &=  \hat{\mathbf{e}}_3 \times \hat{\mathbf{q}} \: ,\\
\hat{\mathbf{e}}_{p,j}^\pm \ofq &= k_j^{-1} \left(\pm \alpha_j\ofq \, \hat{\mathbf{q}} - |\mathbf{q}| \, \hat{\mathbf{e}}_3 \right) \label{basis:ep} \\
k_j &= n_j \frac{\omega}{c} = |\Vie{k}{j}{\pm} (\Vie{q}{}{})| \label{eq:disp_relation} \: .
\end{align}
\label{basis}%
\end{subequations}
In other words, the wave vector $\Vie{k}{j}{\pm} \ofq$ of an elementary plane wave is decomposed into its projection $\mathbf{q}$ in the lateral vector plane $(\hat{\mathbf{e}}_1, \hat{\mathbf{e}}_2)$ and the component $\pm \alpha_j \ofq$ along $\hat{\mathbf{e}}_3$.
The sum over $a=\pm$ takes into account both upwards ($+$) and downwards ($-$) propagating and evanescent (and possibly growing) waves. The field amplitude is decomposed in the \emph{local polarization basis} $(\hat{\mathbf{e}}_{p,j}^a \ofq,\hat{\mathbf{e}}_s \ofq)$, hence $\Cie{E}{j,\alpha}{a} \ofq$ denotes the component of the field amplitude in the polarization state $\alpha$ of the mode characterized by $a$ and $\Vie{q}{}{}$.
In this basis, the directions given by $\hat{\mathbf{e}}_{p,j}^\pm \ofq$, and $\hat{\mathbf{e}}_s \ofq$ are the directions of the p- and s-polarization of the electric field amplitude, respectively. Furthermore, the electromagnetic fields have to satisfy the boundary conditions
\begin{subequations}
\begin{align}
	\mathbf{n}(\mathbf{x}_\parallel) \times \Big[ \mathbf{E}_{2}(\mathbf{s} (\mathbf{x}_\parallel) ) - \mathbf{E}_1(\mathbf{s} (\mathbf{x}_\parallel)) \Big] &= \mathbf{0}\\
	\mathbf{n}(\mathbf{x}_\parallel) \times \Big[ \mathbf{H}_{2}(\mathbf{s} (\mathbf{x}_\parallel)) - \mathbf{H}_1(\mathbf{s} (\mathbf{x}_\parallel)) \Big] &= \mathbf{0} \: ,
\end{align}\label{BC}
\end{subequations}
where $\mathbf{n}(\mathbf{x}_\parallel)$ is a vector normal to the interface at the surface point $\mathbf{s} (\mathbf{x}_\parallel) = \mathbf{x}_\parallel + \zeta(\mathbf{x}_\parallel) \hat{\mathbf{e}}_3$, given by
\begin{equation}
	\mathbf{n}(\mathbf{x}_\parallel) = \hat{\mathbf{e}}_3 - \frac{ \partial \zeta}{\partial x_1} (\mathbf{x}_\parallel)\: \hat{\mathbf{e}}_1 - \frac{\partial \zeta}{\partial x_2} (\mathbf{x}_\parallel) \: \hat{\mathbf{e}}_2	\: \mathrm{.}
\end{equation}
Here, $\partial \cdot /\partial x_k$ denotes the partial derivative along the direction $\hat{\mathbf{e}}_k$.
Following Soubret~\emph{et al.}~\cite{Soubret2001a}, by substituting the field expansion Eq.~(\ref{fieldexpansion}) into Eq.~(\ref{BC}) and by a clever linear integral combination of the boundary conditions inspired by the extinction theorem \cite{Pattanayak1972}, one can show that the upward or downward field amplitudes in medium $2$ can be linked to the upward and downward field amplitudes in medium $1$ via the following integral equation defined for $a_{2} = \pm$, and $\Vie{p}{}{}$ in the vector plane $(\hat{\mathbf{e}}_1,\hat{\mathbf{e}}_2)$:
\begin{align}
  \sum_{a_1=\pm} \int \: \Cie{J}{2,1}{a_{2},a_1}\ofpq \: \Vie{M}{2,1}{a_{2},a_{1}}\ofpq \:  \BCie{E}{1}{a_1} \ofq \dtwopi{q}{}
  \nn\\
  = \frac{2 \, a_{2} \, n_1 n_2 \,\alpha_{2}\ofp}{\epsilon_{2} - \epsilon_{1}} \, \BCie{E}{2}{a_{2}} \ofp	.
  \label{rrefinal}
\end{align}
Here $\BCie{E}{j}{a} \ofq = (\Cie{E}{j,p}{a} \ofq,\Cie{E}{j,s}{a} \ofq)^\mathrm{T}$ denotes a column vector of the polarization components of the field amplitude in medium $j$.
Moreover, $\Vie{M}{l,m}{b,a} \ofpq$ is the 2$\times$2 matrix
\begin{equation}
	\Vie{M}{l,m}{b,a} \ofpq
  = k_1 k_2 \begin{pmatrix}
  \vecUnit{e}_{p,l}^b(\vec{p})\cdot\vecUnit{e}_{p,m}^a\ofq &
  \vecUnit{e}_{p,l}^b\ofp\cdot\vecUnit{e}_{s}\ofq \\
  \vecUnit{e}_{s}\ofp\cdot\vecUnit{e}_{p,m}^a\ofq &
  \vecUnit{e}_{s}\ofp\cdot\vecUnit{e}_{s}\ofq \\
  \end{pmatrix},
  \label{Mdef}
\end{equation}
\sloppy which originates from a change of coordinate system between the local polarization basis  $(\hat{\mathbf{e}}_{p,l}^b \ofp,\hat{\mathbf{e}}_s \ofp)$ and $(\hat{\mathbf{e}}_{p,m}^a \ofq,\hat{\mathbf{e}}_s \ofq)$, defined for $a = \pm$, $b = \pm$, and $l, m \in \{1, 2\}$ with $l \neq m$.
The kernel scalar factor $\Cie{J}{l,m}{b,a}\ofpq$ encodes the surface geometry and is defined as
\begin{align}
  \Cie{J}{l,m}{b,a} &\ofpq = \left[ b \alpha_l\ofp - a \alpha_m\ofq \right]^{-1}
  \nn\\&\times
  \int\exp \left[-i(\klb \ofp - \kma \ofq) \cdot \vec{s}(\pvec{x}) \right] \dtwox .
  \label{Iintdef}
\end{align}
Notice that, as already pointed out in Ref.~\citenum{Soubret2001a}, due to the symmetry of the boundary conditions, one may also show in the same way that
\begin{align}
  \sum_{a_{2}=\pm} \int \: \Cie{J}{1,2}{a_1,a_{2}}&\ofpq \: \Vie{M}{1,2}{a_1,a_{2}}\ofpq \:  \BCie{E}{2}{a_{2}}\ofq \dtwopi{q}{}
  \nn\\
  &= \frac{2 \, a_{1} \, n_1 n_2 \,\alpha_{1}\ofp}{\epsilon_{1} - \epsilon_{2}} \, \BCie{E}{1}{a_{1}}\ofp ,
  \label{rrefinalswitch}
\end{align}
which can be obtained from Eq.~(\ref{rrefinal}) by interchanging the subscripts $1$ and $2$. Typically, Eq.~(\ref{rrefinal}) is appropriate to solve the problem of reflection whereas Eq.~(\ref{rrefinalswitch}) is appropriate to solve the problem of transmission, as we will see later.

\medskip
So far, we have stayed general and simply assumed the presence of an electromagnetic field decomposed in propagating and non-propagating waves in each region.
Therefore, there is no uniqueness in the solutions to the transfer equations, Eqs.~(\ref{rrefinal}) and (\ref{rrefinalswitch}). In fact, the transfer equations simply link plane wave components of eigenmodes of the system on both sides of the interface. To ensure a unique solution, i.e. to really have a scattering problem, one needs to impose some constraints on the field.
First, we need to introduce an incident field to our model. This will split the field expansion into a sum of an incident field, which is given by our model of the problem, and a scattered field.
Note that within this framework, the incident field may be chosen to be in either medium, or to be a combination of excitations incident from different media.

In our case, the incident field will be taken as a plane wave incident from medium 1 and defined as
\begin{align}
  \Vie{E}{0}{} (\mathbf{x})
  = &
  \left[ \Cie{E}{0,p}{} \, \hat{\mathbf{e}}_{p,1}^{\mathnormal{-}} (\Vie{p}{0}{})  + \Cie{E}{0,s}{} \, \hat{\mathbf{e}}_s  (\Vie{p}{0}{}) \right]
  \nn\\&\times
  \exp \left( i \Vie{k}{1}{-} (\Vie{p}{0}{}) \cdot \mathbf{x} \right),
  \label{incField}
\end{align}
where $\Vie{p}{0}{}$ is the projection of the wave vector of the incident wave onto the $(\vecUnit{e}_1,\vecUnit{e}_2)$ plane, with the property $| \Vie{p}{0}{} | \leq k_1$, i.e. we consider an incident propagating wave.
The fact that this is the only incident wave considered, together with the Sommerfeld radiation condition at infinity, gives that the only elementary waves allowed in the scattered field are those with wave vectors of the form $\Vie{k}{1}{+}(\Vie{p}{}{})$ and $\Vie{k}{2}{-}(\Vie{p}{}{})$ in medium~$1$ and $2$, respectively.
This property can be expressed by defining the field amplitudes
\begin{subequations}
\begin{align}
  \BCie{E}{1}{-} (\Vie{q}{}{}) &= (2 \pi)^2 \: \delta (\Vie{q}{}{} - \Vie{p}{0}{}) \: \BCie{E}{0}{}\: ,\\
  \BCie{E}{2}{+} (\Vie{q}{}{}) &= \mathbf{0} \: ,
\end{align}
\label{sommerfeld}%
\end{subequations}
where $\BCie{E}{0}{} = (\Cie{E}{0,p}{},\Cie{E}{0,s}{})^\mathrm{T}$.
Next, we assume that the scattered field amplitudes are linearly related to the incident field amplitude $\BCie{E}{0}{}$ via the reflection and transmission amplitudes, $\Vie{R}{}{} ( \Vie{q}{}{} | \Vie{p}{0}{})$ and $\Vie{T}{}{} ( \Vie{q}{}{} | \Vie{p}{0}{})$, defined as
\begin{subequations}
\begin{align}
  \BCie{E}{1}{+} (\Vie{q}{}{}) &= \Vie{R}{}{} ( \Vie{q}{}{}| \Vie{p}{0}{})  \BCie{E}{0}{},\\
  \BCie{E}{2}{-} (\Vie{q}{}{}) &= \Vie{T}{}{} ( \Vie{q}{}{}| \Vie{p}{0}{}) \BCie{E}{0}{}.
\end{align}
\label{RTdef}%
\end{subequations}
The reflection and transmission amplitudes are therefore described by 2$\times$2 matrices of the form,
\begin{equation}
\Vie{X}{}{} = \begin{pmatrix}
  X_{pp} & X_{ps}  \\
  X_{sp} & X_{ss}
 \end{pmatrix} ,
\end{equation}
with $\Vie{X}{}{} = \Vie{R}{}{}$ or $\Vie{T}{}{}$.
From a physical point of view, the complex amplitude $R_{\alpha \beta}( \Vie{q}{}{}| \Vie{p}{0}{})$ (resp. $T_{\alpha \beta}( \Vie{q}{}{}| \Vie{p}{0}{})$) for $\alpha, \beta \in \{p,s\}$ is the field amplitude for the reflected light (resp. transmitted) with lateral wave vector $\Vie{q}{}{}$ in the polarization state $\alpha$ from a unit incident field with lateral wave vector $\Vie{p}{0}{}$  in the polarization state $\beta$.
The reflection and transmission amplitudes are then the unknowns in our scattering problem. The equations we need to solve are deduced from the transfer equations, Eqs.~(\ref{rrefinal}) and (\ref{rrefinalswitch}), by applying them respectively at $a_2= +$ and $a_1=-$ and by using Eqs.~(\ref{sommerfeld}) and (\ref{RTdef}) for the model of the field expansion.
This yields the following two decoupled integral equations for the reflection and transmission amplitudes, the so-called reduced Rayleigh equations, that can be written in the following general form, for $\Vie{X}{}{} = \Vie{R}{}{}$ or $\Vie{T}{}{}$:
\begin{equation}
\int  \mathbf{M}_\mathbf{X} (\Vie{p}{}{}|\Vie{q}{}{}) \: \mathbf{X} (\Vie{q}{}{}|\Vie{p}{0}{}) \dtwopi{q}{}= - \mathbf{N}_\mathbf{X} (\Vie{p}{}{}|\Vie{p}{0}{}) \: ,
\label{RREint}
\end{equation}
where the matrices $\mathbf{M}_\mathbf{X}$ and $\mathbf{N}_\mathbf{X}$ are given by
\begin{subequations}
\begin{align}
  \mathbf{M}_\mathbf{R} (\Vie{p}{}{}|\Vie{q}{}{}) &= \Cie{J}{2,1}{+,+}\ofpq \: \Vie{M}{2,1}{+,+}\ofpq\\
  \mathbf{M}_\mathbf{T} (\Vie{p}{}{}|\Vie{q}{}{}) &= \Cie{J}{1,2}{-,-}\ofpq \: \Vie{M}{1,2}{-,-}\ofpq\\
  \mathbf{N}_\mathbf{R} (\Vie{p}{}{}|\Vie{q}{}{}) &= \Cie{J}{2,1}{+,-}\ofpq \: \Vie{M}{2,1}{+,-}\ofpq\\
  \mathbf{N}_\mathbf{T} (\Vie{p}{}{}|\Vie{q}{}{}) &= \frac{2 n_1 n_2 \alpha_1(\Vie{p}{}{})}{\epsilon_2 - \epsilon_1} (2 \pi)^2 \delta(\Vie{p}{}{} - \Vie{q}{}{}) \Vie{I}{2}{},
\end{align}
\label{matrices}%
\end{subequations}
with $\Vie{I}{2}{}$ denoting the 2$\times$2 identity matrix.

\subsection{RRE for scalar waves}

The reduced Rayleigh equations can also be derived for scalar waves satisfying the scalar Helmholtz equation and subjected to various boundary conditions at the interfaces. Here, we focus on scalar waves subjected to the continuity of the field and its normal derivative at the interface. Under these hypotheses, one can derive the corresponding reduced Rayleigh equations which read
\begin{equation}
\int M_X (\Vie{p}{}{} | \Vie{q}{}{}) \: X(\Vie{q}{}{}|\Vie{p}{0}{}) \dtwoq = - N_X (\Vie{p}{}{} | \Vie{p}{0}{}) \: ,
\end{equation}
where $X = R$ or $T$ is either the scalar reflection or transmission amplitude, and the scalar kernels and right-hand-sides are given by Eq.~(\ref{matrices}) where all the $\Vie{M}{l,m}{b,a}$ matrices are replaced by the scalar constant $k_1 k_2 = n_1 n_2 \, \omega^2 / c^2$ and $\Vie{I}{2}{}$ is replaced by the scalar constant 1. We would like to stress that the fact that one can go from the electromagnetic RRE to the scalar RRE by the aforementioned changes is only true for the case where the scalar field is subjected to the continuity of the field and its normal derivative at the surface.
The obtained equations could in principle be used for modeling the scattering of a quantum particle by a surface between two regions of constant potential. For other types of boundary conditions, as for the case of acoustic waves for example \cite{Lekner1987}, one would obtain different expressions. The interested reader will find detailed derivations of the scalar RRE for different boundary conditions as well as the electromagnetic RRE in Ref.~\citenum{banon:thesis}. In this paper, we will use the presented scalar RRE, for which the analysis is simplified compared to the case for electromagnetic waves, to explain the fundamental mechanism of the Yoneda effect.
We will show that the Yoneda effect is present for scalar waves and can be decoupled from additional effects induced by the polarization of electromagnetic waves, such as the Brewster scattering effect. The identified mechanism for scalar waves will then be extended to electromagnetic waves.

\subsection{Perturbative method}
Probably the most common approximate solution to Eq.~\eqref{RREint} is based on a perturbative expansion of the reflection and transmission amplitudes in powers of the interface profile function. This approach, often called ``small amplitude perturbation theory'' (SAPT) or ``small perturbation method'' (SPM), has shown that it is capable of obtaining solutions of the RRE of high qualitative and quantitative predictive power, for interfaces with sufficiently small slopes and amplitudes.
To first order in $\zeta$ for the reflection and transmission amplitudes, the method is often interpreted as a \emph{single scattering approximation}. When implemented to the complete fourth order in the surface profile function for the intensity, i.e. involving terms up to third order in the amplitude, the method has been used to obtain reliable results that also correctly include multiple scattering effects, most notably the backscattering peaks observed in reflection from metallic surfaces \cite{Maradudin1993,Maradudin1995,Mcgurn1996,Simonsen2010}.

To first order in the interface profile function $\zeta$, we have for $\vec{X}=\vec{R}$ or $\vec{T}$ that
\begin{align}
  \mathbf{X} \ofuipi{p}{}{p}{0} \approx \: &\Vie{X}{}{(0)} \ofuipi{p}{}{p}{0} - i \Vie{X}{}{(1)} \ofuipi{p}{}{p}{0} \: ,
\end{align}
where
\begin{subequations}
\begin{align}
  \Vie{R}{}{(0)} \ofuipi{p}{}{p}{0} = \: &(2 \pi)^2 \delta(\Vie{p}{}{} -\Vie{p}{0}{}) \Vie{\boldsymbol \rho}{}{(0)} (\Vie{p}{0}{}) \label{eq:sapt0:r} \: ,\\
  \Vie{T}{}{(0)} \ofuipi{p}{}{p}{0} = \: &(2 \pi)^2 \delta(\Vie{p}{}{} -\Vie{p}{0}{})\Vie{\boldsymbol \tau}{}{(0)} (\Vie{p}{0}{}) \label{eq:sapt0:t}  \: ,\\
  \Vie{R}{}{(1)} \ofuipi{p}{}{p}{0} = \: &\hat{\zeta}(\Vie{p}{}{} -\Vie{p}{0}{}) \Vie{\boldsymbol \rho}{}{(1)} \ofuipi{p}{}{p}{0} \nonumber\\
  = \: [\alpha_1(\Vie{p}{}{}) - \alpha_2(\Vie{p}{}{})] \: &\hat{\zeta}(\Vie{p}{}{} -\Vie{p}{0}{}) \, \boldsymbol{\hat{\rho}}^{(1)} \ofuipi{p}{}{p}{0} \label{eq:sapt1:rr} \: ,\\
   \Vie{T}{}{(1)} \ofuipi{p}{}{p}{0} = \: &\hat{\zeta}(\Vie{p}{}{} -\Vie{p}{0}{}) \Vie{\boldsymbol \tau}{}{(1)} \ofuipi{p}{}{p}{0} \nonumber\\
     = \: [\alpha_1(\Vie{p}{}{}) - \alpha_2(\Vie{p}{}{})] \: &\hat{\zeta}(\Vie{p}{}{} -\Vie{p}{0}{}) \, \boldsymbol{\hat{\tau}}^{(1)} \ofuipi{p}{}{p}{0} \: . \label{eq:sapt1:tt}
\end{align}
\label{1storder:amplitude}%
\end{subequations}
\sloppy Here $\hat{\zeta}$ denotes the Fourier transform of $\zeta$, and $\Vie{\boldsymbol \rho}{}{(0)} (\Vie{p}{0}{})$ and $\Vie{\boldsymbol \tau}{}{(0)} (\Vie{p}{0}{})$ are  matrix-valued amplitudes for the zero order reflection and transmission amplitudes respectively, and are given by
\begin{widetext}
\begin{subequations}
\begin{align}
\Vie{\boldsymbol \rho}{}{(0)} (\Vie{p}{0}{}) &= \frac{\alpha_1(\Vie{p}{0}{}) - \alpha_2(\Vie{p}{0}{})}{\alpha_2(\Vie{p}{0}{}) + \alpha_1(\Vie{p}{0}{})} \: \left[\Vie{M}{2,1}{+,+}\ofuipi{p}{0}{p}{0}\right]^{-1}  \: \Vie{M}{2,1}{+,-}\ofuipi{p}{0}{p}{0} \label{eq:sapt0:ref}\\
\Vie{\boldsymbol \tau}{}{(0)} (\Vie{p}{0}{}) &= \frac{2 \, \sqrt{\epsilon_1 \epsilon_2} \alpha_1(\Vie{p}{0}{})}{\epsilon_2 - \epsilon_1} \: [\alpha_2(\Vie{p}{0}{}) - \alpha_1(\Vie{p}{0}{})] \: \left[ \Vie{M}{1,2}{-,-} \ofuipi{p}{0}{p}{0}\right]^{-1} \: . \label{eq:sapt0:tra}
\end{align}
\end{subequations}
The matrix-valued amplitudes $\Vie{\boldsymbol \rho}{}{(1)} \ofuipi{p}{}{p}{0}$, $\boldsymbol{\hat{\rho}}^{(1)} \ofuipi{p}{}{p}{0}$, $\Vie{\boldsymbol \tau}{}{(1)} \ofuipi{p}{}{p}{0}$, and $\boldsymbol{\hat{\tau}}^{(1)} \ofuipi{p}{}{p}{0}$ for the first order terms are by
\begin{subequations}
\begin{align}
\Vie{\boldsymbol \rho}{}{(1)} \ofuipi{p}{}{p}{0}  &= [\alpha_1\of{p} - \alpha_2\of{p}] \:\left[ \Vie{M}{2,1}{+,+}\ofuipi{p}{}{p}{}\right]^{-1} \: \Big[  \Vie{M}{2,1}{+,-}\ofuipi{p}{}{p}{0}  + \Vie{M}{2,1}{+,+}\ofuipi{p}{}{p}{0}  \:  \boldsymbol{\rho}^{(0)} (\Vie{p}{0}{}) \Big] \label{eq:sapt1:ref}\\
\Vie{\boldsymbol \tau}{}{(1)} \ofuipi{p}{}{p}{0} &=  [\alpha_1\of{p} - \alpha_2\of{p}] \: \left[\Vie{M}{1,2}{-,-}\ofuipi{p}{}{p}{} \right]^{-1} \: \Vie{M}{1,2}{-,-}\ofuipi{p}{}{p}{0} \: \boldsymbol{\tau}^{(0)} (\Vie{p}{0}{}) \: . \label{eq:sapt1:tra}
\end{align}
\end{subequations}
\end{widetext}
In Eqs.~(\ref{eq:sapt1:rr}) and (\ref{eq:sapt1:tt}), we have given two alternative factorizations of the first order reflection and transmission amplitudes. The factorization including the caret amplitudes is the most appropriate for physical interpretation, while the factorization including the non-caret amplitudes simply aims at separating $\hat{\zeta}$, which is the only factor depending on the specific realization of the surface profile, from the remaining profile-independent amplitude factor. All the aforementioned amplitudes are derived in detail in Appendix~A.

\subsection{Observables}

\sloppy The observables of interest in this study are the so-called \emph{diffuse} or \emph{incoherent component of the mean differential reflection and transmission coefficients} (MDRC and MDTC) denoted $\left\langle \partial R_{\alpha \beta} (\vec{p}|\vec{p}_0)/\partial\Omega_r\right\rangle_{\mathrm{incoh}}$ and $\left\langle \partial T_{\alpha \beta} (\vec{p}|\vec{p}_0) / \partial \Omega_t \right\rangle_{\mathrm{incoh}}$, respectively.
They are both defined as the ensemble average over realizations of the surface profile of the incoherent component of the radiated reflected or transmitted flux of an $\alpha$-polarized wave around a direction given by $\Vie{k}{1}{+} (\Vie{p}{}{})$ or $\Vie{k}{2}{-} (\Vie{p}{}{})$ per unit incident flux of a $\beta$-polarized plane wave with wave vector $\Vie{k}{1}{-} (\Vie{p}{0}{})$, per unit solid angle. Based on the reflection and transmission amplitudes found to first order in $\zeta$, the incoherent component of the MDRC and MDTC can be expressed as
\begin{align}
&\left\langle \frac{\partial R_{\alpha \beta} }{\partial \Omega_r} (\Vie{p}{}{} | \Vie{p}{0}{}) \right\rangle_{\mathrm{incoh}} = \: \epsilon_1 \left(\frac{\omega}{2 \pi c} \right)^2 \: \frac{\cos^2 \theta_r}{\cos \theta_0} \nonumber \\
&\times \sigma^2 \: g (\Vie{p}{}{} -\Vie{p}{0}{}) \: \left|\rho_{\alpha \beta}^{(1)} \ofuipi{p}{}{p}{0} \right|^2  ,
\label{eq:incoMDRC:sapt}
\end{align}
and
\begin{align}
&\left\langle \frac{\partial T_{\alpha \beta} }{\partial \Omega_t} (\Vie{p}{}{} | \Vie{p}{0}{}) \right\rangle_{\mathrm{incoh}} = \: \frac{\epsilon_2^{3/2}}{\epsilon_1^{1/2}} \left(\frac{\omega}{2 \pi c} \right)^2 \: \frac{\cos^2 \theta_t}{\cos \theta_0} \nonumber\\
&\times \sigma^2 \: g (\Vie{p}{}{} -\Vie{p}{0}{}) \:  \left|\tau_{\alpha \beta}^{(1)} \ofuipi{p}{}{p}{0} \right|^2  .
\label{eq:incoMDTC:sapt}
\end{align}
The detailed derivation of the Eqs.~(\ref{eq:incoMDRC:sapt}) and (\ref{eq:incoMDTC:sapt}) can be found in Appendix~B. 
 The definition of the angles of incidence and scattering can be deduced from Fig.~\ref{fig:system}.

\section{Results and discussion}\label{sec:results}

\begin{figure*}[t]
  \centering
  \includegraphics[width=.325\linewidth , trim= 0.cm 0.cm 0.9cm 0.cm,clip]{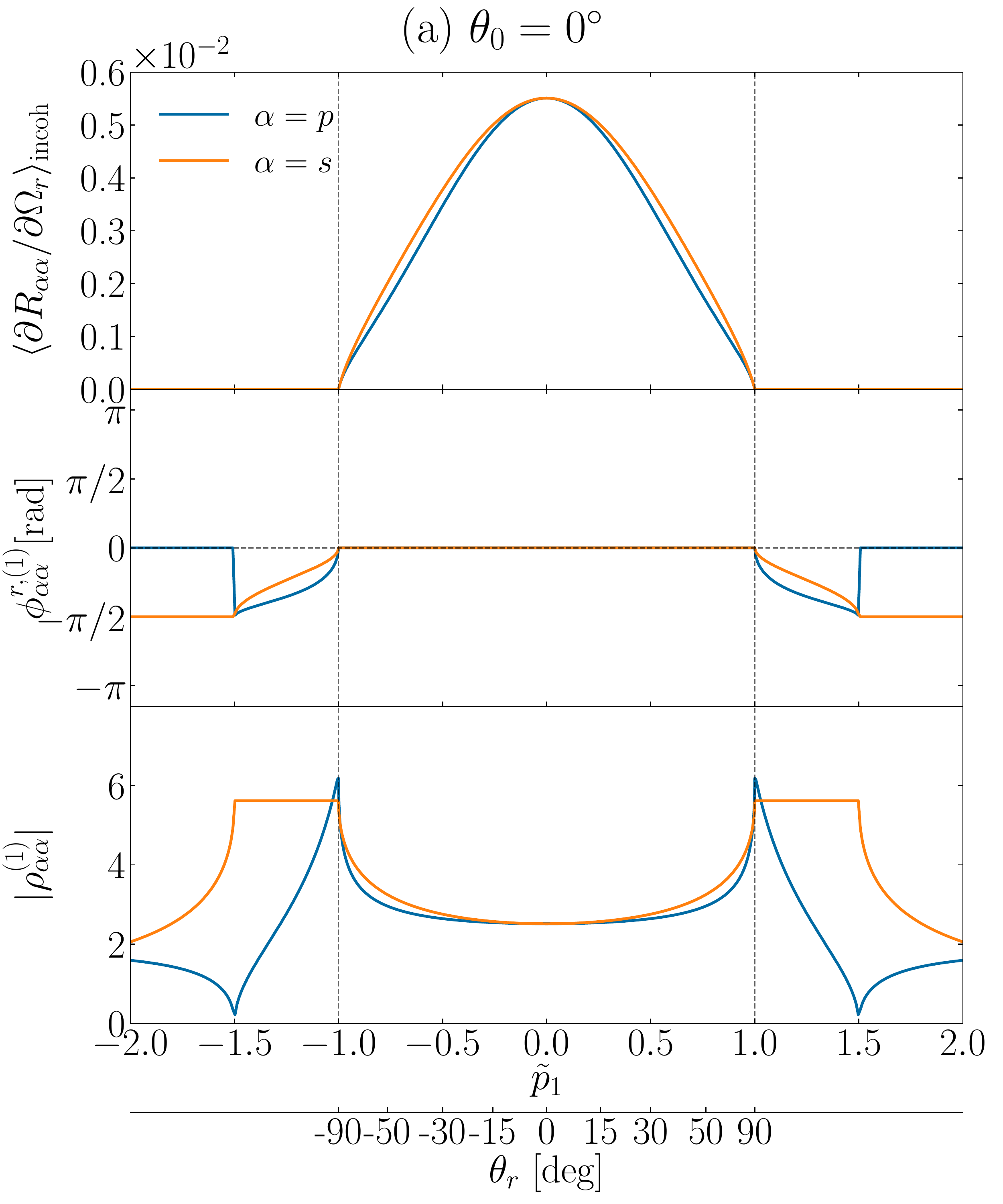}
  \includegraphics[width=.325\linewidth , trim= 0.cm 0.cm 0.9cm 0.cm,clip]{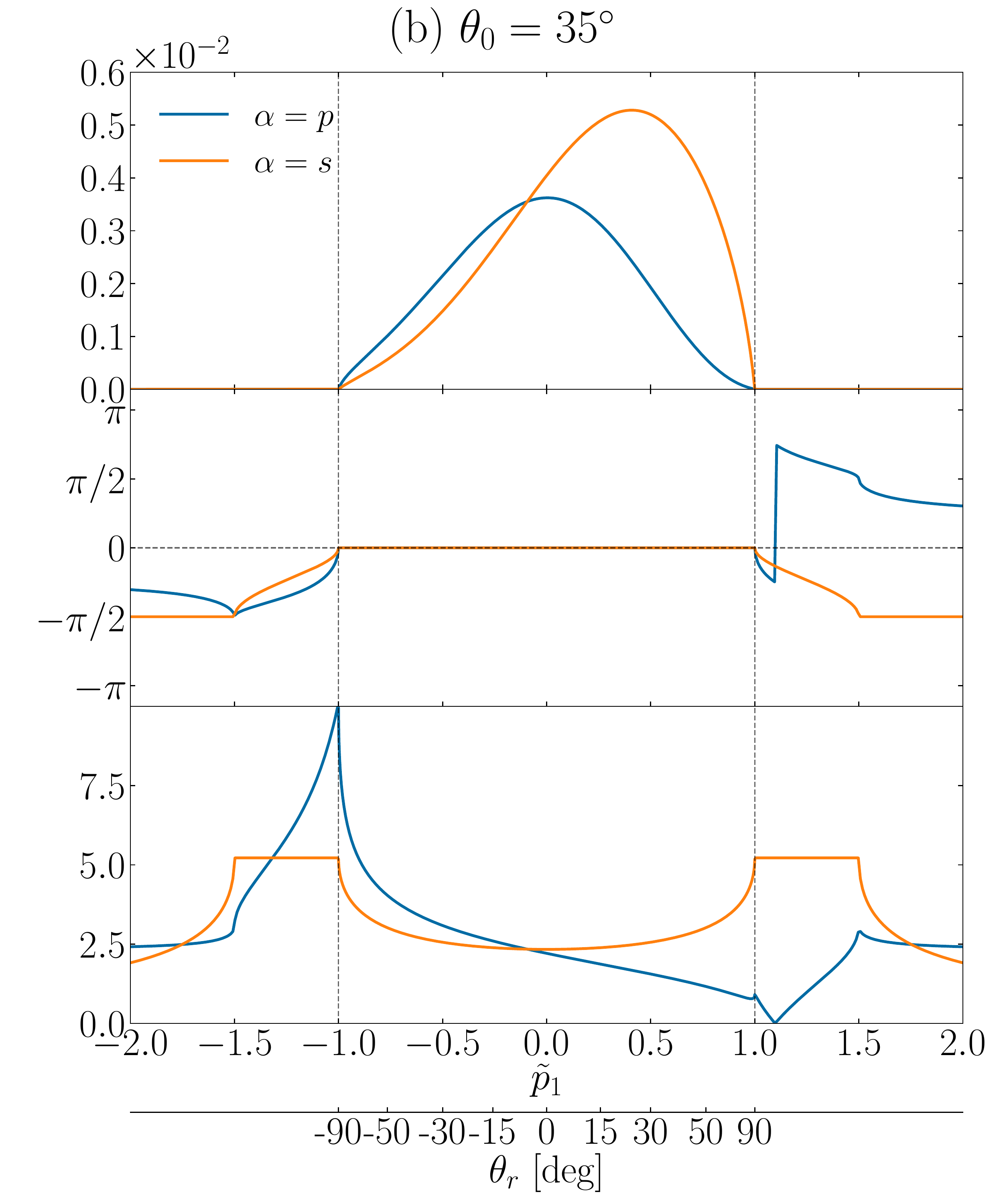}
  \includegraphics[width=.325\linewidth , trim= 0.cm 0.cm 0.9cm 0.cm,clip]{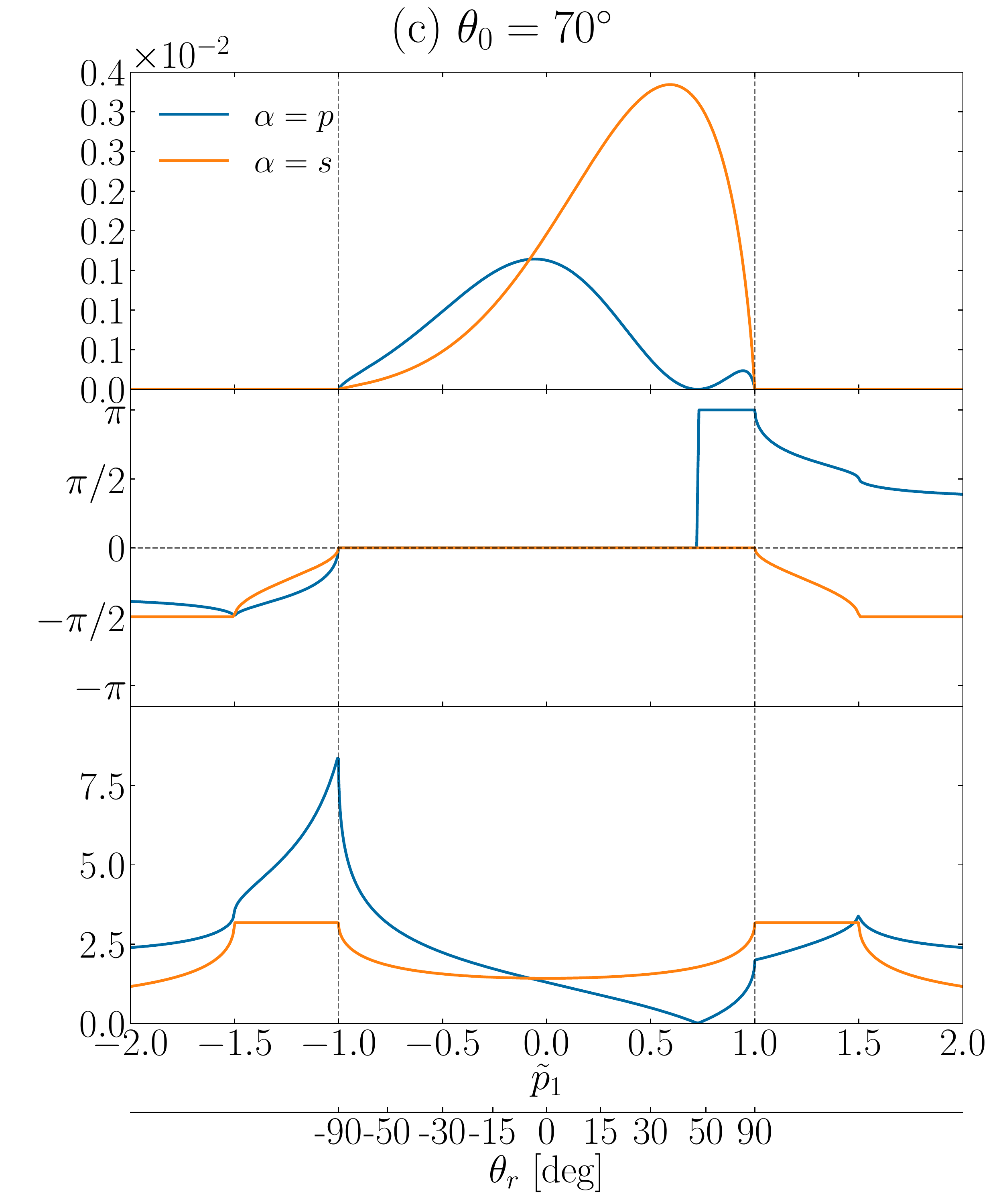}
  \includegraphics[width=.325\linewidth , trim= 0.cm 0.cm 0.9cm 0.cm,clip]{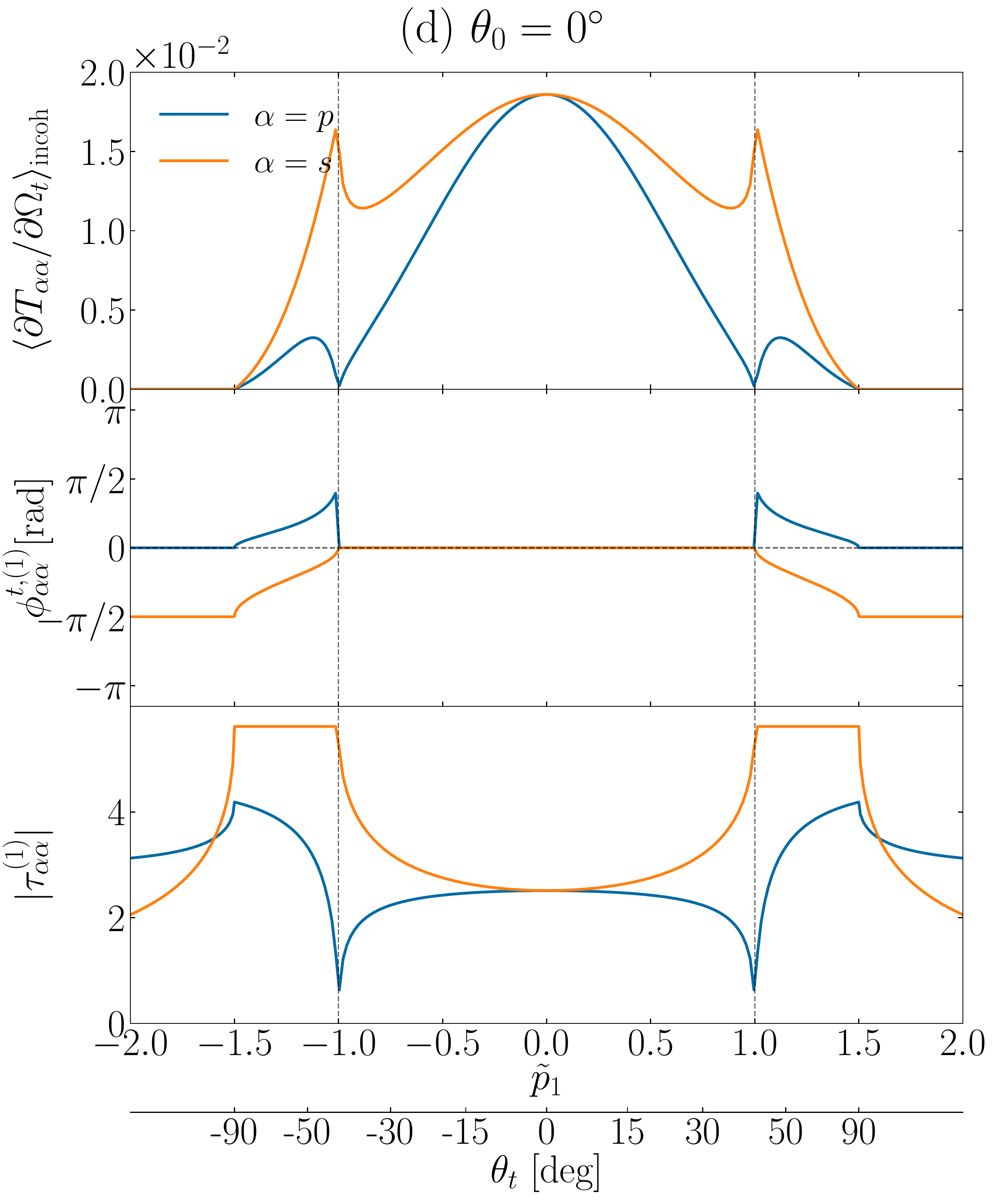}
  \includegraphics[width=.325\linewidth , trim= 0.cm 0.cm 0.9cm 0.cm,clip]{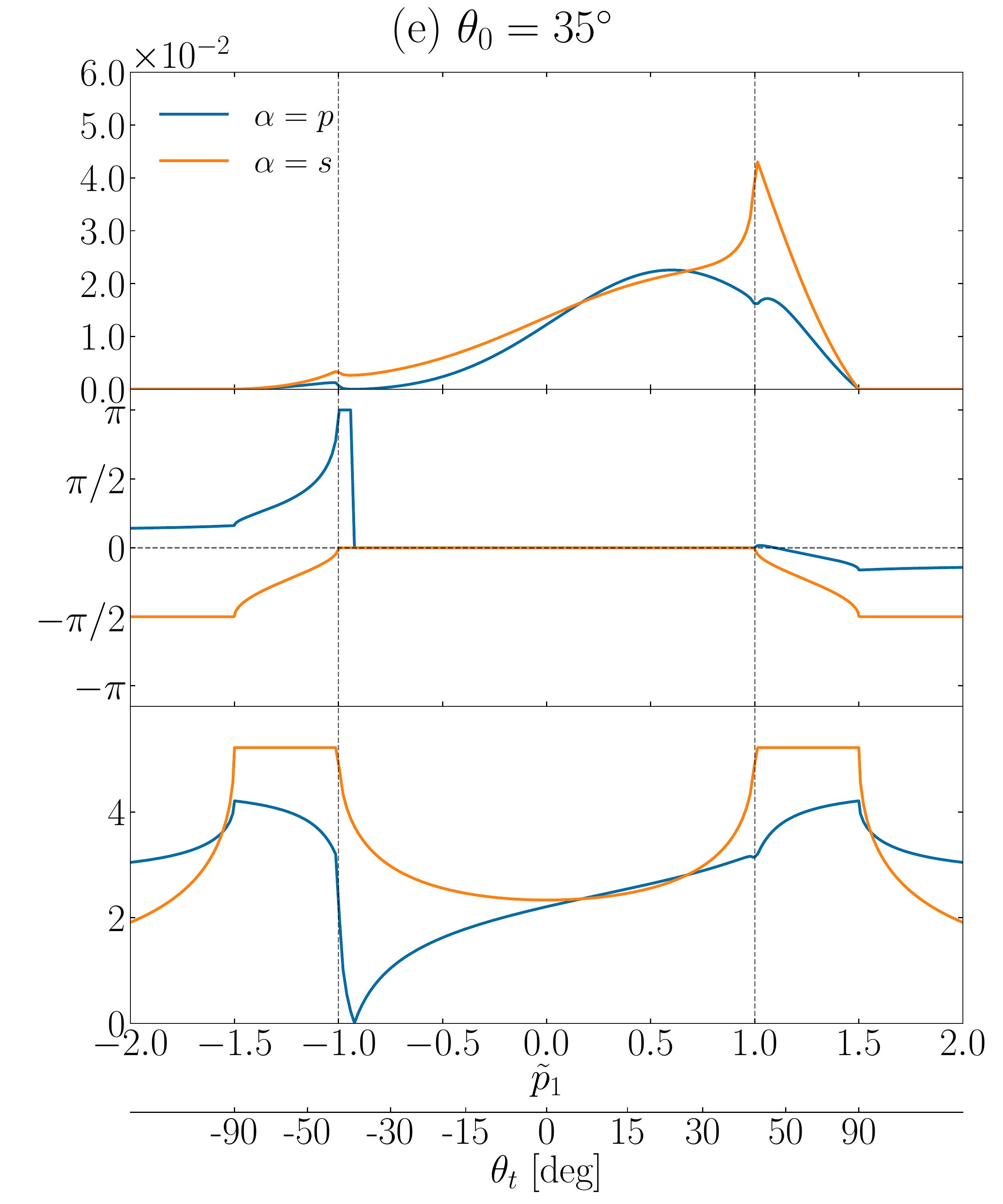}
  \includegraphics[width=.325\linewidth , trim= 0.cm 0.cm 0.9cm 0.cm,clip]{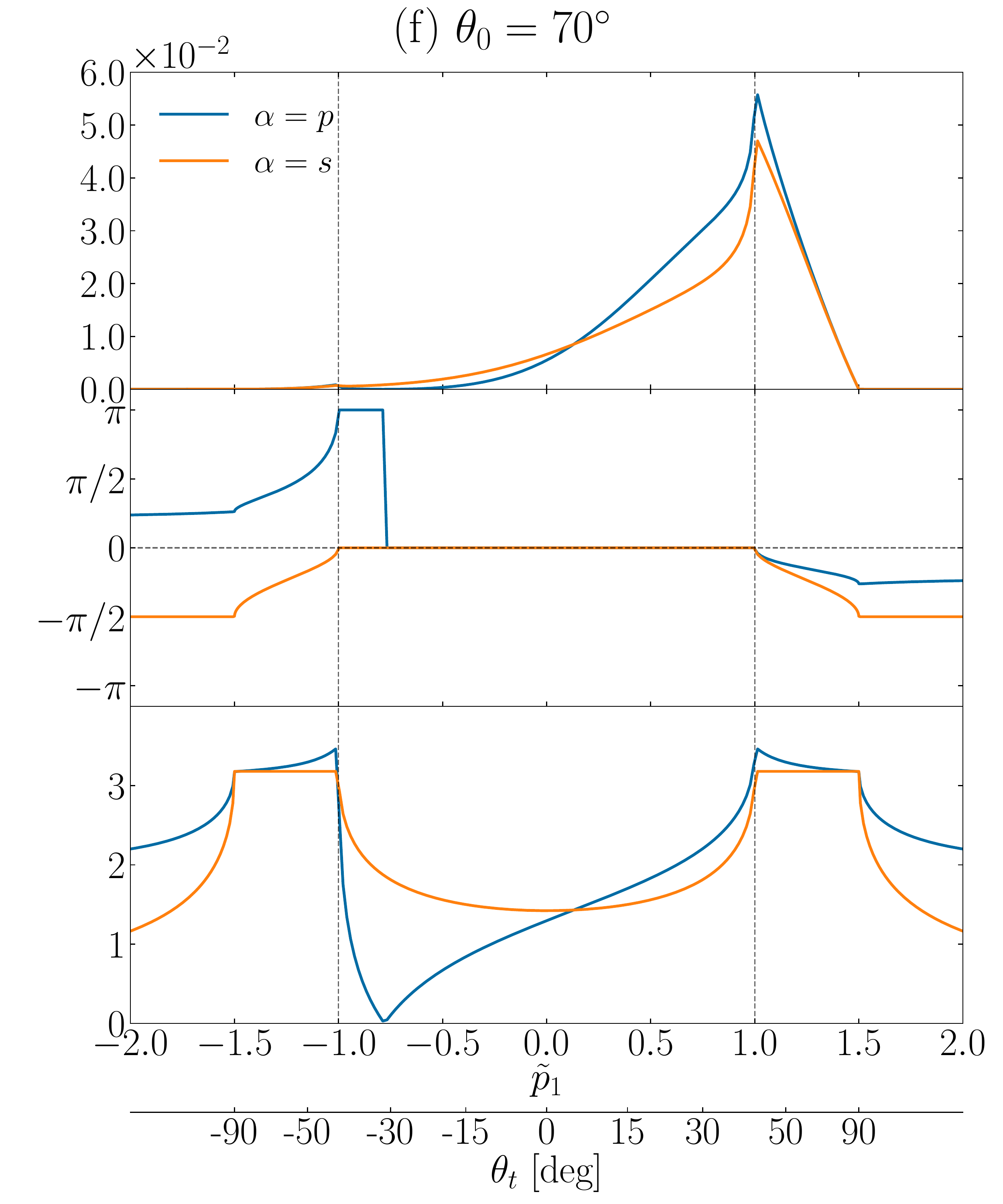}
  \caption{The incoherent component of the MDRC (top row) and MDTC (bottom row) for light incident from vacuum [$\ve_1=1.0$] onto a randomly rough interface with glass [$\ve_2=2.25$], for in-plane co-polarized scattering, as a function of the lateral component of the wave vector of scattering $\tilde{p}_1 = p_1 c / \omega$ (i.e. $p_1$ in unit of $\omega / c$) or polar angle of scattering $\theta_{r,t}$. The polar angle of incidence is indicated on top of each subfigure. The argument and the modulus of $\rho_{\alpha \alpha}^{(1)}$ and $\tau_{\alpha \alpha}^{(1)}$ are indicated in the middle and bottom section of each subfigure respectively. Note that we have adopted here the convention that negative $\theta_{r,t}$ values correspond to $\theta_{r,t} >0$ according Fig.~\ref{fig:system} but for $\phi = \ang{180}$. The vertical lines indicate $| \Vie{p}{}{} | = k_{\min}$.}
  \label{fig:sapt_mdxc_incvacuum}
\end{figure*}

\begin{figure*}[t]
  \centering
  \includegraphics[width=.325\linewidth, trim= 0.cm 0.cm 0.9cm 0.cm,clip]{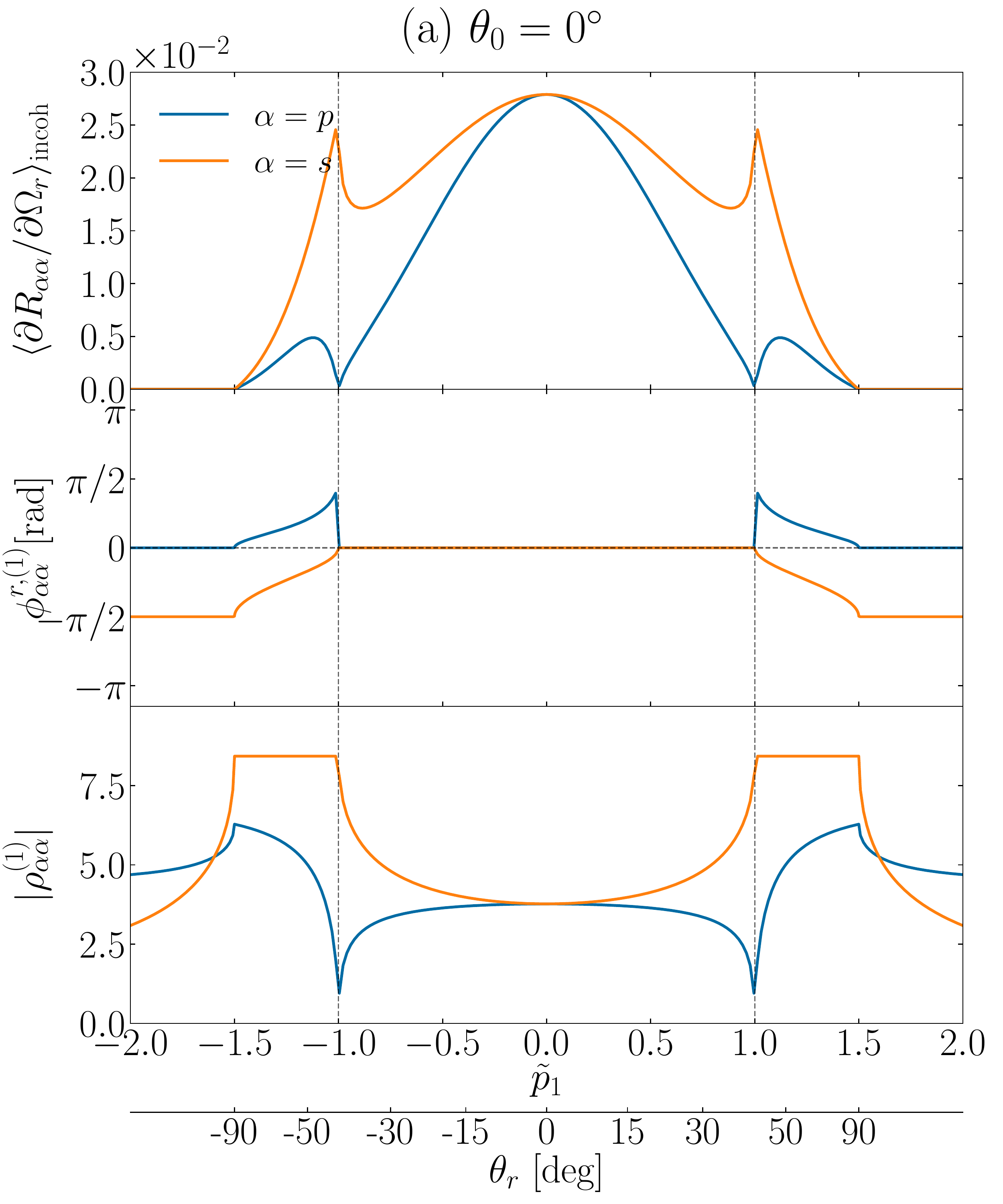}
  \includegraphics[width=.325\linewidth, trim= 0.cm 0.cm 0.9cm 0.cm,clip]{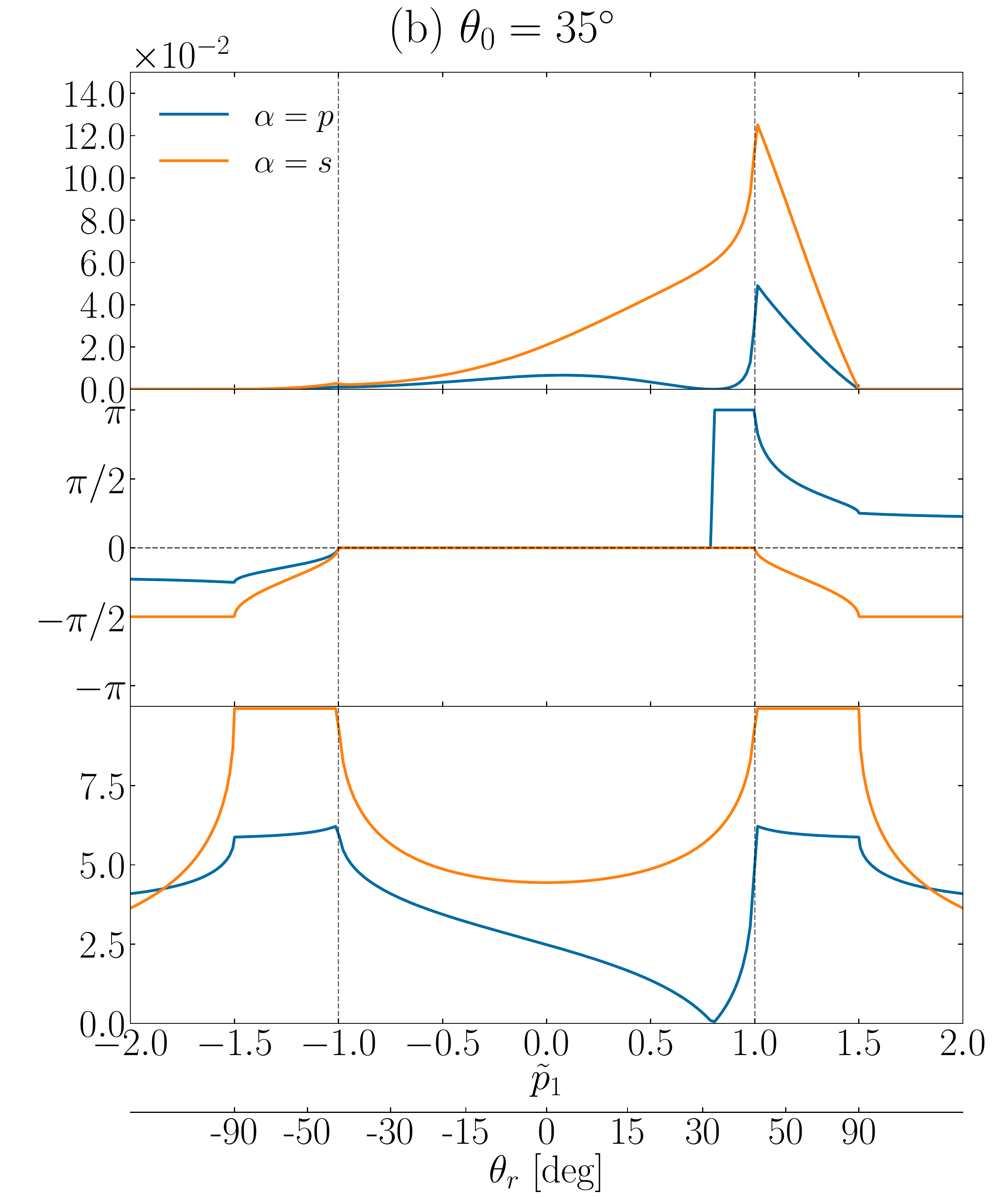}
  \includegraphics[width=.325\linewidth, trim= 0.cm 0.cm 0.9cm 0.cm,clip]{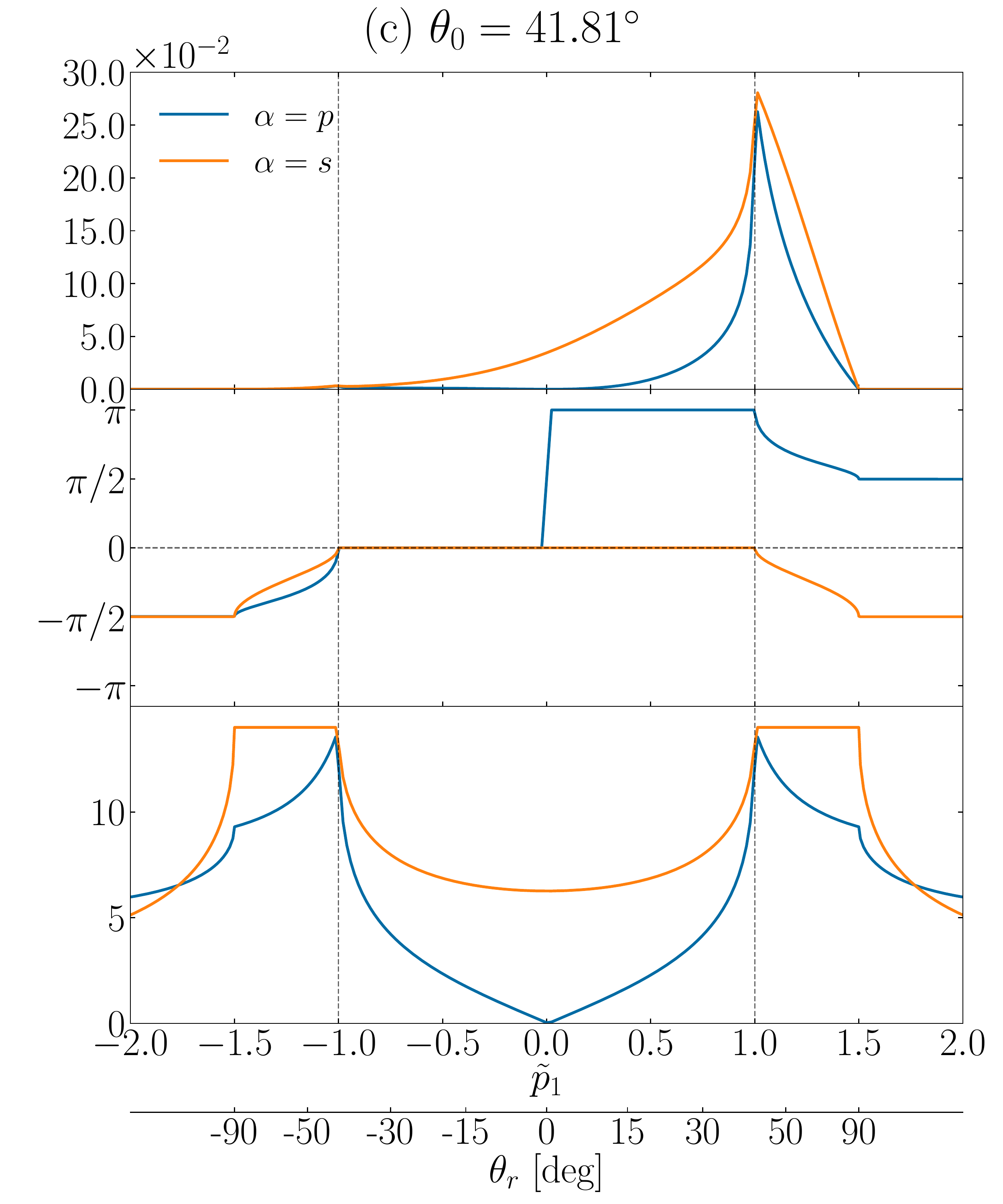}
  \includegraphics[width=.325\linewidth, trim= 0.cm 0.cm 0.9cm 0.cm,clip]{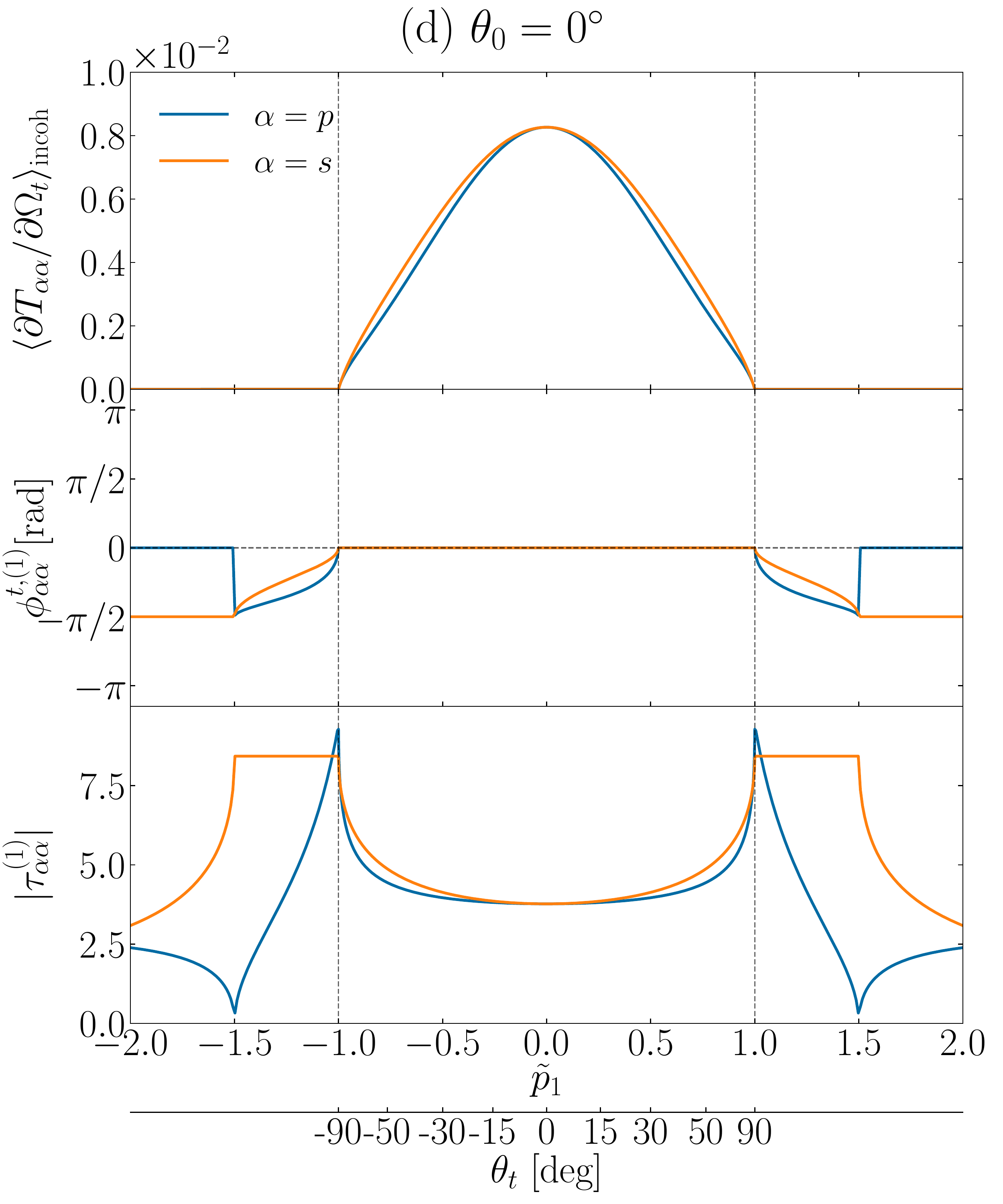}
  \includegraphics[width=.325\linewidth, trim= 0.cm 0.cm 0.9cm 0.cm,clip]{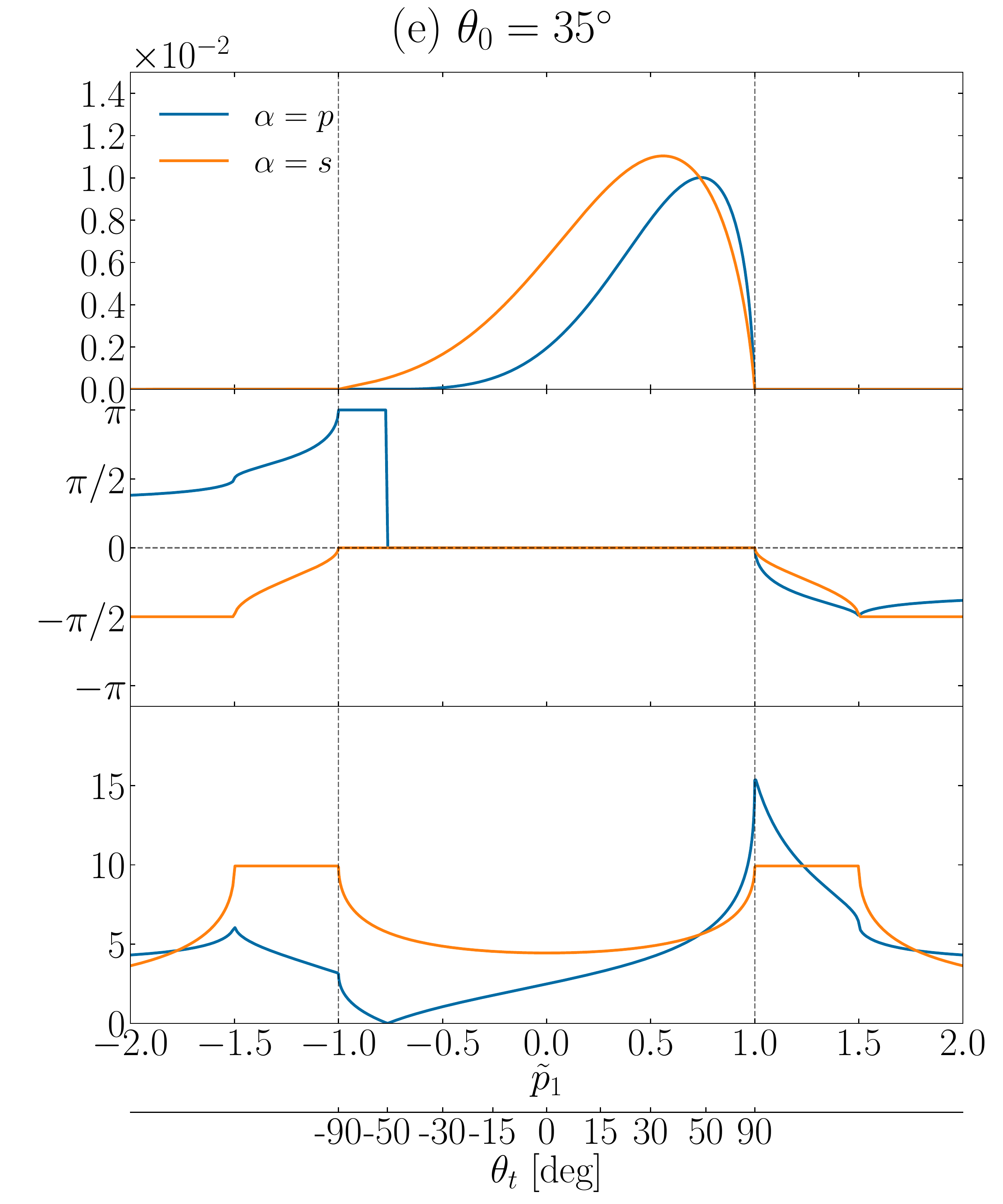}
  \includegraphics[width=.325\linewidth, trim= 0.cm 0.cm 0.9cm 0.cm,clip]{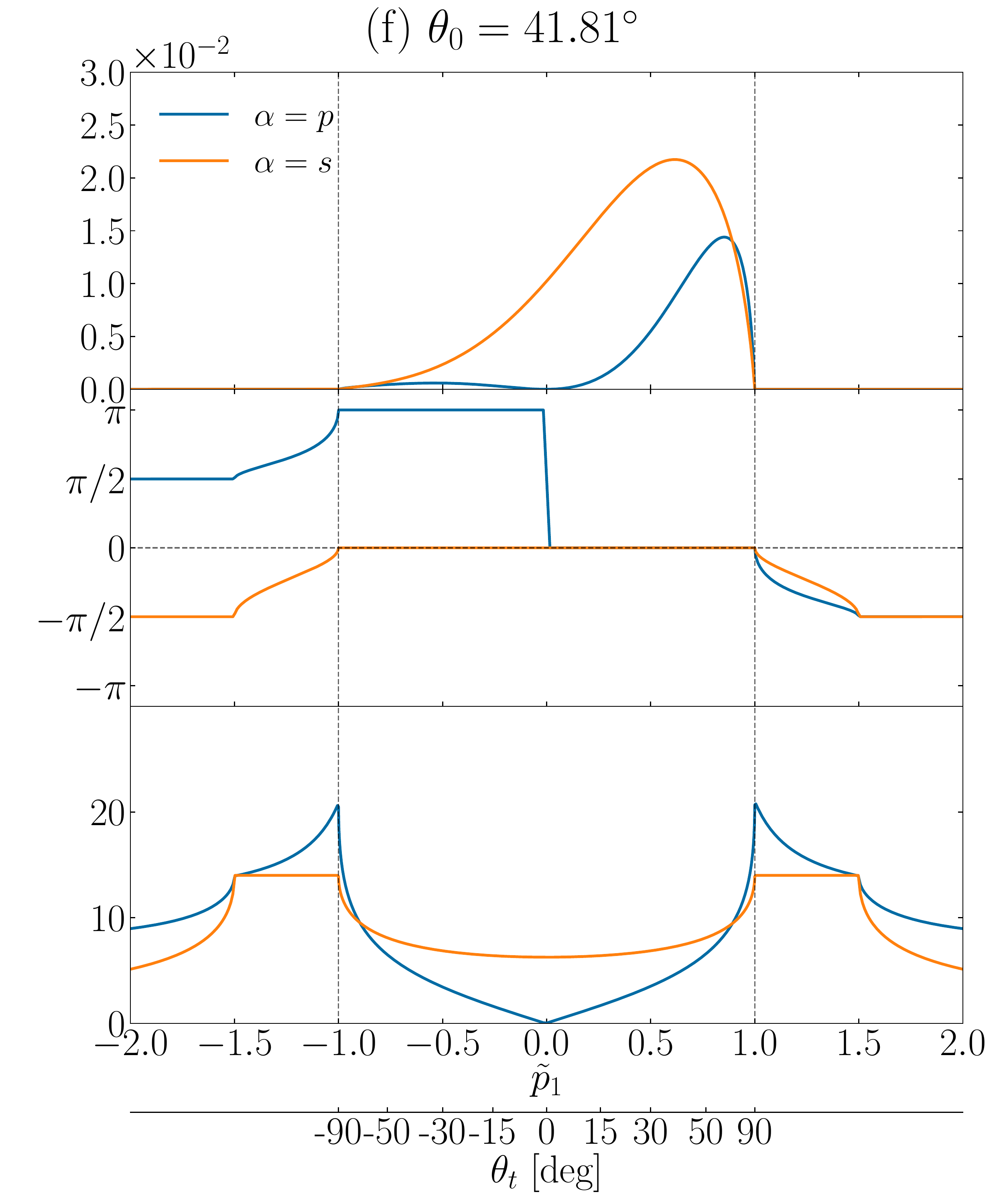}
  \caption{Same as Fig.~\ref{fig:sapt_mdxc_incvacuum}, but for light incident from glass [$\ve_1=2.25$] onto a randomly rough interface with vacuum [$\ve_2=1.0$].}
  \label{fig:sapt_mdxc_incglass}
\end{figure*}

\begin{figure*}[t]
  \centering
  \includegraphics[width=.325\linewidth, trim= 0.cm 0.cm 0.9cm 0.cm,clip]{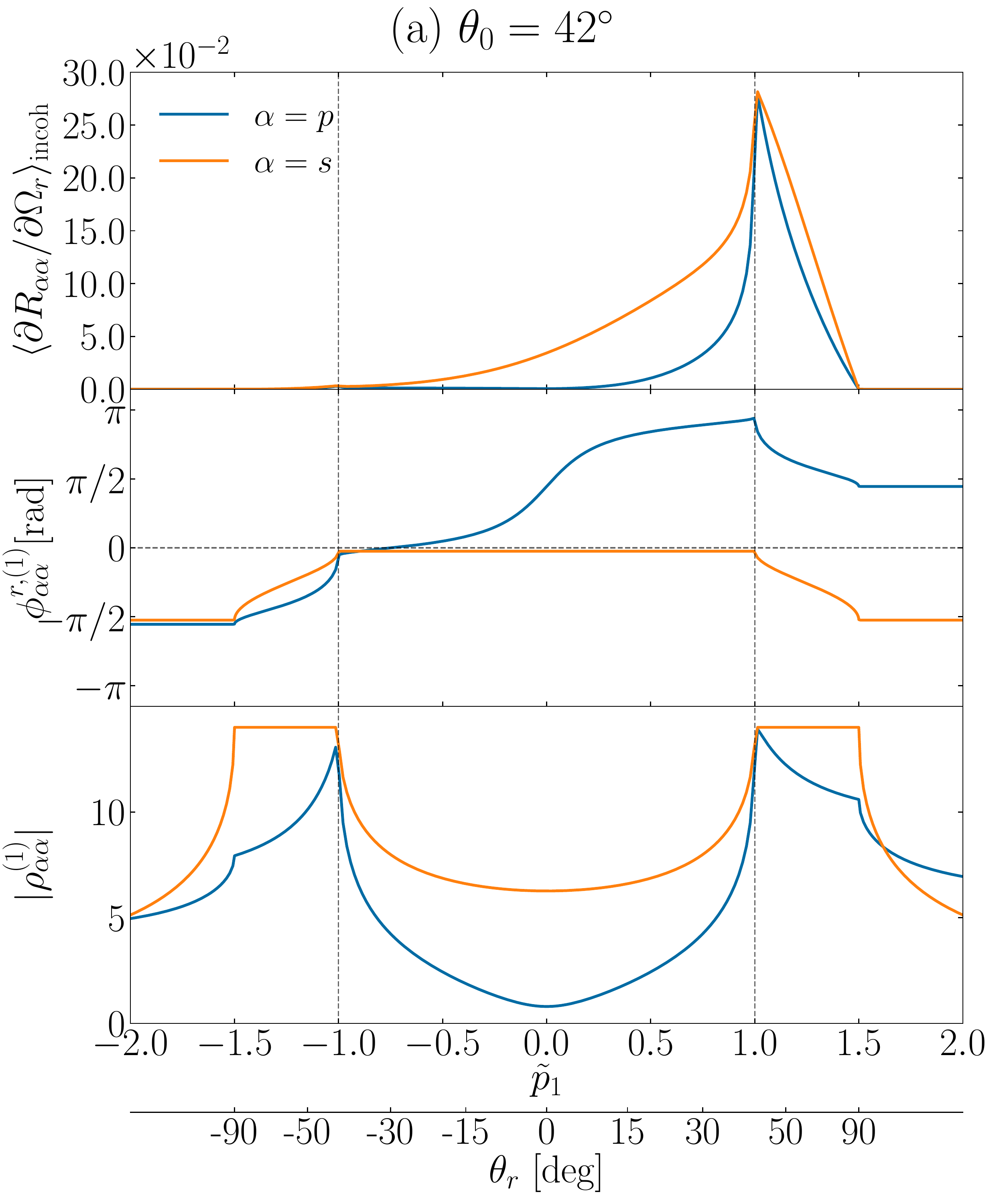}
  \includegraphics[width=.325\linewidth, trim= 0.cm 0.cm 0.9cm 0.cm,clip]{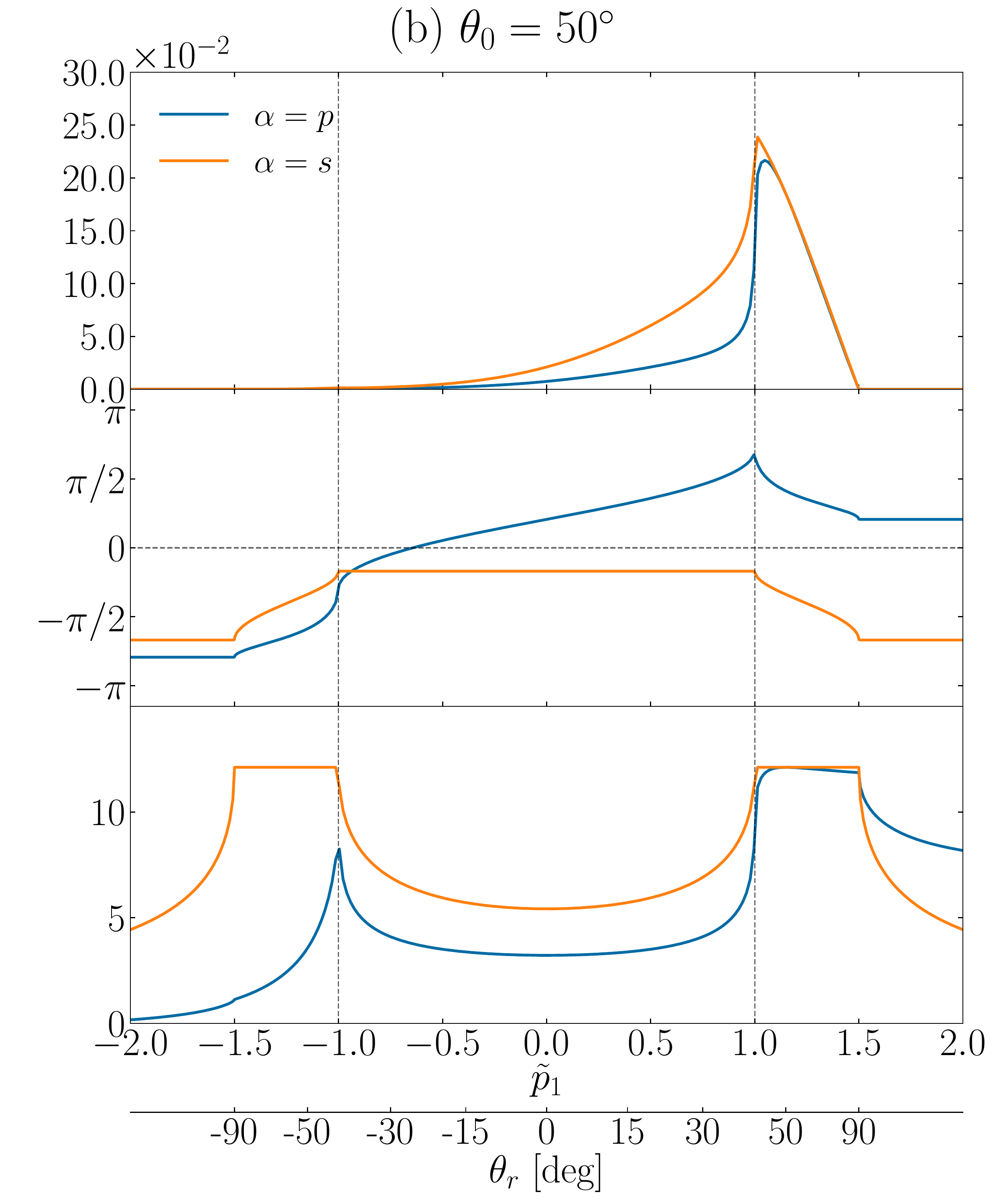}
  \includegraphics[width=.325\linewidth, trim= 0.cm 0.cm 0.9cm 0.cm,clip]{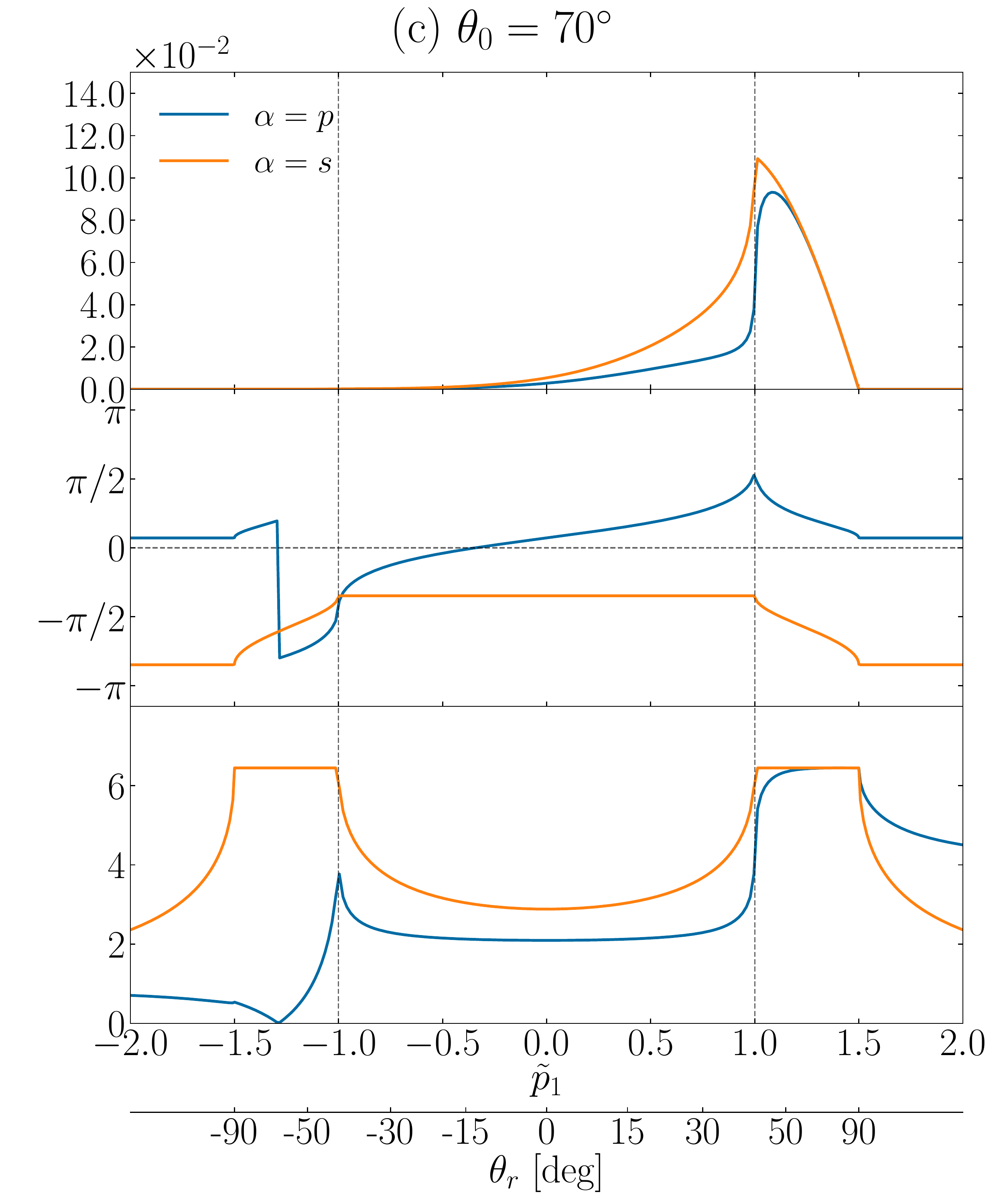}
  \includegraphics[width=.325\linewidth, trim= 0.cm 0.cm 0.9cm 0.cm,clip]{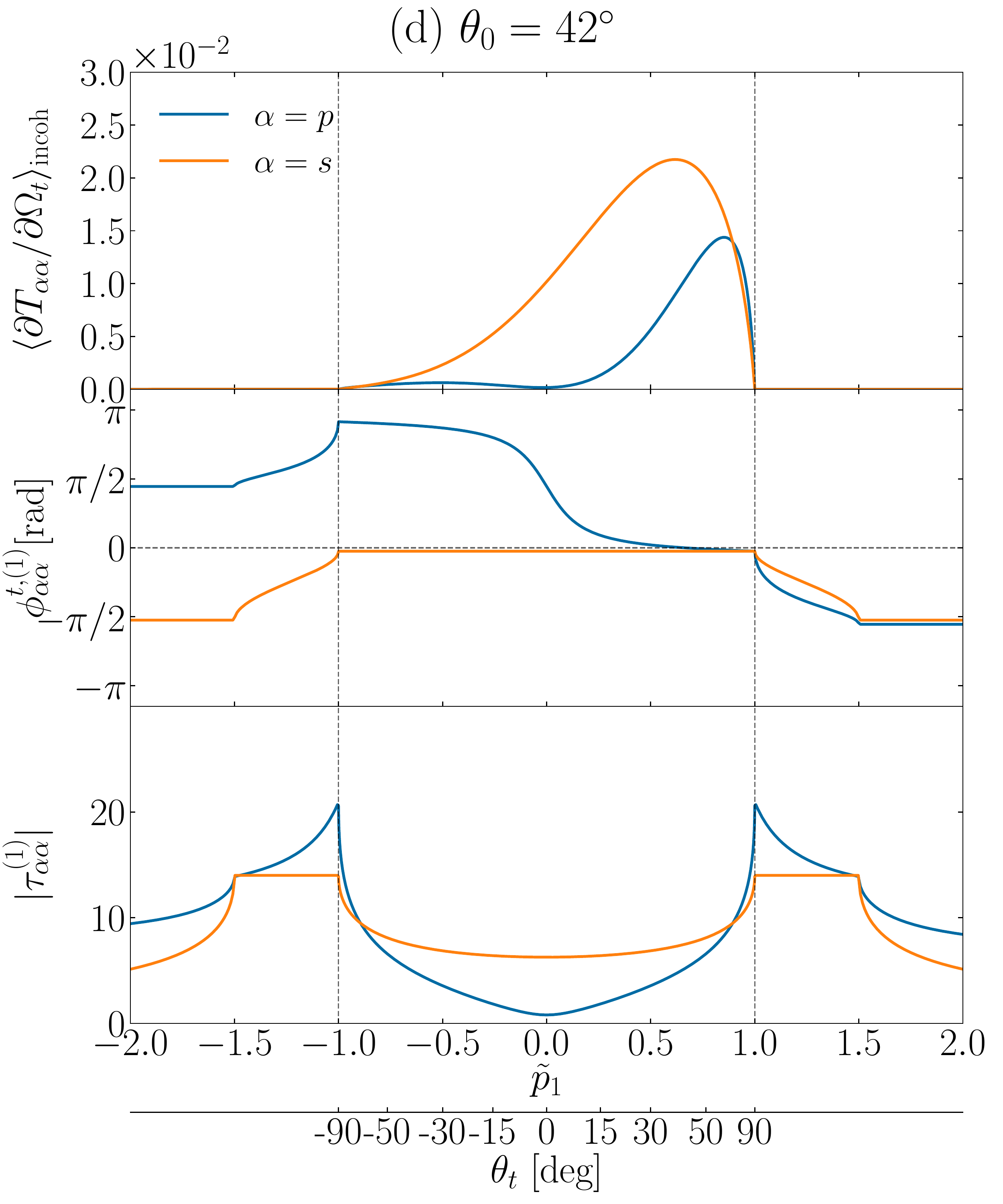}
  \includegraphics[width=.325\linewidth, trim= 0.cm 0.cm 0.9cm 0.cm,clip]{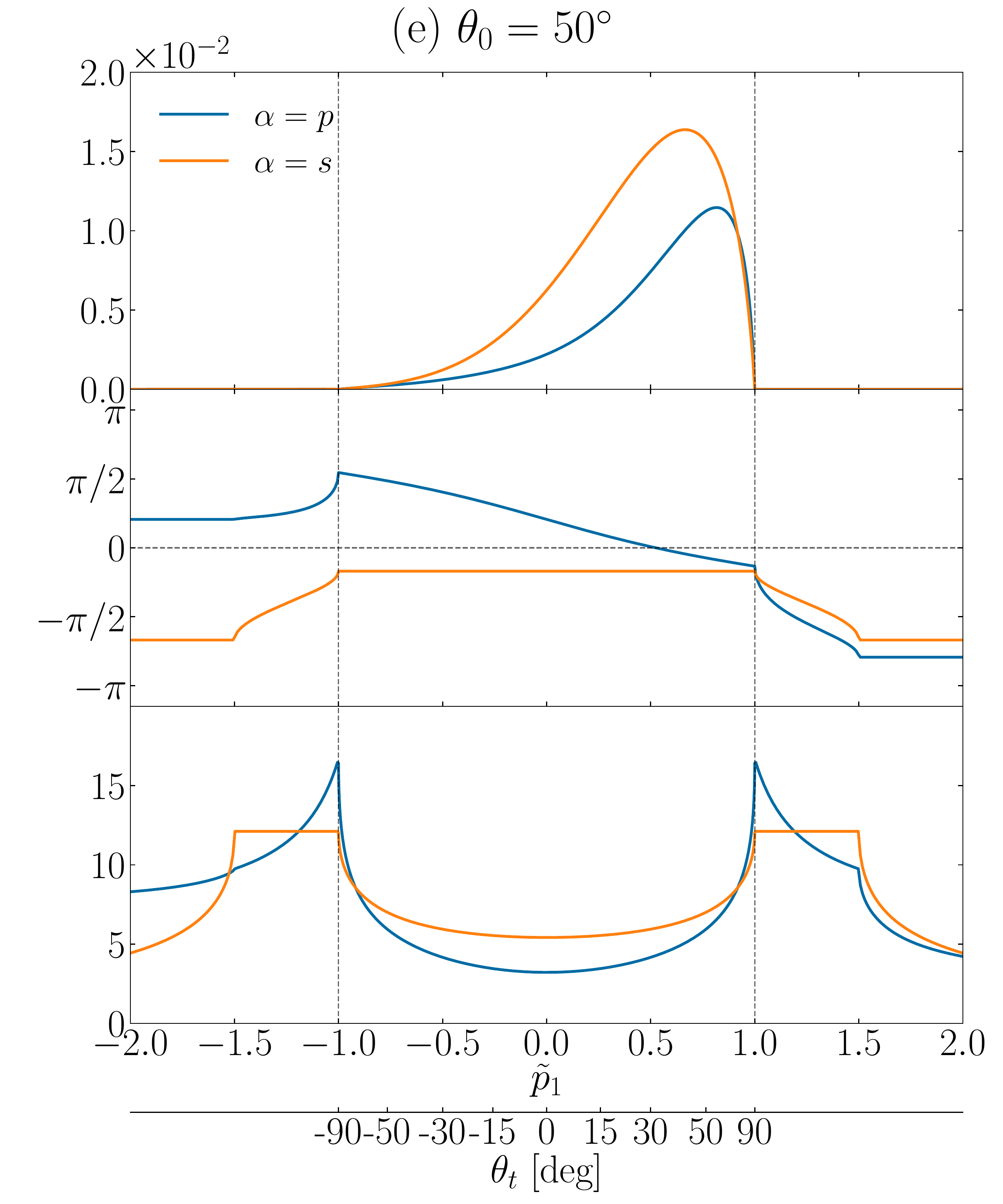}
  \includegraphics[width=.325\linewidth, trim= 0.cm 0.cm 0.9cm 0.cm,clip]{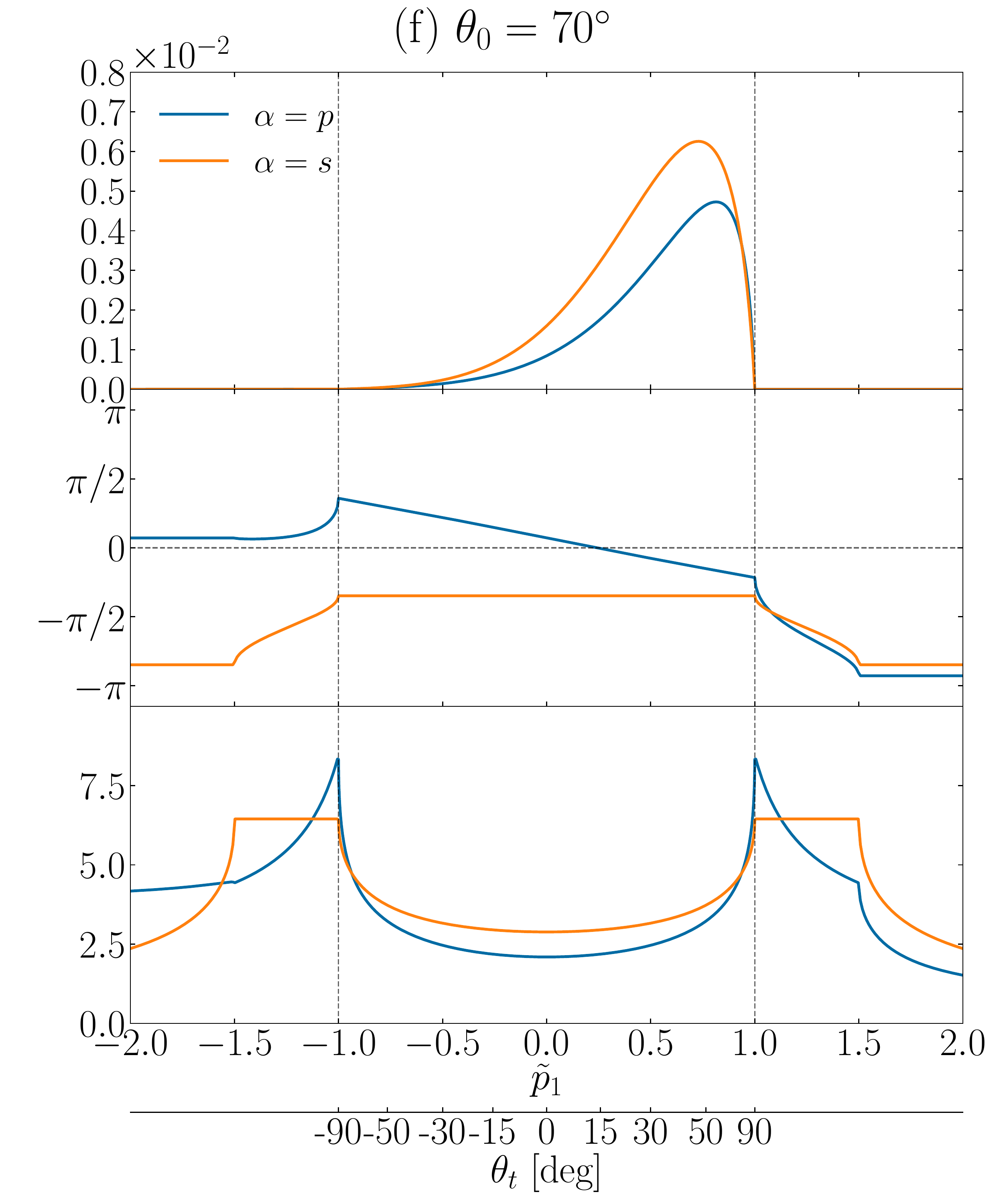}
  \caption{Same as Fig.~\ref{fig:sapt_mdxc_incglass}, but for additional polar angles of incidence $\theta_0$.}
  \label{fig:sapt_mdxc_incglass_contd}
\end{figure*}

In order to study the phenomena observed in the scattering of light from weakly rough dielectric interfaces, we choose to base our discussion on results obtained through small amplitude perturbation theory (SAPT) to lowest non-zero order in the interface profile function, Eqs.~\eqref{1storder:amplitude}. For sufficiently smooth interfaces this approximation has previously been compared to numerical non-perturbative solutions to the reduced Rayleigh equations, where it has been shown to adequately model the phenomena of both the Brewster scattering angles and the Yoneda peaks \cite{Hetland2016a,Hetland2017}.
We will start our investigations with a summary of the features observed in the main physical observables, the MDRC and MDTC [Eqs.~(\ref{eq:incoMDRC:sapt}) and (\ref{eq:incoMDTC:sapt})], followed by more in-depth analyses and discussions from a physics point of view. In figures, the wave vectors are normalized by $\omega / c$, i.e. we define $\tilde{\Vie{p}{}{}} = \Vie{p}{}{} c / \omega$, $\Vie{\tilde{k}}{j}{\pm} = \Vie{k}{j}{\pm} c  / \omega $.

\subsection{Phenomenology of the Yoneda and Brewster scattering effects}\label{sec:pheno}
The top panel of each subfigure in Fig.~\ref{fig:sapt_mdxc_incvacuum} presents results based on  Eqs.~(\ref{eq:incoMDRC:sapt}) and (\ref{eq:incoMDTC:sapt}) for the contribution to the co-polarized diffuse component of the MDRC and MDTC in the plane of incidence~($\vecUnit{p}\parallel\vecUnit{p}_0$), for a configuration where light is incident from vacuum [$\ve_1=1$] onto a two-dimensional randomly rough interface with glass [$\ve_2=2.25$]. The incident light was assumed to be a \textit{p}- or \textit{s}-polarized plane wave of wavelength $\lambda=\SI{632.8}{\nm}$ in vacuum.
In the current work all results presented for randomly rough interfaces consist of interfaces defined by an isotropic Gaussian height distribution with rms height $\sigma=\SI{32}{nm}=\lambda/20$ and an isotropic Gaussian correlation function of transverse correlation length $a=\SI{211}{nm}=\lambda/3$.


For normal incidence [$\theta_0=\ang{0}$, Fig.~\ref{fig:sapt_mdxc_incvacuum}(a)] the MDRC distributions are nearly featureless. The differences in the scattered intensities observed for \textit{p}- and \textit{s}-polarized incident light are very small. Note that the scattered intensity is zero beyond the limit of propagation in the medium of reflection ($|p_1| > k_1$). The overall bell-shape of the distributions can be attributed in part to the Gaussian correlation function for the transverse correlation length in the interface profile together with the $\cos^2 \theta_r$ factor of the MDRC, as seen in  Eq.~(\ref{eq:incoMDRC:sapt}).
The corresponding transmitted intensity (MDTC) shown in Fig.~\ref{fig:sapt_mdxc_incvacuum}(d), however, shows several interesting features. As is detailed in Ref.~\citenum{Hetland2017}, we now observe pronounced peaks in \textit{s}-polarization and narrow dips to zero in \textit{p}-polarization around $|p_1| = k_1$. For normal incidence these features are independent of the azimuthal angle of transmission $\phi$. The peaks have become known as ``Yoneda peaks'', and are always found at the parallel wavevectors $\vec{p}$ along the propagation limit in the less dense medium [i.e. $|\vec{p}| = \min (k_1, k_2)$]. The polar angles corresponding to the dips to zero in the MDTC have been called the ``Brewster scattering angles'' \cite{Kawanishi1997}, and are unique to scattered light which is \textit{p}-polarized.
As the polar angle of incidence is increased ($\theta_0= \ang{35}$ or $\ang{70}$ in Fig.~\ref{fig:sapt_mdxc_incvacuum}), we observe that a Brewster scattering angle also appears in the MDRC. In transmission, the distributions of the MDTC behave very predictably in the \textit{s}-polarized case as the weight of the distribution is shifted to higher polar scattering angles. However, in the case of \textit{p}-polarization the Brewster scattering angle in the direction of $\phi=\ang{180}$ [negative values of $\theta_t$ in Fig.~\ref{fig:sapt_mdxc_incvacuum}(d)] shifts to positions closer to $\theta_t=\ang{0}$ as the angle of incidence is increased, and the dip to zero in the forward scattering direction ($\phi=\ang{0}$) first becomes a non-zero local minimum and is gradually replaced with a Yoneda peak similar to the one found for \textit{s}-polarization.

Figure~\ref{fig:sapt_mdxc_incglass} presents results similar to those in Fig.~\ref{fig:sapt_mdxc_incvacuum}, but for the situation where the media are interchanged; the light is now incident from glass [$\ve=2.25$] onto a two-dimensional randomly rough interface with vacuum [$\ve=1$].
A closer inspection of the distributions of the MDRC for normal incidence reveals that the distributions are reminiscent of the distributions seen in transmission for the MDTC in Fig.~\ref{fig:sapt_mdxc_incvacuum}, and vice versa. This similarity between intensity distributions for which the media of propagation is the same is an expected symmetry, but as the angle of incidence increases these similarities gradually fade.
For light impinging on the interface at $\theta_0=\ang{35}$, the Brewster scattering angle for the MDRC is now in the forward scattering direction, and as documented in Ref.~\citenum{Hetland2016a} it shifts closer to $\theta_r=\ang{0}$ as the angle of incidence increases towards the critical angle given by $\theta_c=\arcsin(n_2 / n_1)$.
Results for an angle of incidence equal to the critical angle, $\theta_0=\theta_c$, are presented in Figs.~\ref{fig:sapt_mdxc_incglass}(c) and \ref{fig:sapt_mdxc_incglass}(f).
For the same system but for polar angles of incidence larger than the critical angle, presented in Fig.~\ref{fig:sapt_mdxc_incglass_contd}, the dip to zero MDRC in the forward scattering direction is gradually overtaken by a Yoneda peak for \textit{p}-polarized light. Contrary to the case for transmission in Fig.~\ref{fig:sapt_mdxc_incvacuum} however, the peak in \textit{p}-polarization never grows beyond the peak in \textit{s}-polarization.
For the intensity distributions of the transmitted light we again observe a gradual shift of the weight of the distributions into the forward scattering direction, but the Brewster scattering angle is now only visible (strictly speaking) for $\theta_0=\ang{35}$ and $\theta_0=\ang{41.81}$, where it is now found in the backward scattering direction and at $\theta_t=\ang{0}$, respectively.


The scattering in both reflection and transmission from such a randomly rough interface has been thoroughly studied in the past, and the distributions of the MDRC and MDTC presented in Figs.~\ref{fig:sapt_mdxc_incvacuum}, \ref{fig:sapt_mdxc_incglass} and \ref{fig:sapt_mdxc_incglass_contd} were partially explained based on the components of the perturbative approximation in Refs.~\citenum{Hetland2016a} and \citenum{Hetland2017}.
However, these publications stopped short of presenting a full physical interpretation of the features seen in these distributions. In the current work we aim to finalize this analysis, and to that end we expand the investigation to include results for the complex amplitudes on which the MDRC and MDTC are based.
The center panel of each subfigure in Figs.~\ref{fig:sapt_mdxc_incvacuum}, \ref{fig:sapt_mdxc_incglass} and \ref{fig:sapt_mdxc_incglass_contd} presents the average phase, $\phi^{r,(1)}_{\alpha \alpha}$, $\phi^{t,(1)}_{\alpha \alpha}$ of the co-polarized scattered light, obtained from the argument of the complex amplitudes $\rho^{(1)}_{\alpha\alpha}$ or $\tau^{(1)}_{\alpha\alpha}$ for $\alpha \in \{p,s\}$ given in Eqs.~(\ref{1storder:amplitude}~d) and (\ref{1storder:amplitude}~e), respectively.
The lower panel of each subfigure shows the modulus of $\rho^{(1)}_{\alpha\alpha}$ and $\tau^{(1)}_{\alpha\alpha}$.

In passing we emphasize that even if the results  presented are based on a perturbation method to lowest non-zero order in the interface profile function, previous studies have demonstrated their validity for the parameters and dielectric constants assumed in obtaining them. In addition, the results for both MDRC and MDTC have been compared against numerical results based on the extinction theorem based method described in Ref.~\citenum{Simonsen2010} for a 1D system, a method known to be rigorous.

\subsection{Physical interpretation of SAPT to first order}\label{sec:interpretation}
\emph{Order zero, Fresnel amplitudes} --- First we revisit the interpretation of the Fresnel coefficients which are encoded in the amplitudes $\boldsymbol{\rho}^{(0)} (\Vie{p}{0}{})$ and $\boldsymbol{\tau}^{(0)} (\Vie{p}{0}{})$ [Eqs.~(\ref{eq:sapt0:ref}) and (\ref{eq:sapt0:tra})].

We start our analysis looking at the case of reflection. The Fresnel amplitudes for s- and p-polarized waves reflected by a planar surface between two dielectrics read \cite{jackson}
\begin{subequations}
\begin{align}
  r_{s}^\mathrm{F}(\Vie{p}{0}{})  &= \frac{\alpha_1(\Vie{p}{0}{}) - \alpha_2(\Vie{p}{0}{})}{\alpha_1(\Vie{p}{0}{}) + \alpha_2(\Vie{p}{0}{})}\\
  r_{p}^\mathrm{F}(\Vie{p}{0}{})  &= \frac{\epsilon_2 \alpha_1(\Vie{p}{0}{}) - \epsilon_1 \alpha_2(\Vie{p}{0}{})}{\epsilon_2 \alpha_1(\Vie{p}{0}{}) + \epsilon_1 \alpha_2(\Vie{p}{0}{})} \: ,
\end{align}
\end{subequations}
which we have here written in a common form in terms of the components of the wave vectors along $\Vie{\hat{e}}{3}{}$. It is easy to show by using straightforward algebra that these expressions are equivalent to $\rho_{ss}^{(0)} (\Vie{p}{0}{})$ and $\rho_{pp}^{(0)} (\Vie{p}{0}{})$ respectively, given by perturbation theory to zero order. An equivalent way of writing the Fresnel amplitudes which follows directly from Eq.~\eqref{eq:sapt0:ref} and the definition of the $\Vie{M}{l,m}{b,a} ( \Vie{p}{}{} \st \Vie{q}{}{})$ matrix in terms of the polarization vectors, Eq.~\eqref{Mdef}, is
\begin{subequations}
\begin{align}
  \rho_{ss}^{(0)} (\Vie{p}{0}{}) &= \rho^{(0)} (\Vie{p}{0}{}) \, \frac{\Vie{\hat{e}}{s}{}(\Vie{p}{0}{}) \cdot\Vie{\hat{e}}{s}{}(\Vie{p}{0}{}) }{\Vie{\hat{e}}{s}{}(\Vie{p}{0}{}) \cdot \Vie{\hat{e}}{s}{}(\Vie{p}{0}{})} \label{eq:rhoss0}\\
    \rho_{pp}^{(0)} (\Vie{p}{0}{}) &= \rho^{(0)} (\Vie{p}{0}{}) \, \frac{\Vie{\hat{e}}{p,2}{+}(\Vie{p}{0}{}) \cdot\Vie{\hat{e}}{p,1}{-}(\Vie{p}{0}{}) }{\Vie{\hat{e}}{p,2}{+}(\Vie{p}{0}{}) \cdot \Vie{\hat{e}}{p,1}{+}(\Vie{p}{0}{})} \label{eq:rhopp0} \\
  \rho^{(0)} (\Vie{p}{0}{}) &= \frac{\alpha_1(\Vie{p}{0}{}) - \alpha_2(\Vie{p}{0}{})}{\alpha_1(\Vie{p}{0}{}) + \alpha_2(\Vie{p}{0}{})} \label{eq:rho0:scalar}
     \: .
\end{align}
\label{eq:order0}%
\end{subequations}
In Eq.~(\ref{eq:rhoss0}), we have intentionally chosen not to simplify the dot products (all equal to 1) to illustrate that the Fresnel amplitudes expressed in the form given by Eqs.~(\ref{eq:rhoss0}) and (\ref{eq:rhopp0}) exhibit a remarkable factorization which reveals two facets of the physics of scattering from a microscopic point of view. First, both Fresnel amplitudes in Eqs.~(\ref{eq:rhoss0}) and (\ref{eq:rhopp0}) share the same first factor, $\rho^{(0)}(\Vie{p}{0}{})$ defined in Eq.~(\ref{eq:rho0:scalar}), which corresponds to the reflection amplitude for a scalar plane wave subjected to the continuity of the scalar field and its normal derivative at the surface.
In other words, this first factor can be interpreted as the coherent response of arrays of individual scatterers (at the atomic level) which scatter the incident wave 
as \emph{spherical waves}. The second factor, which differs for each polarization, is the signature of the dipolar nature of the radiation of each individual scatterer. Indeed, for an s-polarized incident wave, the scattering dipoles are excited along the $\Vie{\hat{e}}{2}{}$-axis and hence re-emit isotropically in the plane of incidence $(\Vie{\hat{e}}{1}{},\Vie{\hat{e}}{3}{})$.
We argue that this is the reason why the second factor is identically equal to 1 for s-polarized light. For p-polarization, the scattering dipoles are excited along some direction in the plane of incidence $(\Vie{\hat{e}}{1}{},\Vie{\hat{e}}{3}{})$ and therefore the reflection amplitude given by the scattering of spherical waves must be weighted with the second factor in Eq.~(\ref{eq:rhopp0}) in order to take into account the dipole radiation pattern.
Such a factorization and interpretation of the Fresnel amplitudes were given and thoroughly discussed by Doyle \cite{Doyle1985} in light of the Ewald-Oseen extinction theorem and its original derivation by Ewald based on microscopic optics \cite{Ewald1916}.

For a planar surface all scattered waves interfere destructively in all directions but the specular, as indicated by the Dirac distribution in Eqs.~(\ref{eq:sapt0:r}) and (\ref{eq:sapt0:t}). This is not the case when the surface is non-planar, and the above interpretation suggests that the spherical-like waves scattered away from the specular direction are then to be weighted by the appropriate dipole factor, even for s-polarized light (as will be the case for the first order term).

From Eq.~(\ref{eq:order0}), we can deduce two properties well known for the reflection of a plane wave at a planar interface between two dielectric media. First, for $\epsilon_1 > \epsilon_2$ there exists a critical polar angle $\theta_c = \arcsin(n_2 / n_1)$, or equivalently a critical norm of the lateral wave vector $p_c = k_2$, such that for all angles of incidence larger than $\theta_c$ (equivalently for all lateral wave vectors where $|\Vie{p}{0}{}| > p_c$), all the incident power is reflected. The phenomenon of total internal reflection is entirely controlled by the factor $\rho^{(0)}(\Vie{p}{0}{})$ present for both polarizations, and hence can be analyzed from a scalar wave picture decoupled from polarization effects.
From a physical point of view, total internal reflection occurs whenever the refracted wave is evanescent in the medium of transmission, and therefore it cannot transport energy away from the surface. It is instructive to analyze the behavior of the reflection amplitude $\rho^{(0)}(\Vie{p}{0}{})$ as the refracted wave turns evanescent in the second medium as one varies the incident lateral wave vector $\Vie{p}{0}{}$. For $|\Vie{p}{0}{}| < p_c$, both $\alpha_1(\Vie{p}{0}{})$ and $\alpha_2(\Vie{p}{0}{})$ are real.
As $|\Vie{p}{0}{}| \to p_c$, $\rho^{(0)}(\Vie{p}{0}{})$ moves on the real line towards 1 when $\alpha_2$ vanishes, $\alpha_1(\Vie{p}{0}{}) = \sqrt{\epsilon_1 -\epsilon_2} \: \omega / c$. When $|\Vie{p}{0}{}| > p_c$, $\alpha_2(\Vie{p}{0}{})$ becomes pure imaginary and $\rho^{(0)}(\Vie{p}{0}{})$ starts to trace a circular arc in the lower half of the complex plane (negative imaginary part) with unit modulus (the fact that $|\rho^{(0)}(\Vie{p}{0}{})| = 1$ for $|\Vie{p}{0}{}|> p_c$ is immediate since then $\rho^{(0)}(\Vie{p}{0}{})$ is of the form $z^* / z$ where $z$ is a non-zero complex number). As $|\Vie{p}{0}{}| \to k_1$, the reflected wave (and the incident wave) reaches the limit of propagation in the first medium and $\alpha_1(\Vie{p}{0}{}) \to 0$ which makes the reflection amplitude real, negative, equal to $-1$.
Thus, as we go from the critical point to grazing incidence the reflection amplitude traces a half circle in the complex plane with unit modulus. The argument of the reflection amplitude, the phase, hence varies from 0 to $-\pi$~rad. This gradual phase shift is known as the Goos-H\"{a}nchen phase shift and can be interpreted as follows. If we regard the reflected and refracted waves as two components of a single mode, then as the wave enters the second medium the wave oscillates while propagating along the surface before it eventually goes back into the first medium where it can continue to propagate to infinity. As the wave propagates along the surface, while being evanescent in medium 2, it acquires a temporal delay which depends on its wave vector. This delay is translated into a phase shift in the harmonic regime as the wave oscillates back into medium 1.
Geometrically this process is often interpreted as if the wave is reflected from the second medium only after a slight penetration into it \cite{jackson}.

The second phenomenon of interest is that of the polarization angle, or Brewster's angle, which, as the name indicates, requires us to analyze the polarization dependent factor in the reflection amplitudes. For an s-polarized wave the polarization factor is identically equal to 1 and no polarization angle is observed.
However, the Fresnel amplitude for p-polarized light, Eq.~(\ref{eq:rhopp0}), is shown to be proportional to $\Vie{\hat{e}}{p,2}{+}(\Vie{p}{0}{}) \cdot \Vie{\hat{e}}{p,1}{-}(\Vie{p}{0}{})$ i.e. it is proportional to the component of the incident electric field (given by $E_0 \Vie{\hat{e}}{p,1}{-}(\Vie{p}{0}{}) $) along the direction given by $\Vie{\hat{e}}{p,2}{+}(\Vie{p}{0}{})$.
We recall that the direction given by $\Vie{\hat{e}}{p,2}{+}(\Vie{p}{0}{})$ corresponds to the local p-polarization direction for a wave whose wave vector is given by $\Vie{k}{2}{+} (\Vie{p}{0}{})$, in other words, a wave which propagates upwards in the second medium. These factors of the dot products in Eq.~\eqref{eq:rhopp0} therefore seem to indicate that the reflected amplitude for p-polarized light depends on a projection of the incident field along the polarization vector of a seemingly \emph{nonexisting} wave, propagating along the wave vector $\Vie{k}{2}{+} (\Vie{p}{0}{})$.
However, such a seemingly virtual wave does have a physical interpretation, based on the mutual interaction between waves propagating in dielectric media.
Doyle \cite{Doyle1985} provided an explanation based on the concept of the \emph{wave triad} originally introduced by Ewald \cite{Ewald1916}. Ewald considered a dense array of dipole scatterers (the entire dielectric medium) situated in a half space and excited by an incident plane wave incident from the vacuum half-space, filling the whole space between the scatterers. He showed that the dipole scatterers would respond to the excitation in such a way that there exist planes of scatterers of coherent response, meaning that all dipoles within such a plane oscillate in phase. As a consequence of this fact and that the array of scatterers is bounded within a half space by a planar interface, the superposition of all elementary wavelets emitted by each individual scatterer results in the propagation of three plane waves: two waves called \emph{vacuum waves} propagating with a phase velocity equal to $c$ and one wave propagating with phase velocity $c/n$ called  \emph{polarization wave}, where $n$ corresponds to the refractive medium made of scatterers within the macroscopic picture.
The wave propagating with phase velocity $c/n$ corresponds to the transmitted wave in the macroscopic picture, while one of the waves propagating with phase velocity $c$ serves to exactly cancel the incident wave within the dielectric medium. The other wave propagating with phase velocity $c$ exits the medium and corresponds to the reflected wave. The wave vectors of the different waves are naturally given by Snell's law, and Ewald's derivation can be viewed as a microscopic validation of Snell's law.

When two half-spaces are filled with dipole scatterers of different dipole moments, similar arguments apply with the difference that the superposition of all wavelets emitted by all scatterers (i.e. from both sides of the interface) must be taken into account. This results in three wave triads: one triad associated with the incident wave, one for the reflected wave and one for the refracted wave, which all satisfy the so-called dynamical conditions \cite{Doyle1985}. The incident polarization wave in medium 1, propagating with phase velocity $c/n_1$ and wave vector $\Vie{k}{1}{-}(\Vie{p}{0}{})$, is associated with two waves propagating with phase velocity $c/n_2$ and with wave vectors $\Vie{k}{2}{\pm}(\Vie{p}{0}{})$.
Similarly, the reflected (refracted) polarization wave, whose wave vector is given by $\Vie{k}{1}{+}(\Vie{p}{0}{})$ [$\Vie{k}{2}{-}(\Vie{p}{0}{})$], is associated with two waves propagating with phase velocity $c/n_2$ ($c/n_1$) and wave vectors $\Vie{k}{2}{\pm}(\Vie{p}{0}{})$ [$\Vie{k}{1}{\pm}(\Vie{p}{0}{})$]. The dynamical conditions state that the amplitudes of the different waves are such that (i) the wave associated with the refracted polarization wave and propagating along $\Vie{k}{1}{-}(\Vie{p}{0}{})$ in medium 2 cancels the incident wave and (ii) that the superposition of waves associated with the incident and reflected polarization waves and propagating along $\Vie{k}{2}{+}(\Vie{p}{0}{})$ vanishes (more details can be found in Refs.~\citenum{Doyle1985} and \citenum{Ewald1916}).

In the following, we will refer to the wave vectors $\Vie{k}{1}{a}(\Vie{p}{}{})$ and $\Vie{k}{2}{a}(\Vie{p}{}{})$, i.e. wave vectors sharing the same projection in the $(\Vie{\hat{e}}{1}{},\Vie{\hat{e}}{2}{})$-plane and pointing either both upward or downward, as \emph{Snell-conjugate} wave vectors.
When it comes to the polarization dependence of the reflection amplitudes, the fact that $\rho_{pp}^{(0)}(\Vie{p}{0}{})$ is proportional to $\Vie{\hat{e}}{p,2}{+}(\Vie{p}{0}{}) \cdot \Vie{\hat{e}}{p,1}{-}(\Vie{p}{0}{})$ indicates that the amplitude of the reflected wave is controlled by the component of the incident field along the p-polarization vector associated with the Snell-conjugate wave vector of the wave vector of the reflected wave. This indicates that the direction of the dipole oscillation is intimately linked to waves in the aforementioned triad.

Equation~\eqref{eq:rhopp0} provides an interesting condition for the well known Brewster's angle. The Brewster phenomenon for dielectric media is commonly defined as the extinction of the \ppol-polarized reflected wave in the case of a planar interface.
From Eq.~\eqref{eq:rhopp0}, it is clear that, assuming $\epsilon_1 \neq \epsilon_2$, the Fresnel amplitude vanishes if and only if $\Vie{\hat{e}}{p,2}{+}(\Vie{p}{0}{}) \cdot \Vie{\hat{e}}{p,1}{-}(\Vie{p}{0}{}) = 0$. In other words, \emph{the condition for Brewster's law can be stated as the orthogonality between the polarization vector of the incident wave} ($\Vie{\hat{e}}{p,1}{-}(\Vie{p}{0}{})$) \emph{and the polarization vector of the Snell-conjugate wave associated with the reflected wave} ($\Vie{\hat{e}}{p,2}{+}(\Vie{p}{0}{})$). Since $\Vie{\hat{e}}{p,2}{+}(\Vie{p}{0}{})$ is orthogonal to
$\Vie{k}{2}{+} (\Vie{p}{0}{})$, we can restate the condition for Brewster's angle as the polar angle where $\Vie{k}{2}{+} (\Vie{p}{0}{})\parallel\Vie{\hat{e}}{p,1}{-}(\Vie{p}{0}{})$.
This means that we can also define Brewster's angle as the angle of reflection ensuring colinearity between the incident field amplitude and the wave vector which is Snell-conjugate to that of the reflected wave. Note that we take here a slight change of point of view compared to the common phrasing. One usually refer to Brewster's angle as an angle of incidence, while we prefer to refer to the angle of reflection. Obviously, the two are the same for a planar interface, but the latter point of view is the one which will hold true for non-planar interfaces. Nevertheless, it is convenient to use the term Brewster's incidence for a planar interface and we can define it as the angle of incidence $\theta_\mathrm{B} = \arctan (n_2 / n_1)$ which yields a Brewster (non-)reflected wave. We will see that this angle of incidence, $\theta_0 = \theta_\mathrm{B}$, has a remarkable property in the case of scattering by a non-planar interface.

Brewster's angle in the case of non-magnetic media is often said to be the angle of incidence that results in a right angle ($\ang{90}$) between the wave vector of the transmitted wave and that of the (non-)reflected wave. In the case of a planar interface, our new definition of Brewster angle agrees with this explanation.
Indeed, if $\Vie{\hat{e}}{p,1}{-}(\Vie{p}{0}{}) \cdot \Vie{\hat{e}}{p,2}{+}(\Vie{p}{0}{}) = 0$ it is immediate that $\Vie{k}{1}{+}(\Vie{p}{0}{}) \cdot \Vie{k}{2}{-}(\Vie{p}{0}{}) = 0$. However, we will see below that the new geometrical criterion proposed in the above holds when applied with Snell-conjugate wave vector associated with a non-specularly scattered wave, while the ``right angle'' criterion between wave vectors breaks down.

The Fresnel amplitudes for the \emph{refracted} wave for s- and p-polarized light expressed in terms of polarization vectors, presented in a similar fashion as Eq.~\eqref{eq:order0}, read
\begin{subequations}
\begin{align}
  \tau_{ss}^{(0)} (\Vie{p}{0}{}) &= \frac{\tau^{(0)} (\Vie{p}{0}{})}{\Vie{\hat{e}}{s}{}(\Vie{p}{0}{}) \cdot \Vie{\hat{e}}{s}{}(\Vie{p}{0}{})} \label{eq:tauss0}\\
  \tau_{pp}^{(0)} (\Vie{p}{0}{}) &=
  \frac{\tau^{(0)}(\Vie{p}{0}{})}{\Vie{\hat{e}}{p,1}{-} (\Vie{p}{0}{}) \cdot \Vie{\hat{e}}{p,2}{-} (\Vie{p}{0}{})}\\
  \tau^{(0)} (\Vie{p}{0}{}) &=
  \frac{c^2}{\omega^2}\frac{2\alpha_1(\Vie{p}{0}{})}{(\epsilon_1 -\epsilon_2)}  [\alpha_1(\Vie{p}{0}{}) - \alpha_2(\Vie{p}{0}{})] \label{eq:tau0:scalar} \: .
\end{align}
\label{eq:tau0}%
\end{subequations}
From Eq.~(\ref{eq:tau0}), it is readily observed that neither the s- nor p-polarized zero order transmitted wave vanishes in general, which relates to the common experience that no Brewster angle is known for transmission through a planar interface. This fact does not, however, prevent the existence of Brewster \emph{scattering} angles in the diffusely transmitted light. Equation~\eqref{eq:tau0} will therefore be important in the remainder of this paper.
Note the presence of the factor $\alpha_1(\Vie{p}{0}{}) - \alpha_2(\Vie{p}{0}{})$ in the transmission amplitude of the scalar wave in Eq.~\eqref{eq:tau0:scalar} which is identical to the numerator of the reflection amplitude in Eq.~\eqref{eq:rho0:scalar}. The analysis of this term on total internal reflection hence leads to a similar behavior for the transmission amplitude, in the sense that $\tau^{(0)}(\Vie{p}{0}{})$ leaves the real line and traces a path in the complex plane when total internal reflection occurs. This fact illustrates the coupling between the reflected and the transmitted waves, which may be interpreted as two components of the same mode.

\medskip

\emph{First order} --- We analyze now the first order amplitudes given by SAPT. Let us consider first the case of a \emph{scalar wave} subjected to the continuity of the field and its normal derivative with respect to the surface. We have seen in Sec.~\ref{sec:theory} that it suffices to replace the $\Vie{M}{lm}{ba}$ matrices by $k_1 k_2$ and the identity  matrix by $1$ in Eqs.~(\ref{eq:sapt1:ref}) and (\ref{eq:sapt1:tra}) to obtain the first order reflection and transmission amplitudes for the corresponding scalar problem. This yields the following first order scalar amplitudes
%
\begin{subequations}
\begin{align}
  R^{(1)} (\Vie{p}{}{}|\Vie{p}{0}{}) = \: &S (\Vie{p}{}{}|\Vie{p}{0}{}) \, \big[1 + \rho^{(0)}(\Vie{p}{0}{})\big] \label{eq:R1:scalar}\\
  T^{(1)} (\Vie{p}{}{}|\Vie{p}{0}{}) = \: &S (\Vie{p}{}{}|\Vie{p}{0}{}) \,  \tau^{(0)} (\Vie{p}{0}{}) \label{eq:T1:scalar} \\
  S (\Vie{p}{}{}|\Vie{p}{0}{}) = \: &(\alpha_1(\Vie{p}{}{}) - \alpha_2(\Vie{p}{}{})) \: \hat{\zeta}(\Vie{p}{}{} -\Vie{p}{0}{}) \: .
\end{align}
\label{eq:RT1:scalar}%
\end{subequations}
The first point to notice in Eqs.~(\ref{eq:R1:scalar}) and (\ref{eq:T1:scalar}) is that \emph{both} the reflection \emph{and} transmission amplitudes are proportional to $S (\Vie{p}{}{}|\Vie{p}{0}{}) = (\alpha_1(\Vie{p}{}{}) - \alpha_2(\Vie{p}{}{})) \: \hat{\zeta}(\Vie{p}{}{} -\Vie{p}{0}{})$. Second, this common factor encodes to whole $\Vie{p}{}{}$-dependence of the amplitudes. Indeed, the remaining factors $1 + \rho^{(0)}(\Vie{p}{0}{})$ and $\tau^{(0)} (\Vie{p}{0}{})$ respectively, only depend on the incident in-plane wave vector $\Vie{p}{0}{}$. How should we interpret such a factorization? Since the reflection and transmission amplitudes vary in the same way with $\Vie{p}{}{}$, we can say that both the elementary reflected wave with in-plane wave vector $\Vie{p}{}{}$ and the elementary transmitted wave with the \emph{same} wave vector $\Vie{p}{}{}$ are coupled. They can be viewed as two pieces of a single scattered mode defined in the whole space. In fact, for scalar waves, we have $1+\rho^{(0)} (\Vie{p}{0}{}) = \tau^{(0)} (\Vie{p}{0}{})$, which gives $R^{(1)}(\Vie{p}{}{}|\Vie{p}{0}{}) = T^{(1)}(\Vie{p}{}{}|\Vie{p}{0}{})$; the first order reflection and transmission amplitudes are equal! This strengthen the coupled mode interpretation. Indeed, it means that in the plane wave expansion of the scalar  scattered field, which can be written
\begin{subequations}
\begin{align}
E_1 (\Vie{r}{}{}) &= E_0 \int R(\Vie{q}{}{} | \Vie{p}{0}{}) \, \exp \left( i \Vie{k}{1}{+}(\Vie{q}{}{}) \cdot \Vie{r}{}{} \right) \, \dtwoq \\
E_2 (\Vie{r}{}{}) &= E_0 \int T(\Vie{q}{}{} | \Vie{p}{0}{}) \, \exp \left( i \Vie{k}{2}{-}(\Vie{q}{}{}) \cdot \Vie{r}{}{} \right) \, \dtwoq  \, ,
\end{align}
\end{subequations}
where one has a priori two amplitudes $R$ and $T$ to be determined, the first order result states that there is actually fewer degree of freedom since (to first order) $R = T$. By defining a \emph{coupled plane wave mode} as
\begin{equation}
\mathcal{C}^{a,b}(\Vie{p}{}{} , \Vie{r}{}{}) = \begin{cases}
\exp \left( i \Vie{k}{1}{a}(\Vie{p}{}{}) \cdot \Vie{r}{}{} \right) & \mathrm{for} \: x_3 > 0 \\
\exp \left( i \Vie{k}{2}{b}(\Vie{p}{}{}) \cdot \Vie{r}{}{} \right) & \mathrm{for} \: x_3 < 0
\end{cases} \: ,
\label{eq:coupled:mode}
\end{equation}
the plane wave expansion can be recast into a coupled plane wave mode expansion with only one amplitude $X = R = T$ for the field in the whole space as
\begin{equation}
E (\Vie{r}{}{}) = E_0 \, \int X(\Vie{q}{}{} | \Vie{p}{0}{}) \, \mathcal{C}^{+, -} (\Vie{q}{}{}, \Vie{r}{}{}) \: \dtwoq \: .
\end{equation}
This is an important conceptual step which needs some comments. The coupled plane wave modes defined in Eq.~(\ref{eq:coupled:mode}) are such that they satisfy the Helmholtz equation in both media and are continuous across the interface $x_3 = 0$. They are then continous eigenmodes of the Helmholtz equation where the interface is planar between the two media, in the same way plane waves are eigenmodes of the Helmholtz equation in a homogenous medium. In addition, taking $a=+$ and $b=-$ select the eigenmodes which satisfy an outgoing wave radiation condition. The coupled plane wave mode expansion thus seems to be, in some sense, a more natural choice than the plane wave expansion to describe the scattering problem at hand. Note, however, that the normal derivative of the coupled mode across the interface, $x_3 = 0$, is discontinous. The jump is given by $i (\alpha_1(\Vie{p}{}{}) - \alpha_2(\Vie{p}{}{})) e^{i \Vie{p}{}{} \cdot \Vie{x}{\parallel}{}}$. This indicates that the problem of scattering by a rough surface can be mapped to a problem of scattering by planar surface with surface sources, at least to first order in the surface profile function. This is to be related to the so-called equivalent surface-current model \cite{kroger}.

 Let us analyze further the scattering amplitudes in Eq.~(\ref{eq:RT1:scalar}). Since there is no summation over intermediate wave vectors $\Vie{q}{}{}$, we can consider that a scattered mode characterized by a wave vector $\Vie{p}{}{}$ is decoupled from a scattered mode characterized by a wave vector $\Vie{q}{}{} \neq \Vie{p}{}{}$. The scattered modes are, however, coupled to the incident wave which acts as a source. It is also fruitful to go back to the derivation of SAPT to get a intuitive understanding of the meaning of the factor $S(\Vie{p}{}{} | \Vie{p}{0}{}) = (\alpha_1(\Vie{p}{}{}) - \alpha_2(\Vie{p}{}{})) \: \hat{\zeta}(\Vie{p}{}{} -\Vie{p}{0}{})$. This factor originates from the Taylor expansion of the exponential factor $\exp [-i (b \alpha_2(\cdot) - \alpha_1(\cdot)) \zeta(\cdot)]$ in the definition of the $\Cie{J}{lm}{ba}$-integral (see Appendix A).
  What the $\Cie{J}{lm}{ba}$-integral encodes is intuitively a sum of complex amplitudes along the surface where each scattered path will experience different phase shift depending on their scattering event along the surface. The factor $\exp [-i (b \alpha_2(\cdot) - \alpha_1(\cdot)) \zeta(\cdot)]$ can be thought as a phase factor varying along the surface. The factor  $(\alpha_1(\Vie{p}{}{}) - \alpha_2(\Vie{p}{}{})) \: \hat{\zeta}(\Vie{p}{}{} -\Vie{p}{0}{})$ originates from the linear approximation of this phase factor. Thus it corresponds to the approximation of the \emph{interference} pattern, or \emph{speckle field}, resulting from single scattering events with phases linearly approximated. Taking one step further in our interpretation, $S(\Vie{p}{}{} | \Vie{p}{0}{})$ can then be roughly said to be a \emph{probability amplitude} for a change of in-plane wave vector (or momentum) from $\Vie{p}{0}{}$ to $\Vie{p}{}{}$. This is not the entire reflection nor transmission amplitude yet. We still have to give an interpretation to the remaining factors in Eq.~(\ref{eq:RT1:scalar}). The factor $1 + \rho^{(0)}(\Vie{p}{0}{})$ is the sum of the incident unit field and the corresponding reflected zero order field, hence it describes the \emph{total zero order field} in medium 1. Similarly, $\tau^{(0)} (\Vie{p}{0}{})$ is the zero order transmitted amplitude and it describes the total zero order field in medium 2. Thus the first order reflected and transmitted amplitudes are proportional, respectively, to the total zero order field in the medium of reflection and transmission. In other words, we can say that it is not only the incident wave that acts as a source for the first order waves, but the sum of the incident and scattered zero order field. What does this mean microscopically? It means that to first order in the surface profile, the state of oscillation of elementary point sources of spherical waves constituting the media is the same as that of the corresponding planar interface problem. The surface surface roughness is so small that it does not influence significantly the coherent response of the array of point sources in the media compared to the planar case. Nevertheless, since the surface is rough, the point sources in the volume of the selvedge region interfere in the far field in such a way that they  produce a speckle pattern in contrast to the planar case where the sum of all point source radiations interfere destructively in all direction but the specular.


Now that we have given a physical interpretation to the first order amplitudes for scalar waves, let us consider the case of \emph{electromagnetic waves}. The first order amplitudes for electromagnetic waves are given by Eqs.~(\ref{eq:sapt1:ref}) and (\ref{eq:sapt1:tra}). We have seen when discussing the Fresnel amplitudes that, in view of a physical interpretation, it was beneficial to express the $\Vie{M}{lm}{ba}$ matrices in terms of the polarization vectors according to Eq.~(\ref{Mdef}). The first order electromagnetic amplitudes read as a function of the polarization vectors as
\begin{subequations}
\begin{align}
\Vie{R}{}{(1)} (\Vie{p}{}{}|\Vie{p}{0}{}) = S(\Vie{p}{}{} | \Vie{p}{0}{}) \, \hat{\boldsymbol \rho}^{(1)} (\Vie{p}{}{}|\Vie{p}{0}{}) \\
\Vie{T}{}{(1)} (\Vie{p}{}{}|\Vie{p}{0}{}) =  S(\Vie{p}{}{} | \Vie{p}{0}{}) \, \hat{\boldsymbol \tau}^{(1)} (\Vie{p}{}{}|\Vie{p}{0}{})
\end{align}
\label{eq:RT1:EM}
\vspace*{-0.3cm}
\end{subequations}
with
\begin{subequations}
\begin{align}
  \hat{\rho}_{s s}^{(1)} ( \Vie{p}{}{} | \Vie{p}{0}{}) &= \frac{ \Vie{\hat{e}}{s}{}(\Vie{p}{}{}) \cdot \Vie{E}{1,s}{(0)}(\Vie{p}{0}{})}{\Vie{\hat{e}}{s}{} (\Vie{p}{}{}) \cdot \Vie{\hat{e}}{s}{} (\Vie{p}{}{})}   \label{eq:rho1ss}\\
  \hat{\rho}_{p s}^{(1)} ( \Vie{p}{}{} | \Vie{p}{0}{}) &= \frac{\Vie{\hat{e}}{p,2}{+}(\Vie{p}{}{}) \cdot \Vie{E}{1,s}{(0)}(\Vie{p}{0}{})}{\Vie{\hat{e}}{p,2}{+}(\Vie{p}{}{}) \cdot \Vie{\hat{e}}{p,1}{+}(\Vie{p}{}{})}   \label{eq:rho1ps}\\
  \hat{\rho}_{s p}^{(1)} ( \Vie{p}{}{} | \Vie{p}{0}{}) &= \frac{ \Vie{\hat{e}}{s}{}(\Vie{p}{}{}) \cdot \Vie{E}{1,p}{(0)}(\Vie{p}{0}{}) }{\Vie{\hat{e}}{s}{} (\Vie{p}{}{}) \cdot \Vie{\hat{e}}{s}{} (\Vie{p}{}{})}  \label{eq:rho1sp}\\
  \hat{\rho}_{p p}^{(1)} ( \Vie{p}{}{} | \Vie{p}{0}{}) &= \frac{\Vie{\hat{e}}{p,2}{+}(\Vie{p}{}{}) \cdot \Vie{E}{1,p}{(0)}(\Vie{p}{0}{})}{\Vie{\hat{e}}{p,2}{+}(\Vie{p}{}{}) \cdot \Vie{\hat{e}}{p,1}{+}(\Vie{p}{}{})}  \label{eq:rho1pp}
\end{align}
\label{eq:rho1:all}%
\end{subequations}
for the reflection amplitudes and
\begin{subequations}
\begin{align}
  \hat{\tau}_{s s}^{(1)} ( \Vie{p}{}{} | \Vie{p}{0}{}) &= \frac{\Vie{\hat{e}}{s}{} (\Vie{p}{}{}) \cdot \Vie{E}{2,s}{(0)} (\Vie{p}{0}{})}{\Vie{\hat{e}}{s}{} (\Vie{p}{}{}) \cdot \Vie{\hat{e}}{s}{} (\Vie{p}{}{})}   \label{eq:tau1ss}\\
  \hat{\tau}_{p s}^{(1)} ( \Vie{p}{}{} | \Vie{p}{0}{}) &= \frac{\Vie{\hat{e}}{p,1}{-} (\Vie{p}{}{}) \cdot \Vie{E}{2,s}{(0)} (\Vie{p}{0}{})}{\Vie{\hat{e}}{p,1}{-} (\Vie{p}{}{}) \cdot \Vie{\hat{e}}{p,2}{-} (\Vie{p}{}{})} \label{eq:tau1ps}\displaybreak[0]\\
  \hat{\tau}_{s p}^{(1)} ( \Vie{p}{}{} | \Vie{p}{0}{}) &= \frac{\Vie{\hat{e}}{s}{} (\Vie{p}{}{}) \cdot \Vie{E}{2,p}{(0)} (\Vie{p}{0}{})}{\Vie{\hat{e}}{s}{} (\Vie{p}{}{}) \cdot \Vie{\hat{e}}{s}{} (\Vie{p}{}{})}  \label{eq:tau1sp} \\
  \hat{\tau}_{p p}^{(1)} ( \Vie{p}{}{} | \Vie{p}{0}{}) &= \frac{\Vie{\hat{e}}{p,1}{-} (\Vie{p}{}{}) \cdot \Vie{E}{2,p}{(0)} (\Vie{p}{0}{})}{\Vie{\hat{e}}{p,1}{-} (\Vie{p}{}{}) \cdot \Vie{\hat{e}}{p,2}{-} (\Vie{p}{}{})}  \label{eq:tau1pp}
\end{align}
\label{eq:tau1:all}%
\end{subequations}
for the transmission amplitudes. Here we have defined the total zero order field amplitudes in media 1 and 2, for $s$- and $p$-polarized incident light, as
\begin{subequations}
\begin{align}
  \Vie{E}{1,s}{(0)}(\Vie{p}{0}{}) &= \left[ 1 + \rho_{ss}^{(0)} (\Vie{p}{0}{}) \right] \, \Vie{\hat{e}}{s}{}(\Vie{p}{0}{}) \label{eq:E1s}\\
  \Vie{E}{1,p}{(0)}(\Vie{p}{0}{}) &= \Vie{\hat{e}}{p,1}{-}(\Vie{p}{0}{}) + \rho_{pp}^{(0)} (\Vie{p}{0}{}) \Vie{\hat{e}}{p,1}{+}(\Vie{p}{0}{}) \label{eq:E1p}\\
  \Vie{E}{2,s}{(0)}(\Vie{p}{0}{}) &= \tau_{ss}^{(0)} (\Vie{p}{0}{}) \, \Vie{\hat{e}}{s}{}(\Vie{p}{0}{}) \label{eq:E2s}\\
  \Vie{E}{2,p}{(0)}(\Vie{p}{0}{}) &= \tau_{pp}^{(0)} (\Vie{p}{0}{}) \Vie{\hat{e}}{p,2}{-}(\Vie{p}{0}{}) \label{eq:E2p}\: .
\end{align}
\label{eq:Eall}%
\end{subequations}
The factorization suggested by Eq.~(\ref{eq:RT1:EM}) is very similar to that given by Eq.~(\ref{eq:RT1:scalar}) for scalar waves. We have already discussed the speckle factor $ S(\Vie{p}{}{} | \Vie{p}{0}{})$ for scalar waves and it keeps its interpretation for electromagnetic waves. The difference between electromagnetic and scalar waves resides in the last factor, $\hat{\boldsymbol \rho}^{(1)} (\Vie{p}{}{}|\Vie{p}{0}{})$ and $\hat{\boldsymbol \tau}^{(1)} (\Vie{p}{}{}|\Vie{p}{0}{})$ respectively for the reflection and transmission amplitude. In the case of scalar waves, the last factor encoded the state of the scalar zero order field. The factors $\hat{\boldsymbol \rho}^{(1)} (\Vie{p}{}{}|\Vie{p}{0}{})$ and $\hat{\boldsymbol \tau}^{(1)} (\Vie{p}{}{}|\Vie{p}{0}{})$ also encodes the state of the electromagnetic zero order field. Indeed this is readily seen from Eq.~(\ref{eq:Eall}) where $\Vie{E}{j,\beta}{(0)}(\Vie{p}{0}{})$ is the sum of the $\beta$-polarized unit incident (if $j=1$) field amplitude and the corresponding zero order scattered field. We will hence call $\Vie{E}{j,\beta}{(0)}(\Vie{p}{0}{})$ \emph{the total zero order field amplitude}. Microscopically, this factor contains the information of the polarization state of the media for the planar system, i.e. information about the unperturbed collective oscillation of dipoles constituting the media. As for the scalar case, the radiation emitted by dipoles in the volume of the selvdege region interfere in the far field (factor $S(\Vie{p}{}{} | \Vie{p}{0}{})$) but now we need to take into account that dipole radiation is anisotropic. This is the reason why the amplitudes $\hat{\boldsymbol \rho}^{(1)} (\Vie{p}{}{}|\Vie{p}{0}{})$ and $\hat{\boldsymbol \tau}^{(1)} (\Vie{p}{}{}|\Vie{p}{0}{})$ also depend on $\Vie{p}{}{}$ and not only on $\Vie{p}{0}{}$. By a close inspection of the expressions given in  Eqs.~(\ref{eq:rho1:all}) and (\ref{eq:tau1:all}), we note that all these amplitudes can be written as
\begin{equation}
\hat{x}_{\alpha \beta}^{(1)} (\Vie{p}{}{}|\Vie{p}{0}{}) = \frac{\Vie{\hat{e}}{\alpha,\bar{j_x}}{a_x} (\Vie{p}{}{}) \cdot \Vie{E}{j_x,\beta}{(0)} (\Vie{p}{0}{})}{\Vie{\hat{e}}{\alpha,\bar{j_x}}{a_x} (\Vie{p}{}{}) \cdot \Vie{\hat{e}}{\alpha,j_x}{a_x} (\Vie{p}{}{})}
\end{equation}
with $x = \rho$ or $\tau$, $j_x = 1$ or $2$, and $a_x = \pm$ respectively for $x = \rho$ and $x = \tau$, and $\bar{j}_x = 1$ if $j_x = 2$ and $\bar{j}_x = 2$ if $j_x = 1$. We also adopt the convention that $ \Vie{\hat{e}}{s,j}{a} (\Vie{p}{}{}) = \Vie{\hat{e}}{s}{} (\Vie{p}{}{})$ independently of $j$ and $a$. Stated in a sentence, the caret amplitude is given by the \emph{the total zero order field amplitude projected on the Snell-conjugate polarization vector of the measured polarization normalized by the similar projection as if the total zero order field were replaced by the measured polarization vector}. The normalization can be puzzling at first, so let us start by interpreting the numerator in some particular cases. In the case of $s \to s$ reflection we have that $\hat{\rho}_{ss}^{(1)} (\Vie{p}{}{}|\Vie{p}{0}{})$ is proportional to $\Vie{\hat{e}}{s}{}(\Vie{p}{}{}) \cdot \Vie{E}{1,s}{(0)}(\Vie{p}{0}{})$. This seems quite intuitive, since it means that the probability of scattering to a $s$-polarized wave with wave vector $\Vie{p}{}{}$ is proportional to the projection of the zero order field along the outgoing polarization state vector. Consider now the first order reflection amplitude $\hat{\rho}_{pp}^{(1)} (\Vie{p}{}{}|\Vie{p}{0}{})$. It is proportional to $\Vie{\hat{e}}{p,2}{+}(\Vie{p}{}{}) \cdot \Vie{E}{1,p}{(0)}(\Vie{p}{0}{})$, which means that the probability for scattering to a $p$-polarized wave with wave vector $\Vie{p}{}{}$ is proportional to the projection of the zero order field along the polarization vector of the \emph{Snell-conjugate wave} [$\Vie{\hat{e}}{p,2}{+}(\Vie{p}{}{})$] associated with the outgoing polarization vector [$\Vie{\hat{e}}{p,1}{+}(\Vie{p}{}{})$]. This is slightly less intuitive than the $s \to s$ case. What this means is that, in virtue of Ewald's interpretation discussed for the Fresnel amplitudes, it is the Snell-conjugate wave that controls the polarization coupling. This fact is the analog of what we found for the reflection of a $p$-polarized wave at a planar interface, Eq.~(\ref{eq:rhopp0}). For $s$-polarized waves, this interpretation still holds since Snell-conjugate pairs always have the same $s$ polarization vector.\\

Let us attempt to summarize the overall interpretation of the first order amplitudes for electromagnetic waves in a few sentences. \emph{The first order reflection and transmission amplitudes for polarized waves are the product of a common probability amplitude for a single change of in-plane momentum from $\Vie{p}{0}{}$ to $\Vie{p}{}{}$, which is of a scalar nature, with a probability amplitude for a polarization coupling between the outgoing wave and the zero order state. The probability amplitude for a change of momentum is microscopically linked to the interference of point sources from the volume of the selvedge region. The probability amplitude for polarization coupling, i.e. to scatter to a $\alpha$-polarized wave with in-plane wave vector $\Vie{p}{}{}$ from an incident $\beta$-polarized wave with in-plane wave vector $\Vie{p}{0}{}$ (given $\Vie{p}{0}{}$ and $\Vie{p}{}{}$), is proportional to the projection coefficient of the total zero order field $\Vie{E}{j_x,\beta}{(0)}(\Vie{p}{0}{})$ along the polarization vector of the Snell-conjugate wave associated with the measured wave. Microscopically, this last factor is the trace of the polarization state of the media, i.e. the coherent oscillation of dipoles in the planar interface case, indicating that the dipoles in the selvedge region are not perturbed and oscillate accordingly to the zero order state.}\\

We are now ready for a more in-depth analysis of the Yoneda and Brewster scattering effects based on the physical interpretation and concepts we have developed in the present section.

\subsection{The physical origin of the Yoneda effect}\label{sec:yoneda}

Our observations on the Yoneda and Brewster scattering effects in Section~\ref{sec:pheno} led us to the conclusion that the two effects can be explained independently. The fact that the Brewster scattering angle coincides with the Yoneda critical angle for normal incidence can, for the time being, be considered a simple coincidence.
Since the Yoneda phenomenon seems to be independent of polarization we can attempt an explanation solely based on scalar waves and consider Eq.~(\ref{eq:RT1:scalar}) a relevant simplified model, in an analogous fashion as Eq.~(\ref{eq:rho0:scalar}) was sufficient to explain total internal reflection from a planar surface. In fact, for the scattering of s-polarized waves restricted to the plane of incidence ($\Vie{p}{}{} \parallel \Vie{p}{0}{}$) the reflection and transmission amplitudes are exactly given by Eq.~(\ref{eq:RT1:scalar}). We will therefore keep to scalar waves for the main analysis, but we will also illustrate our conclusions with results obtained for s-polarized waves.
In the following, it will be convenient to refer to the smallest and largest dielectric constant by $\epsilon_{\min}$ and $\epsilon_{\max}$ respectively, and more generally we will index by $\min$ and $\max$ the quantities corresponding to these media. Our analysis will be independent of the configurations of the media but will require us to distinguish the optically denser medium from the less dense medium for the scattered waves.

We have seen in Sec.~\ref{sec:interpretation} that we can consider that the incident wave impinges on the surface with an in-plane lateral wave vector $\Vie{p}{0}{}$ and, within a single scattering point of view, gives rise to a scattered elementary wave reflected with the in-plane lateral wave vector $\Vie{p}{}{}$ and a scattered elementary wave transmitted with the in-plane lateral wave vector $\Vie{p}{}{}$. 
The two elementary waves are two components of the same coupled mode characterized by the in-plane wave vector $\Vie{p}{}{}$. As argued in Sec.~\ref{sec:interpretation}, the probability for a change of lateral wave vector from $\Vie{p}{0}{}$ to $\Vie{p}{}{}$ is controlled by the factor $[\alpha_1(\Vie{p}{}{}) - \alpha_2(\Vie{p}{}{})] \: \hat{\zeta}(\Vie{p}{}{} -\Vie{p}{0}{})$ (up to the zero order state amplitude).
%
What does determine whether the intensity is enhanced in the optically denser medium for a given elementary scattered wave? Since the elementary reflected and transmitted waves are coupled into a single coupled mode, we can also interpret the probability amplitude for a change of momentum as a way to allocate part of the energy from the incident wave, to be shared between, and radiated away by, the two scattered waves gathered in the coupled mode with the shared lateral wave vector $\Vie{p}{}{}$. This allocated energy to the coupled mode must then be shared between the reflected and transmitted wave components.

Let us first consider the situation where the shared lateral wave vector of the scattered waves is restricted to $|\Vie{p}{}{}| < k_{\min} = p_c$, which means that both waves are allowed to propagate to infinity in their respective medium. Under this assumption, the total energy of the two waves will be shared \emph{a priori} non-trivially between the two waves.
However, if now the shared lateral wave vector is such that $k_{\min} < |\Vie{p}{}{}| < k_{\max}$, the wave scattered in the optically less dense medium will be evanescent. Therefore the total energy for the coupled mode will be carried away solely by the wave which can propagate, namely the one scattered into the dense medium, resulting in the apparent sudden increase of intensity at the transition between propagation and evanescence of the wave scattered in the optically less dense medium.
An illustrative way of seeing that the intensity needs to be enhanced is by analyzing the factor $\alpha_1(\Vie{p}{}{}) - \alpha_2(\Vie{p}{}{})$ assuming $|\hat{\zeta}|$ to vary slowly. For $|\Vie{p}{}{}| < p_c$ both $\alpha_1(\Vie{p}{}{})$ and $\alpha_2(\Vie{p}{}{})$ are real. As $|\Vie{p}{}{}| \to p_c$ from below, $\alpha_{\min}(\Vie{p}{}{}) \to 0$ and $\alpha_1(\Vie{p}{}{}) - \alpha_2(\Vie{p}{}{}) \to \pm \alpha_c$, with $\alpha_c = \sqrt{\epsilon_{\max} - \epsilon_{\min}} \, \omega / c$. By writing $p = |\Vie{p}{}{}| = p_c - \Delta p$, with $\Delta p > 0$, we can make an asymptotic analysis of $|\alpha_1(\Vie{p}{}{}) - \alpha_2(\Vie{p}{}{})|$ as $p \to p_c$ from below. In this way we obtain the following result
\begin{align}
  &|\alpha_1(\Vie{p}{}{}) - \alpha_2(\Vie{p}{}{})| \frac{c}{\omega} = [\alpha_{\max}(\Vie{p}{}{}) - \alpha_{\min}(\Vie{p}{}{})] \frac{c}{\omega} \nonumber \\
  &= [\epsilon_{\max} - (\tilde{p}_c - \Delta \tilde{p})^2]^{1/2} - [\epsilon_{\min} - (\tilde{p}_c - \Delta \tilde{p})^2]^{1/2} \nonumber \\
  &= [\epsilon_{\max} - \epsilon_{\min} + 2 \epsilon_{\min}^{1/2} \Delta \tilde{p} - \Delta \tilde{p}^2 ]^{1/2} - [2 \epsilon_{\min}^{1/2} \Delta \tilde{p} - \Delta \tilde{p}^2 ]^{1/2} \nonumber \\
  &= \alpha_c \, \frac{c}{\omega} - [2 \epsilon_{\min}^{1/2} \Delta \tilde{p}]^{1/2} + o(\Delta \tilde{p}^{1/2}) \: .
  \label{eq:asymptotic:below:pc}
\end{align}
Here we have chosen to work with unit-less quantities and denoted $\tilde{p} = p \, c /\omega$ for conciseness. From Eq.~(\ref{eq:asymptotic:below:pc}) it then follows that as $\Delta p \to 0$, $|\alpha_1(\Vie{p}{}{}) - \alpha_2(\Vie{p}{}{})|$ must increase towards $\alpha_c$ in an inner-neighborhood of the circle $p = p_c$. Furthermore, the asymptotic expansion reveals that the critical point will be reached with a sharp edge (infinite slope) for $p < p_c$ as can be deduced from the square root behavior in $\Delta p$.
Note that both the reflection and transmission amplitudes exhibit the same behavior independently of which medium is denser. This is due to the fact that the two waves are part of the same mode. However, as the wave propagating in the less dense medium becomes a grazing wave, the corresponding differential scattering coefficient is forced to vanish due to the angular dependence in $\cos^2 \theta_s$ ($\theta_s = \theta_r$ or $\theta_t$ depending on the context). The complex amplitude is nevertheless enhanced for both the reflected and transmitted wave.
This is illustrated for example in Figs.~\ref{fig:sapt_mdxc_incvacuum}(a) and \ref{fig:sapt_mdxc_incvacuum}(d), which corresponds to a case for which the medium of incidence is vacuum. From the results presented in these figures, we can see that while the incoherent component of the MDRC is forced to go to zero when $p_1 \to p_c = \omega / c$, the surface-independent part of the reflection amplitude $\rho_{ss}^{(1)}$ exhibits a sharp increase in modulus. Simultaneously, the surface-independent part of the transmission amplitude $\tau_{ss}^{(1)}$ also exhibits a similar sharp increase in modulus as $p_1$ approaches $p_c$. Consequently, since the wave can propagate away from the surface in the second medium (which consists of glass in this specific case), the corresponding incoherent component of the MDTC exhibits a similar increase. Note that both the phases associated with $\rho_{ss}^{(1)}$ and $\tau_{ss}^{(1)}$ remain constant and equal to 0 for $p_1 < p_c$ for all $\theta_0$ in Fig.~\ref{fig:sapt_mdxc_incvacuum}, since the complex amplitude stays on the real line in the case where $\epsilon_1 < \epsilon_2$ independent of the angle of incidence.

Figures~\ref{fig:sapt_mdxc_incglass} and \ref{fig:sapt_mdxc_incglass_contd} support the same conclusion but by interchanging the role of the media. The only difference worth noting is that the phases $\phi_{ss}^{r,(1)}$ and $\phi_{ss}^{t,(1)}$ have a constant plateau for $p_1 < p_c$ which is equal to 0 only for $\theta_0 < \theta_c$. The plateau is offset for $\theta_0 > \theta_c$. This overall phase offset is due to the Goos-H\"{a}nchen phase shift associated with total internal reflection of the zero order wave. Indeed, recall that the first order amplitudes are proportional to the total zero order field amplitudes. As a consequence, if the zero order waves exhibit a phase shift, it will affect the first order amplitudes in the form of a constant phase offset for all $\Vie{p}{}{}$.

When $|\Vie{p}{}{}| > p_c$, $\alpha_{\min}$ becomes purely imaginary and $\alpha_1(\Vie{p}{}{}) - \alpha_2(\Vie{p}{}{})$ thus moves off the real line. For $p_c < |\Vie{p}{}{}| < k_{\max}$, we find that in this regime $\alpha_1(\Vie{p}{}{}) - \alpha_2(\Vie{p}{}{})$ keeps a constant modulus equal to $\alpha_c$. Indeed, by writing $\alpha_{\min}(\Vie{p}{}{}) = i \beta_{\min}(\Vie{p}{}{})$ we have
\begin{align}
|\alpha_1(\Vie{p}{}{}) - \alpha_2(\Vie{p}{}{})| &= |\alpha_{\max}(\Vie{p}{}{}) - i \beta_{\min}(\Vie{p}{}{})| \nonumber\\
&= \Big[ \alpha_{\max}^2(\Vie{p}{}{}) + \beta_{\min}^2(\Vie{p}{}{}) \Big]^{1/2} \nonumber\\
&= \Big[ \epsilon_{\max} - \tilde{p}^2 + \tilde{p}^2 - \epsilon_{\min} \Big]^{1/2} \, \omega / c \nonumber \\
&= \alpha_c
 \: . \label{eq:asymptotic:above:pc}
\end{align}
The complex number $\alpha_1(\Vie{p}{}{}) - \alpha_2(\Vie{p}{}{})$ thus traces a circular arc of radius $\alpha_c$ in the complex plane. Finally, when $|\Vie{p}{}{}| > k_{\max}$, both the reflected and transmitted waves are evanescent, $\alpha_{\max}$ becomes pure imaginary and hence $\alpha_1(\Vie{p}{}{}) - \alpha_2(\Vie{p}{}{})$ moves along the imaginary axis.
The constant value of $|\rho_{ss}^{(1)}|$ and $|\tau_{ss}^{(1)}|$ in the regime $k_{\min} < \Vie{p}{}{} < k_{\max} $ can be appreciated for all angles of incidence illustrated in Figs.~\ref{fig:sapt_mdxc_incvacuum} -- \ref{fig:sapt_mdxc_incglass_contd}, while the phases exhibit a smooth variation from their plateau value and decay by a total amount of $-\pi/2$ when reaching $p_1 = k_{\max}$. Once the threshold of $k_{\max}$ has been passed, the phases remain constant and the modulii decay towards zero as $|\Vie{p}{}{}| \to \infty$ (which can easily be deduced from a straightforward asymptotic analysis leading to $\alpha_1(\Vie{p}{}{}) - \alpha_2(\Vie{p}{}{}) \sim i (\epsilon_{\max}-\epsilon_{\min}) \omega^2 /(2 c^2 p)$).
The phase change associated with the transition from the real line to the imaginary line in the complex plane is therefore $- \pi / 2$. This gradual phase change is similar to that of the Goos-H\"{a}nchen phase shift discussed for the reflection by a planar surface. The difference of absolute total phase change, of $\pi$ for the case of the Fresnel amplitude and $\pi / 2$ in the case of the scattered waves, comes mathematically from the fact that in the former case the amplitude is written as the ratio of a complex number and its complex conjugate, while in the latter case there is no such ratio. The phase consequently turns twice as fast in the former case than in the latter.
A physical interpretation of this difference is that for the Fresnel amplitude both the incident and outgoing wave vector must vary simultaneously (since they are the same), while in the case of a scattered wave the incident wave vector is fixed while only the outgoing wave vector is allowed to vary. In fact, we have only analyzed the phase associated with the factor $\alpha_1(\Vie{p}{}{}) - \alpha_2(\Vie{p}{}{})$ in Eq.~(\ref{eq:RT1:scalar}). The phase of the overall complex amplitude will be the sum of the aforementioned phase, that given by the argument of $\hat{\zeta}(\Vie{p}{}{} - \Vie{p}{0}{})$, and the phase given by the argument of the total zero order amplitude $[1 + \rho^{(0)}(\Vie{p}{0}{})]$ or $\tau^{(0)}(\Vie{p}{0}{})$.
In particular, if the angle of incidence is such that total internal reflection occurs for the zero order field, the overall phase of the scattered amplitude will contain a signature of the Goos-H\"{a}nchen phase shift associated with the total internal reflection of the zero order field in addition to the corresponding Goos-H\"{a}nchen phase shift associated with the Yoneda effect. Note that when averaged over surface realizations, the phase contribution coming from $\hat{\zeta}$ averages to zero. This supports our choice of limiting the detailed investigation to the surface-independent factors in Eq.~(\ref{eq:RT1:scalar}).\\

To summarize, let us gather some important results and answer some of the questions which were left unanswered in previous studies. First, we would like to stress that the above analysis predicts a critical angle for the Yoneda phenomenon which is independent both of the angle of incidence and of which medium the incident wave came from. The Yoneda transition will therefore always occur at the same polar angle of scattering: the one given by $|\Vie{p}{c}{}| = k_{\min}$. We also want to emphasize that the approximate solution of the reduced Rayleigh equations obtained via SAPT to first order in the surface profile is commonly accepted as a single scattering approximation.
In light of our analysis of the Yoneda phenomenon, it is clear that the analogy of the Yoneda phenomenon with that of total internal reflection put forward in the literature may seem a valid one. There are, however, some comments to be made about this analogy. It is important to emphasize the underlying cause of total internal reflection, namely the impossibility of an evanescent wave to carry energy away from the surface, given the assumed scattering system. Indeed, trying to directly and naively apply the total internal reflection argument would lead one to expect an absence of the Yoneda effect in transmission into the dense medium based on a single scattering picture, as this would require multiple scattering events.
Indeed, one could imagine that the incident wave would need to scatter once to a transmitted grazing or evanescent wave and then a second time to be scattered in reflection in the dense medium and therefore follow the rule of total internal reflection. Such a naive picture would be in contradiction with results from numerical experiments based on first order perturbation theory \cite{Hetland2016a,Hetland2017}, or at least contradict the common single-scattering picture associated with it, and we believe that our interpretation in terms of coupled mode resolves this issue.
The results presented in Refs.~\citenum{Hetland2016a} and \citenum{Hetland2017} validated the qualitative use of SAPT in describing the Yoneda phenomenon, for the roughness parameters assumed in these studies, when compared to numerical results obtained through a non-perturbative solution of the reduced Rayleigh equations. Similar non-perturbative solutions were found to match experimental results showing the Yoneda phenomenon in Ref.~\citenum{Gonzalez-Alcalde2016}.

In fact, the Yoneda phenomenon for weakly rough surfaces originates from the same physical mechanism as the Rayleigh anamolies for periodic dielectric gratings. The continuous set of scattered wave vectors in the case of a randomly rough surface can be viewed as probing a diffracted order scattered from a periodic surface with continuously changing lattice constant. It is easy to show numerically and with SAPT to first order, that the behavior of the efficiency of a given diffractive order as the lattice constant is changed exhibits the same characteristic peak as the Yoneda peak when its counter part in the less dense medium becomes evanescent. The perturbative analysis in the case of a periodic grating is exactly the same as in the case of a randomly rough surface with the only difference being that $\Vie{p}{}{}$ must be replaced by the in-plane wave vector of the diffractive order of interest and make the lattice constant vary instead.

As a remark, we would like to point out that since the analysis was carried out for the scattering of a scalar wave subjected to the continuity of the field and its normal derivative with respect to the surface, we predict that the Yoneda phenomenon should also be observed for the scattering of a quantum particle by a rough interface between two regions of constant potential.

In studying the results from Figs.~\ref{fig:sapt_mdxc_incvacuum} -- \ref{fig:sapt_mdxc_incglass_contd} we avoided a direct discussion for p-polarized waves, for which the results put forward by the scalar wave analysis seem to be invalidated. The analysis done for scalar waves is, in fact, still valid but must be complemented with additional effects, due to polarization, not only for p-polarized light but also for s-polarized light when the scattering direction is out of the plane of incidence as suggested by Eqs.~(\ref{eq:rho1:all}, \ref{eq:tau1:all}). This is the subject of the following section.

\subsection{Physical and geometrical explanations of the Brewster scattering effect}
\label{sec:brewster_1storder}

For a randomly rough surface, we have seen in Figs.~\ref{fig:sapt_mdxc_incvacuum} -- \ref{fig:sapt_mdxc_incglass_contd} that we may find a Brewster scattering angle for a wide range of angles of incidence if we look at both the reflected and transmitted light (the MDRC and MDTC). We will now see that the general Brewster scattering phenomenon, roughly defined as a wave scattered with zero amplitude in a single scattering approximation, also extends to scattered waves in the evanescent regime. To this end we will continue our dissection of the phenomenon through perturbative theory.

\emph{In-plane reflection} --- Let us focus first on the case of co-polarized scattering in the plane of incidence to fix the ideas.
Equation~\eqref{eq:rho1ss} shows that $\hat{\rho}_{s s}^{(1)} ( \Vie{p}{}{} \st \Vie{p}{0}{})$ is proportional to $\Vie{\hat{e}}{s}{}(\Vie{p}{}{}) \cdot \Vie{E}{s,1}{(0)} (\Vie{p}{0}{})$, where $\Vie{E}{s,1}{(0)} (\Vie{p}{0}{})$ is the \emph{total zero order field amplitude} in medium 1 given by the sum of the unit incident field amplitude and the reflected field amplitude given by the Fresnel coefficient for an \spol-polarized wave.
This relation indicates that the field amplitude of the \emph{first order} reflected amplitude for the wave scattered with lateral wave vector $\Vie{p}{}{}$ is proportional to the projection of its polarization vector on the total zero order field. For scattering in the plane of incidence $\Vie{\hat{e}}{s}{}(\Vie{p}{}{}) = \Vie{\hat{e}}{s}{}(\Vie{p}{0}{})$ and therefore the first order reflection amplitude reduces to that of the scalar wave Eq.~\eqref{eq:R1:scalar}. Consequently, there is no extinction for $\spol\to\spol$ scattering in the plane of incidence for any angle of incidence. The same analysis and conclusion hold for the transmitted s-polarized wave.

\begin{figure*}[t]
  \centering
  \includegraphics[width=.95\linewidth, trim= 0.cm 0.cm 0.cm 0.cm,clip]{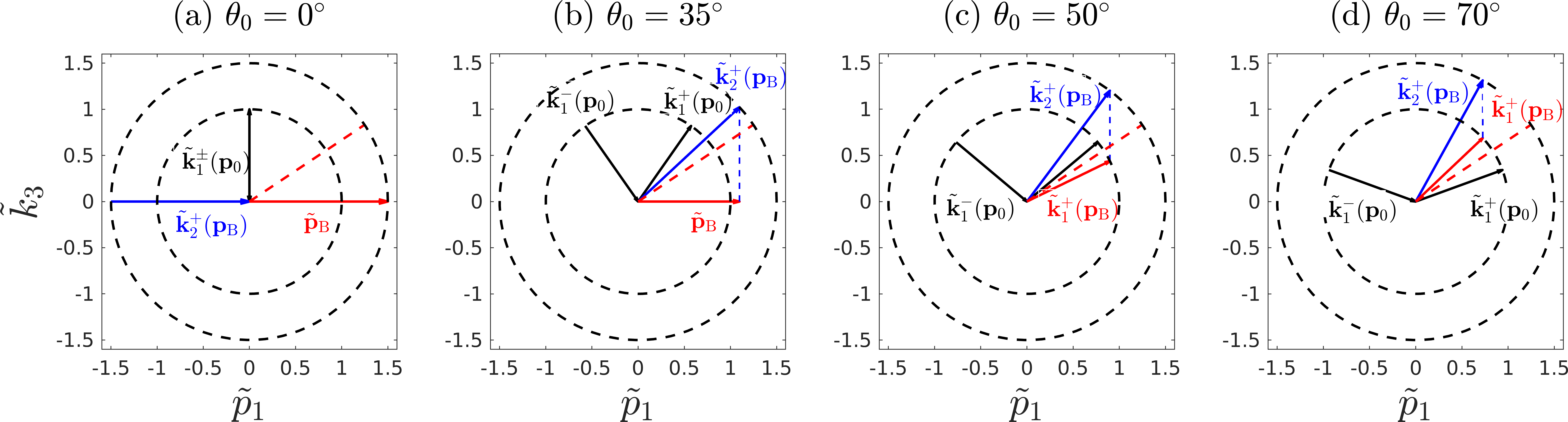}
  \caption{Illustration of the geometrical criterion for in-plane Brewster scattering for different polar angles of incidence: (a) $\theta_0 = \ang{0}$, (b) $\theta_0 = \ang{35}$, (c) $\theta_0 = \ang{50}$ and (d) $\theta_0 = \ang{70}$. The dashed circles represent the norm of the full wave vectors, given by the dispersion relations ($|\Vie{k}{j}{\pm}| = k_j = n_j \omega / c$), in vacuum ($\epsilon_1 = 1$ inner circle) and glass ($\epsilon_2 = 2.25$ outer circle).
  The black arrows represent respectively the incident wave vector $\Vie{k}{1}{-} (\Vie{p}{0}{})$, which is drawn as pointing towards the origin for clarity, and the wave vector of the reflected zero order wave, $\Vie{k}{1}{+} (\Vie{p}{0}{})$. The blue arrow represents the wave vector of the virtual wave, $\Vie{k}{2}{+} (\Vie{p}{\mathrm{B}}{})$, from which the lateral wave vector of the Brewster wave, $\Vie{p}{\mathrm{B}}{}$, is deduced by projection along $\hat{\mathbf{e}}_1$.
  From $\Vie{p}{\mathrm{B}}{}$, the full wave vector for the Brewster wave, $\Vie{k}{1}{+} (\Vie{p}{\mathrm{B}}{})$, can be drawn (provided propagation in medium 1) as a red arrow. Note that if the Brewster wave is evanescent, only $\Vie{p}{\mathrm{B}}{}$ is draw in red as the out-of-plane component of $\Vie{k}{1}{+} (\Vie{p}{\mathrm{B}}{})$ is purely imaginary. The red dashed line indicates the Brewster angle for a planar surface approximately equal to $\ang{56.3}$ in this case. Unit-less wavectors, $\tilde{k} = k c / \omega$ , are used in the illustration.}
  \label{fig:wavevectors_vg}
\end{figure*}

Similarly, for p-polarized light, Eq.~\eqref{eq:rho1pp} shows that the first order reflection amplitude is proportional to $\Vie{\hat{e}}{p,2}{+}(\Vie{p}{}{}) \cdot \Vie{E}{1,p}{(0)}(\Vie{p}{0}{})$, where we recall that $\Vie{E}{1,p}{(0)}(\Vie{p}{0}{})$ is the total zero order field amplitude given by the sum of the unit incident field amplitude and the reflected field amplitude given by the Fresnel coefficient for \ppol-polarized waves. Equation~(\ref{eq:rho1pp}) states that the \emph{first order} field amplitude is proportional to the projection of the Snell-conjugate wave's polarization vector $\Vie{\hat{e}}{p,2}{+}(\Vie{p}{}{})$ along the direction of the \emph{total zero order field}. Note the similarity with what was found for the Fresnel coefficient for p-polarized light in Eq.~(\ref{eq:rhopp0}).
From Eq.~(\ref{eq:rho1pp}) we can deduce a simple geometrical criterion for Brewster scattering within first order perturbation theory:\emph{ The lateral wave vector(s) $\Vie{p}{\mathrm{B}}{}$ of the elementary Brewster scattered wave(s), for which the reflection amplitude for a p-polarized reflected wave vanishes given a p-polarized incident wave with lateral wave vector $\Vie{p}{0}{}$ is given by the condition of orthogonality between the p-polarization vector of the Snell-conjugate scattered wave(s) and the total zero order field in medium 1, i.e.}
\begin{equation}
 \Vie{\hat{e}}{p,2}{+}(\Vie{p}{\mathrm{B}}{}) \cdot \Vie{E}{p,1}{(0)} (\Vie{p}{0}{}) = 0 \: .
 \label{eq:brewster_criterion}
\end{equation}
As a direct consequence, in the case of co-polarized scattering in the plane of incidence, the geometrical condition can be re-stated as a requirement on the colinearity between the Snell-conjugate wave vector and the total zero order field, which is exactly the same geometrical criterion found in the case of reflection from a planar interface.
%
A second corollary is that for in-plane scattering $\Theta_{\mathrm{B}} (\theta_\mathrm{B}) = \theta_\mathrm{B}$: the \emph{Brewster scattering angle} is equal to the \emph{Brewster angle for a planar interface} when the angle of incidence is equal to the Brewster angle for a planar interface, $\theta_0 = \theta_\mathrm{B}$ (or so-called Brewster incidence).
In other words, the Brewster angle for a planar interface, $\theta_\mathrm{B}$, is a fixed point for the mapping which associates the angle of incidence to the Brewster scattering angle: $\Theta_\mathrm{B} : \theta_0 \mapsto \Theta_\mathrm{B}(\theta_0)$.
This is readily understood from the geometrical criterion expressed by Eq.~(\ref{eq:brewster_criterion}). At Brewster incidence the zero order reflected wave vanishes (by definition of Brewster incidence). Thus the total zero field amplitude is simply the incident field amplitude, $\Vie{E}{p,1}{(0)} (\Vie{p}{0}{}) = \Vie{\hat{e}}{p,1}{-}(\Vie{p}{0}{})$, and consequently, the Brewster scattering angle is necessarily equal to $\theta_\mathrm{B}$.

\begin{figure*}[t]
  \centering
  \includegraphics[width=.75\linewidth, trim= 0.cm 0.cm 0.cm 0.cm,clip]{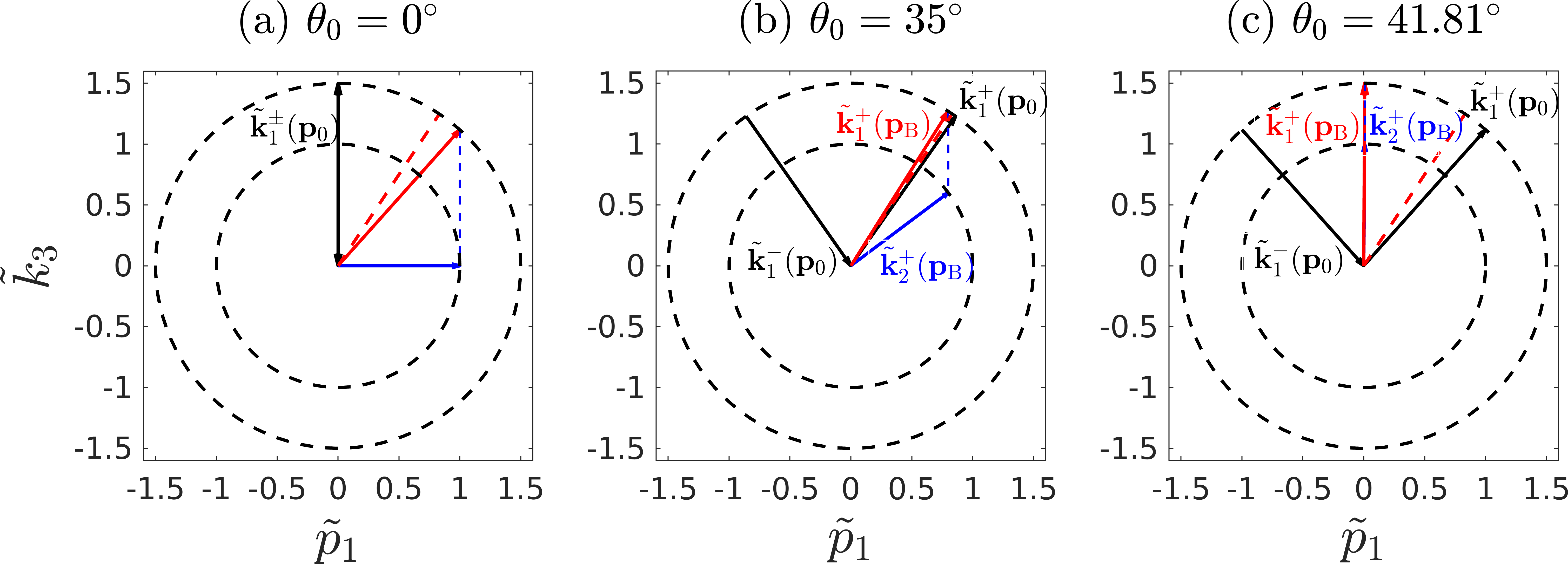}
  \caption{Same as Fig.~\ref{fig:wavevectors_vg} but for incidence in glass ($\epsilon_1 = 2.25$ and $\epsilon_2 = 1$) and for polar angles of incidence: (a) $\theta_0 = \ang{0}$, (b) $\theta_0 = \ang{35}$ and (c) $\theta_0 = \ang{41.81}$. The Brewster angle for a planar surface is approximately equal to $\ang{33.7}$ in this case.}

  \label{fig:wavevectors_gv}
\end{figure*}

Let us now apply the above criterion for tracking the Brewster scattering direction while the angle of incidence varies. We start with the case where the incident plane wave is approaching the rough interface from vacuum, and is reflected from a glass substrate [$\epsilon_1 = 1$ and $\epsilon_2 = 2.25$].
Figure~\ref{fig:wavevectors_vg} presents selected wave vectors for different polar angles of incidence $\theta_0$, highlighting the geometrical relations leading to the Brewster scattering direction. 
 The general construction rules go as follows. First, the wave vectors of the incident and the reflected zero order waves are drawn in black. Second, the direction of the total zero order field given by Eq.~(\ref{eq:E1p}) is determined and the wave vector of the Brewster Snell-conjugate wave, which is colinear to the total zero order field (not represented), is drawn as the blue wave vector $\Vie{k}{2}{+} (\Vie{p}{\mathrm{B}}{})$.
Note that $\Vie{k}{2}{+} (\Vie{p}{\mathrm{B}}{})$ lies on the circle of radius $k_2$. The projection of $\Vie{k}{2}{+} (\Vie{p}{\mathrm{B}}{})$ along $\hat{\mathbf{e}}_1$ gives the Brewster lateral wave vector $\Vie{p}{\mathrm{B}}{}$ from which we deduce $\Vie{k}{1}{+} (\Vie{p}{\mathrm{B}}{})$ in red. Note that the reflected wave associated with $\Vie{k}{1}{+} (\Vie{p}{\mathrm{B}}{})$ may be evanescent, and in that case we simply represent its lateral component $\Vie{p}{\mathrm{B}}{}$ as its component along $\hat{\mathbf{e}}_3$ is pure imaginary.

For normal incidence [Fig.~\ref{fig:wavevectors_vg}(a)] the total zero order electric field lies along $\hat{\mathbf{e}}_1$, and consequently, so does $\Vie{k}{2}{+} (\Vie{p}{\mathrm{B}}{})$. In fact, for normal incidence, due to the fact that the total zero order field lies along $\hat{\mathbf{e}}_1$, there are \emph{two} Brewster waves in the plane of incidence with opposite wave vectors $p_1 = \pm k_2$, but we focus on the one pointing to the right for clarity in Fig.~\ref{fig:wavevectors_vg}(a).
It follows from $\Vie{k}{2}{+} (\Vie{p}{\mathrm{B}}{})$ that $|\Vie{p}{\mathrm{B}}{}| > k_1$ and the corresponding Brewster (non-)~reflected wave is therefore evanescent. Such a case could not be revealed in previous work which focused on the diffusely scattered intensity radiated away from the surface. Nevertheless, the theory suggests that the notion of Brewster scattering should be extended to evanescent waves. This effect is indeed visible by inspection of the modulus of the amplitude $\rho_{pp}^{(1)}$ in Fig.~\ref{fig:sapt_mdxc_incvacuum}(a).
Indeed, we observe that for $p_1 = \pm k_2$, $\rho_{pp}^{(1)}$ vanishes. The corresponding phase $\phi_{pp}^{\mathrm{r},(1)}$ exhibits a jump which is characteristic of the Brewster effect. The phase jump is equal to $\pi/2$ in this case, while in general the phase jump associated with the Brewster effect is equal to $\pi$. The $\pi/2$ jump seems to happen only when two Brewster waves with opposite lateral wave vectors are solutions of the criterion Eq.~(\ref{eq:brewster_criterion}), which as far as we can see only occurs at normal incidence for the systems studied in this paper. It is tempting to interpret the $\pi/2$ jump as actually a $\pi$ jump evenly shared by the two Brewster waves (although this is over interpreted as we will see later).
By progressively increasing the polar angle of incidence, the direction of the total zero order field changes, and so does the wave vector of the Brewster Snell-conjugate wave (which now is unique). For a polar angle of incidence equal to $\ang{35}$, as sketched in Fig.~\ref{fig:wavevectors_vg}(b), we can observe that the projection of $\Vie{k}{2}{+} (\Vie{p}{\mathrm{B}}{})$ along $\hat{\mathbf{e}}_1$ still yields an evanescent Brewster wave, but the lateral wave vector is now closer to the propagation limit. This case corresponds to the parameters assumed in obtaining the results in Fig.~\ref{fig:sapt_mdxc_incvacuum}(b) and we can observe that $\rho_{pp}^{(1)}$ vanishes indeed for $p_1$ just above $k_1$, and that the corresponding phase exhibits a $\pi$ jump.
By further increasing the polar angle of incidence the Brewster wave is found in the propagating region as $|\Vie{p}{\mathrm{B}}{}| < k_1$, and its full wave vector can now be represented as following the inner dashed circle. As the polar angle of incidence increases towards the Brewster angle for a planar surface, the wave vector associated with the reflected zero order wave $\Vie{k}{1}{+} (\Vie{p}{0}{})$, drawn in black in Fig.~\ref{fig:wavevectors_vg}, and the wave vector of the Brewster scattered wave both approach the red dashed line from either sides and cross it at the same angle of incidence, namely the Brewster angle for a planar interface, $\theta_0 = \theta_\mathrm{B}$. Figure~\ref{fig:wavevectors_vg}(c) shows the case where $\theta_0 = \ang{50}$ at a slightly lower angle than the Brewster angle of incidence (approximately equal to $\ang{56.3}$), i.e. just before the cross-over.
When $\theta_0$ is further increased the lateral component of the Brewster wave vector continues to decrease. Figure~\ref{fig:wavevectors_vg}(d) assumes $\theta_0 = \ang{70}$ which corresponds to Fig.~\ref{fig:sapt_mdxc_incvacuum}(c) where we now observe that the Brewster wave is indeed in the propagating region as can be seen both from $\rho_{pp}^{(1)}$ and the extinction of the incoherent component of the MDRC. Note also the $\pi$ jump in the phase. Finally, as the polar angle of incidence approaches $\ang{90}$, the Brewster Snell-conjugate wave does \emph{not} approach the vertical direction as one might naively expect. Indeed, the total zero order field does not become oriented along $\hat{\mathbf{e}}_3$ but along the direction given by the critical angle for total internal reflection.

We now repeat the analysis but for an incident wave approaching the surface in the denser medium ($\ve_1 = 2.25$, $\ve_2 = 1.0$). For normal incidence, the total zero order field is along $\hat{\mathbf{e}}_1$, and yet again we recover two Brewster waves. However, since now the the Snell-conjugate waves are lying on the inner circle ($\epsilon_2 = 1$), the wave vectors $\Vie{k}{1}{+} (\pm \Vie{p}{\mathrm{B}}{})$ correspond to propagating waves in glass, and coincide with the Yoneda threshold. This situation is illustrated in Figs.~\ref{fig:wavevectors_gv}(a) and ~\ref{fig:sapt_mdxc_incglass}(a).
Due to the presence of two Brewster waves, the phase jump is $\pi /2$ [see Fig.~\ref{fig:sapt_mdxc_incglass}(a)]. The coincidence of the Yoneda threshold and the Brewster scattering angle for internal reflection for normal incidence is now explained, and we see that although the two effects are of different nature and decoupled, they occur simultaneously in this case simply as a consequence of the geometry imposed by the dispersion relations.
As the polar angle of incidence is increased, only one Brewster wave remains, and the corresponding lateral wave vector shrinks [see Figs.~\ref{fig:wavevectors_gv}(b) and \ref{fig:sapt_mdxc_incglass}(b)]. The wave vectors of the reflected zero order wave and of the Brewster wave cross each other at the Brewster angle of incidence ($\approx\ang{33.7}$). Now comes an interesting effect which was not present when the wave was incident from the less dense medium.
As the polar angle of incidence approaches the critical angle of total internal reflection of the zero order reflected wave, the Snell-conjugate wave vector and that of the Brewster wave approach the vertical direction and reach it for $\theta_0 = \theta_c$, as displayed in Fig.~\ref{fig:sapt_mdxc_incglass}(c). Then a sudden transition occurs when $\theta_0$ is increased beyond $\theta_c$. In Fig.~\ref{fig:sapt_mdxc_incglass_contd}(a), which shows results for $\theta_0$ just above the critical angle, it seems that the Brewster scattering angle is nowhere to be found. However, the Brewster scattering angle now comes back from the left (backscattering) side, visible in the evanescent region of Fig.~\ref{fig:sapt_mdxc_incglass_contd}(c) where the polar angle of incidence is $\ang{70}$.
What happened? The overall behavior of the phase in Figs.~\ref{fig:sapt_mdxc_incglass_contd}(a)--(c) gives us a good indication. We have mentioned earlier that for s-polarized light, when the zero order reflected wave undergoes total internal reflection, the central phase plateau must undergo a Goos-H\"{a}nchen shift with $\theta_0$ (in fact it is the whole graph which undergoes the shift). Similarly, the p-polarized zero order reflected wave undergoes a Goos-H\"{a}nchen shift and, as a consequence, the two terms in Eq.~(\ref{eq:E1p}) are \emph{not} any longer in phase. In the case where $\epsilon_1 < \epsilon_2$, the arguments of the two complex amplitudes in Eq.~(\ref{eq:E1p}) are always either in phase or are separated by a phase difference of $\pi$.
Therefore, as time progresses, the \emph{real} total zero order field keeps a fixed direction. When $\epsilon_1 > \epsilon_2$, the Goos-H\"{a}nchen phase shift makes the real total zero order field change direction and turn in the plane of incidence as time progresses (it describes an ellipse). Intuitively, this seems to indicate that the corresponding dipole radiation is not expected to be that of an oscillating dipole anymore but that of a \emph{rotating} dipole. It is therefore understandable that the measurement of a propagating p-polarized wave does not yield any direction of extinction when the radiation is emitted from a rotating dipole.
Stated in an equivalent way, $\rho_{pp}^{(0)}$ now draws a lower half circle in the complex plane from $1$ to $-1$ as the angle of incidence is varied from the critical angle to $\ang{90}$, while it previously stayed on the real line. It follows that $\Vie{E}{p,1}{(0)} (\Vie{p}{0}{})$ has a complex amplitude with non-zero imaginary part.
It is therefore not possible for a propagating Snell-conjugate wave to satisfy the requirement of Eq.~(\ref{eq:brewster_criterion}) since its p-polarization vector would be real. Hence, in order to satisfy Eq.~(\ref{eq:brewster_criterion}) the polarization vector $\Vie{\hat{e}}{p,2}{+}(\Vie{p}{\mathrm{B}}{})$ must itself be complex, and the Snell-conjugate wave is now found in the \emph{evanescent} region of medium 2. This is the reason why the Brewster scattering lateral wave vector seems to disappear at the transition $\theta_0 = \theta_c + \Delta \theta_0$ and then come back from the negative $p_1$ side as the angle of incidence is increased, which reveals the evanescent nature of the Snell-conjugate wave.

Note that what we have defined as a p-polarized wave, according to the polarization vector $\Vie{\hat{e}}{p,j}{\pm}(\Vie{p}{}{})$ given in Eq.~\eqref{basis:ep}, takes a rather interesting structure when it is evanescent. For an evanescent wave, $\alpha_j(\Vie{p}{}{})$ is pure imaginary and the polarization vector $\Vie{\hat{e}}{p,j}{\pm}(\Vie{p}{}{})$ hence has a real component along $\Vie{\hat{e}}{3}{}$ and a pure imaginary component along the transverse wave vector direction.
This means that the corresponding real electric field is the sum of a wave polarized along $\Vie{\hat{e}}{3}{}$ and a longitudinal wave (\emph{longitudinal} with respect to the lateral wave vector) dephased by $\pi / 2$ radians with respect to the first wave. The resulting field therefore describes an ellipse in the $(\Vie{\hat{e}}{3}{},\hat{\Vie{p}{}{}})$-plane.\\

\begin{figure*}[ht]
\includegraphics[width=.9\linewidth , trim= 3.5cm 2.5cm 3.5cm 4.cm,clip]{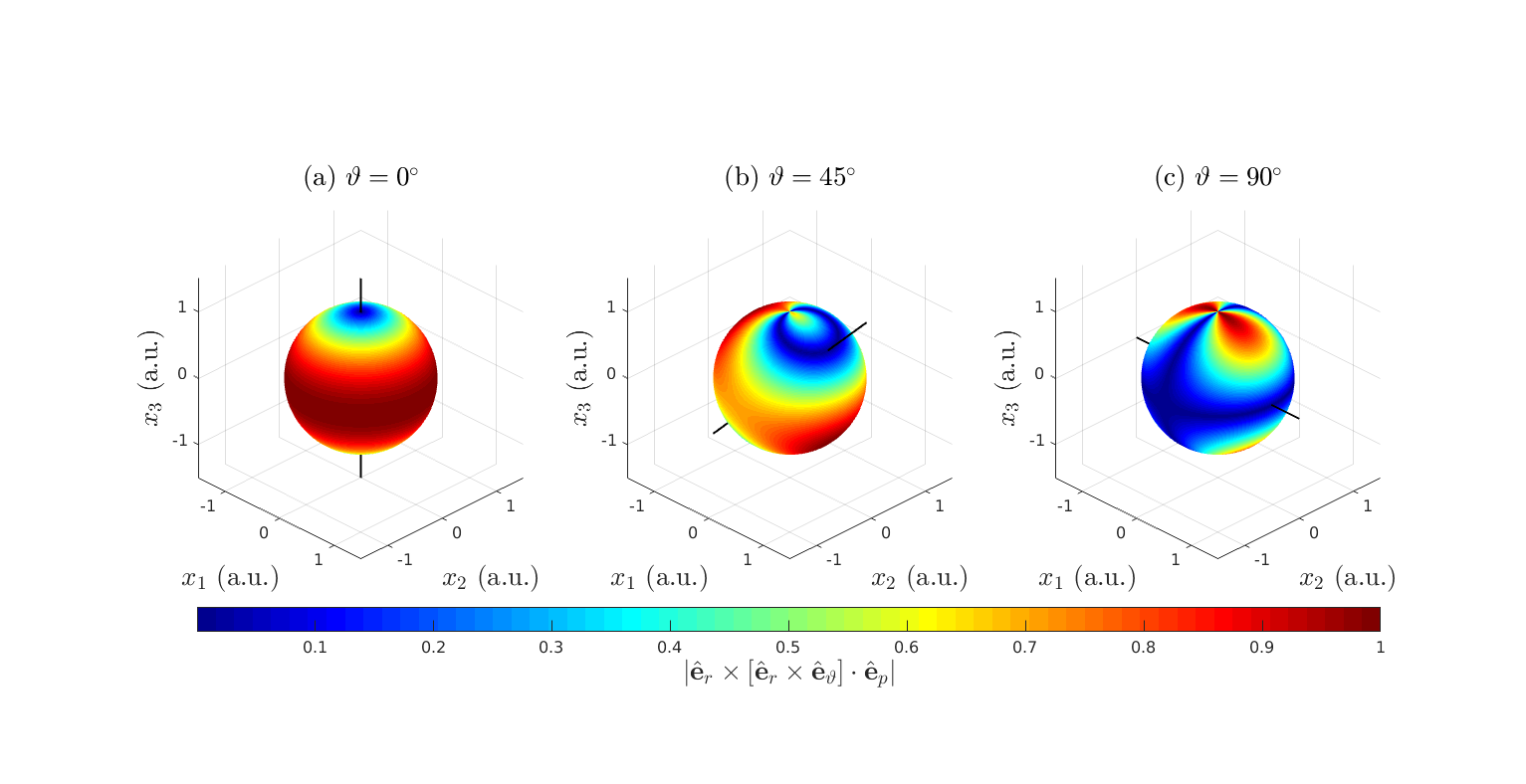}
\includegraphics[width=.9\linewidth , trim= 3.5cm 2.5cm 3.7cm 4.cm,clip]{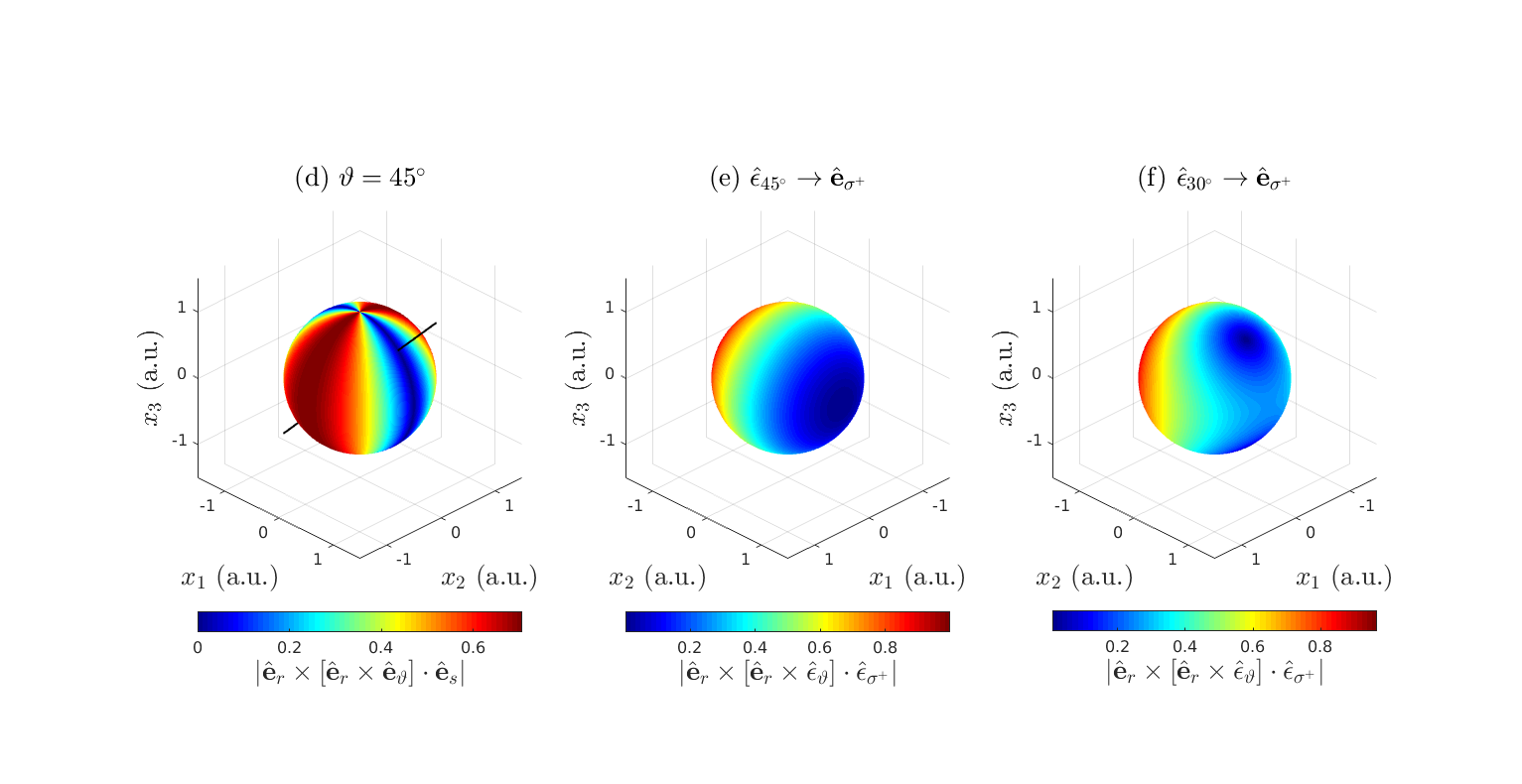}
\caption{
  (a-c) Dependence of the p-polarized radiation of a tilted dipole in free space, $|\Vie{\hat{e}}{r}{} \times [\Vie{\hat{e}}{r}{} \times \Vie{\hat{e}}{\vartheta}{}] \cdot \Vie{\hat{e}}{p}{}|$, on the direction of $\Vie{\hat{e}}{r}{}$ as it runs over the unit sphere for different dipole tilting angles $\vartheta \in \{\ang{0}, \ang{45}, \ang{90}\}$. (d) Similar dependence of the s-polarized radiation of a tilted dipole in free space on $\Vie{\hat{e}}{r}{}$ for $\vartheta = \ang{45}$.
  The black line in panels (a-d) indicates the direction of the dipole moment. (e-f) Dependence of the $\sigma^+$-polarized radiation of a rotating dipole in free space parametrized by $\vartheta = \ang{45}$ and $\vartheta = \ang{30}$ respectively (note the orientation of the coordinate system).
}
\label{fig:dipole}
\end{figure*}

\smallskip
\emph{In-plane transmission} --- The analysis for the Brewster scattering effect in the transmitted light is similar to that of the reflected light and will not be described in details. One difference worth mentioning, however, is that the Brewster scattering direction is generally found in the backscattering region when the corresponding Brewster scattering for reflection is found in the forward scattering direction. Intuitively, this effect can be related to the emission of an oscillating dipole which yields zero emitted power along the direction of oscillation, hence producing two antipodal zero intensity points when the intensity is mapped onto a sphere. This fact is better illustrated in the next section.

\subsection{Polarization of the radiation of oscillating and rotating dipoles in free space}\label{sec:dipole}

Before treating the full angular distribution of the light scattered diffusely by a randomly rough surface, we allow ourselves a detour via the analysis of the polarization properties of the radiation emitted by an oscillating dipole or a rotating dipole in free space. The study of the polarization of the radiation in these two cases gives remarkable insight and intuition into the qualitative physical mechanisms at play for the case of the scattering from a random interface, for which a more quantitative analysis requiring Snell-conjugate waves will be given in the next section.\\

\emph{Polarization of the radiation from an oscillating dipole in free space with respect to the local $(\Vie{\hat{e}}{p}{},\Vie{\hat{e}}{s}{})$ basis} -- 
We consider first the radiation emitted by a single oscillating dipole in free space. We let this dipole, of dipole moment $\Vie{D}{}{}(\vartheta) = d(\sin \vartheta \, \Vie{\hat{e}}{1}{} + \cos \vartheta \, \Vie{\hat{e}}{3}{}) = d \Vie{\hat{e}}{\vartheta}{}$, be tilted from the $x_3$-axis by an angle of $\vartheta \in [0,\pi/2]$ radians. The dipole is placed in free space at the origin of the coordinate system, where it oscillates with angular frequency $\omega$ and radiates the following electric field in the far-field \cite{jackson}:
\begin{equation}
  \Vie{E}{\mathrm{dip}}{} (\Vie{r}{}{},t) = - \frac{\omega^2}{4 \pi \varepsilon_0 c^2} \: \frac{\Vie{\hat{e}}{r}{} \times [\Vie{\hat{e}}{r}{} \times \Vie{D}{}{}(\vartheta)]}{r} \: e^{-i\omega(t-r/c)} \: , \label{eq:dipole}
\end{equation}
where $\Vie{r}{}{} = r \, \Vie{\hat{e}}{r}{} = r (\sin \theta \cos \phi \Vie{\hat{e}}{1}{} + \sin \theta \sin \phi \Vie{\hat{e}}{2}{} + \cos \theta \Vie{\hat{e}}{3}{})$ is the point of observation, and $r = |\Vie{r}{}{}|$. It is well known that no power is radiated along the axis of oscillation of the dipole ($\Vie{\hat{e}}{r}{} \times \Vie{D}{}{}(\vartheta)$ vanishes in Eq.~\eqref{eq:dipole} when $\Vie{\hat{e}}{r}{} \parallel \Vie{\hat{e}}{\vartheta}{}$) and that the radiation is polarized in accordance with the cross products in Eq.~\eqref{eq:dipole}.
The electric field is polarized along the vector $\Vie{\hat{e}}{\theta}{}'$, which is the basis vector tangent to a meridian in a spherical coordinate system $(r,\theta', \phi')$ attached to the dipole direction. We are, however, interested in analyzing the polarization of the dipole radiation with respect to the local polarization basis given in Eq.~\eqref{basis}, which is defined with respect to the propagation direction of the radiation \emph{and} the plane $x_3 = 0$.
Thus we study the following dot products: $\Vie{\hat{e}}{r}{} \times [\Vie{\hat{e}}{r}{} \times \Vie{\hat{e}}{\vartheta}{}] \cdot \Vie{\hat{e}}{s}{}$ and $\Vie{\hat{e}}{r}{} \times [\Vie{\hat{e}}{r}{} \times \Vie{\hat{e}}{\vartheta}{}] \cdot \Vie{\hat{e}}{p}{}$ where $\Vie{\hat{e}}{s}{} = \frac{\Vie{\hat{e}}{3}{} \times \Vie{\hat{e}}{r}{}}{|\Vie{\hat{e}}{3}{} \times \Vie{\hat{e}}{r}{}|}$ and $\Vie{\hat{e}}{p}{} = \frac{\Vie{\hat{e}}{s}{} \times \Vie{\hat{e}}{r}{}}{|\Vie{\hat{e}}{s}{} \times \Vie{\hat{e}}{r}{}|}$ are defined with respect to $\Vie{\hat{e}}{r}{}$ in order to mimic the local s- and p-polarization vectors attached to a scattering direction along $\Vie{\hat{e}}{r}{}$.
The unit vectors $\Vie{\hat{e}}{p}{} = \Vie{\hat{e}}{\theta}{} = \mathrm{d} \Vie{\hat{e}}{r}{} / \mathrm{d}\theta$ and $\Vie{\hat{e}}{s}{} = \Vie{\hat{e}}{\phi}{} = 1/ \sin \theta \: \mathrm{d} \Vie{\hat{e}}{r}{} / \mathrm{d}\phi$ are also the conventional basis vectors in spherical coordinates.
First we observe that $\Vie{\hat{e}}{r}{} \times [\Vie{\hat{e}}{r}{} \times \Vie{\hat{e}}{\vartheta}{}] \cdot \Vie{\hat{e}}{s}{}$ and $\Vie{\hat{e}}{r}{} \times [\Vie{\hat{e}}{r}{} \times \Vie{\hat{e}}{\vartheta}{}] \cdot \Vie{\hat{e}}{p}{}$ are invariant under the transformation $\Vie{\hat{e}}{r}{} \mapsto - \Vie{\hat{e}}{r}{}$, and so the s- and p-polarized distributions of the dipole radiation are symmetric with respect to the origin as $\Vie{\hat{e}}{r}{}$ runs over the unit sphere.
Second, for $\vartheta \in (0,\pi /2]$ radians the identity $\Vie{a}{}{} \times [ \Vie{b}{}{} \times \Vie{c}{}{}] = (\Vie{a}{}{} \cdot \Vie{c}{}{}) \Vie{b}{}{} - (\Vie{a}{}{} \cdot \Vie{b}{}{}) \Vie{c}{}{}$ leads to
\begin{equation}
  \Vie{\hat{e}}{r}{} \times [\Vie{\hat{e}}{r}{} \times \Vie{\hat{e}}{\vartheta}{}] = (\Vie{\hat{e}}{r}{} \cdot \Vie{\hat{e}}{\vartheta}{}) \Vie{\hat{e}}{r}{} - \Vie{\hat{e}}{\vartheta}{} \: ,
\end{equation}
%
hence the projection of the dipole radiation on the local s-polarization basis reads
\begin{equation}
  \Vie{\hat{e}}{r}{} \times [\Vie{\hat{e}}{r}{} \times \Vie{\hat{e}}{\vartheta}{}] \cdot \Vie{\hat{e}}{s}{} = - \Vie{\hat{e}}{\vartheta}{} \cdot  \Vie{\hat{e}}{\phi}{} = - \sin \vartheta \sin \phi \: .
\label{eq:dipole_s}
\end{equation}
A direct consequence of Eq.~\eqref{eq:dipole_s} is that $\Vie{\hat{e}}{r}{} \times [\Vie{\hat{e}}{r}{} \times \Vie{\hat{e}}{\vartheta}{}] \cdot \Vie{\hat{e}}{s}{}$ vanishes for all $\Vie{\hat{e}}{r}{}$ in the $(\Vie{\hat{e}}{1}{},\Vie{\hat{e}}{3}{})$-plane [see Fig.~\ref{fig:dipole}(d)].
The corresponding projection on the local p-polarization basis reads
\begin{equation}
  \Vie{\hat{e}}{r}{} \times [\Vie{\hat{e}}{r}{} \times \Vie{\hat{e}}{\vartheta}{}] \cdot \Vie{\hat{e}}{p}{} = - \Vie{\hat{e}}{\vartheta}{} \cdot  \Vie{\hat{e}}{\theta}{} \: ,
\label{eq:pdipole}
\end{equation}
which is a quantity that depends on $\vartheta$, $\theta$, and $\phi$. In the particular case where $\Vie{\hat{e}}{\theta}{}$ belongs to the $(\Vie{\hat{e}}{1}{},\Vie{\hat{e}}{3}{})$-plane, there are two solutions for Eq.\eqref{eq:pdipole} equal to zero: $\Vie{\hat{e}}{r}{} = \pm \Vie{\hat{e}}{\vartheta}{}$, which correspond to the two intersections of the dipole moment direction with the unit sphere.
This is not surprising since we already know that no power is emitted along the direction of oscillation of the dipole, independent of polarization. More interesting are cases for which $\Vie{\hat{e}}{\theta}{}$, and hence $\Vie{\hat{e}}{r}{}$, does not belong to the $(\Vie{\hat{e}}{1}{},\Vie{\hat{e}}{3}{})$-plane. Expanding the dot product in Eq.~\eqref{eq:pdipole} in terms of the angles $\vartheta$, $\theta$, and $\phi$ we obtain the following implicit equation for the set of points on the unit sphere where the p-polarization component of the dipole radiation vanishes:
\begin{equation}
  \sin \vartheta \, \cos \theta \, \cos \phi - \cos \vartheta \, \sin \theta = 0 \: ,
\end{equation}
or equivalently for non-pathologic cases
\begin{equation}
  \frac{\tan \vartheta}{\tan \theta} = \frac{1}{\cos \phi} \: .
\label{eq:tan_dipole}
\end{equation}
We verify that for the cases $\phi = 0$ and $\phi = \pi$ radians, we recover that $\theta = \vartheta$ and $\theta = \pi - \vartheta$, i.e. the points of intersection of the dipole moment direction and the unit sphere. For $\phi \in (-\pi/2,\pi/2)$, $\cos \phi > 0$, which implies that $\tan \theta > 0$ (recall that $0 < \vartheta < \pi/2$ hence $\tan \vartheta > 0$) and $\tan \vartheta > \tan \theta$.
By the monotony of the tangent function, and the continuity of Eq.~\eqref{eq:tan_dipole} with respect to the variables, we thus deduce that when $\phi$ varies in $(-\pi/2,\pi/2)$ the set of the points of zero traces a curve on the unit sphere latitude-bounded by $\theta < \vartheta$.
By the aforementioned symmetry of the polarization dependence of the dipole radiation with respect to the origin we immediately deduce that when $\phi$ varies in $(\pi/2,3\pi/2)$ the set of the points of zero traces a curve on the unit sphere latitude-bounded by $\theta > \pi - \vartheta$. This is well illustrated in Fig.~\ref{fig:dipole}(b) where $|\Vie{\hat{e}}{r}{} \times [\Vie{\hat{e}}{r}{} \times \Vie{\hat{e}}{\vartheta}{}] \cdot \Vie{\hat{e}}{p}{}|$ is shown as a color map on a unit sphere. For $\vartheta = \ang{45}$ we here observe that the curve of zero p-polarized radiation passes through both the north pole and the intersection point of the dipole moment direction on the northern hemisphere.
The degenerate cases $\vartheta = \ang{0}$ [Fig.~\ref{fig:dipole}(a)] and $\vartheta = \ang{90}$ [Fig.~\ref{fig:dipole}(c)] are also illustrated. For these cases the curves of zero p-polarized radiation reduces to two points (the poles) in the former case, and the equator ($\theta = \pi /2$) and meridians $\phi = \pm \pi/2$ in the latter. Indeed, $\theta$ must go to zero when $\vartheta \to 0$ as $\tan \vartheta$ vanishes, and, \emph{either} $\phi$ must go towards $\pm \pi / 2$ \emph{or} $\theta$ must go towards $\pi/2$ when $\vartheta \to \pi /2$ as $\tan \vartheta$ diverges.\\

\emph{Polarization of the radiation from a rotating dipole in free space with respect to the local $(\Vie{\hat{e}}{\sigma^+}{},\Vie{\hat{e}}{\sigma^-}{})$ basis} -- We now consider the radiation of a dipole rotating in the $(\Vie{\hat{e}}{1}{},\Vie{\hat{e}}{3}{})$-plane. Equation~(\ref{eq:dipole}) still holds, but we need to modify the dipole moment which now reads
\begin{equation}
  \Vie{D}{}{}(\vartheta) = d \: ( \sin \vartheta \, \Vie{\hat{e}}{1}{} + i \cos \vartheta \, \Vie{\hat{e}}{3}{}) = d \, \hat{\boldsymbol{\epsilon}}_{\vartheta} \: .
\end{equation}
The real vector $\mathrm{Re} [\hat{\boldsymbol{\epsilon}}_{\vartheta} \exp(-i \omega t) ]$ hence describes an ellipse in the $(\Vie{\hat{e}}{1}{},\Vie{\hat{e}}{3}{})$-plane whose excentricity is parametrized by $\vartheta$. In the limiting cases $\vartheta = 0$ and $\vartheta = \pi / 2$ radians we obtain an oscillating dipole along $\Vie{\hat{e}}{3}{}$ and $\Vie{\hat{e}}{1}{}$ respectively.
For $\vartheta = \pi/4$ we obtain a circularly rotating dipole. We now consider the polarization of the radiation from such an elliptically rotating dipole with respect to the local left and right circularly polarized basis $\Vie{\hat{e}}{\sigma^+}{}$ and $\Vie{\hat{e}}{\sigma^-}{}$ defined as
\begin{equation}
\Vie{\hat{e}}{\sigma^\pm}{} = \frac{1}{\sqrt{2}} \, ( \Vie{\hat{e}}{p}{} \pm i  \Vie{\hat{e}}{s}{}) \: .
\end{equation}
The $\sigma^+$ polarization component of the rotating dipole radiation is then measured by
\begin{equation}
\Vie{\hat{e}}{r}{} \times [\Vie{\hat{e}}{r}{} \times \hat{\boldsymbol{\epsilon}}_{\vartheta}] \cdot \Vie{\hat{e}}{\sigma^+}{} = - \hat{\boldsymbol{\epsilon}}_{\vartheta} \cdot  \Vie{\hat{e}}{\sigma^+}{} \: ,
\label{eq:rot:dipole}
\end{equation}
which when expressed in terms of the angles reads\footnote{The dot product here must be taken as the Hermitian inner product for complex vectors $\mathbf{a} \cdot \mathbf{b} = \sum_j a_j^* b_j$.}
\begin{align}
&\Vie{\hat{e}}{r}{} \times [\Vie{\hat{e}}{r}{} \times \hat{\boldsymbol{\epsilon}}_{\vartheta}] \cdot \Vie{\hat{e}}{\sigma^+}{} = - \frac{1}{\sqrt{2}} \: \sin \vartheta \cos \theta \cos \phi \nonumber \\ &- \frac{i}{\sqrt{2}} \: ( \cos \vartheta \sin \theta - \sin \vartheta \sin \phi ) \: .
\label{eq:rot:dipole:angles}
\end{align}
The modulus square of Eq.~(\ref{eq:rot:dipole:angles}) yields
\begin{align}
&|\Vie{\hat{e}}{r}{} \times [\Vie{\hat{e}}{r}{} \times \hat{\boldsymbol{\epsilon}}_{\vartheta}] \cdot \Vie{\hat{e}}{\sigma^+}{} |^2 =  \frac{1}{2} \: \sin^2 \vartheta \cos^2 \theta \cos^2 \phi \nonumber \\ &+ \frac{1}{2} \: ( \cos \vartheta \sin \theta - \sin \vartheta \sin \phi )^2 \: .
\label{eq:rot:dipole:angles:square}
\end{align}
The directions of zero $\sigma^+$-polarized light radiation are obtained if and only if both terms on the right-hand side of Eq.~(\ref{eq:rot:dipole:angles:square}) are zero. The first term vanishes if at least $\sin \vartheta$, $\cos \theta$ or $\sin \phi$ is zero. If we first assume that $\vartheta = 0$, then the second term is zero if and only if the condition $\sin \theta = 0$ is satisfied. Such a case corresponds to a dipole oscillating along the $x_3$-axis and its radiation vanishes at the poles of the unit sphere. More interesting are the cases for which $\vartheta \neq 0$ and either $\theta = \pi / 2$ (recall that $\theta \in (0,\pi)$) or $\phi = \pm \pi / 2$.
Let us first assume that $\phi = \pm \pi /2$. The second term in Eq.~(\ref{eq:rot:dipole:angles:square}) then vanishes if and only if
\begin{equation}
  \sin \theta = \pm \tan \vartheta  \: .
\end{equation}
This last condition imposes a constraint on $\vartheta$, which must then have a value between $0$ and $\pi / 4$ in order for $\tan \vartheta$ (and hence $\sin \theta$) to be less than unity. Since $\theta \in (0,\pi)$, only the case $\phi = \pi / 2$ yields two solutions, $\theta_1$ and $\theta_2$, that are symmetric with respect to $\theta = \pi /2$. This is illustrated in Fig.~\ref{fig:dipole}(f). Assuming now that $\theta = \pi / 2$, the second term in Eq.~(\ref{eq:rot:dipole:angles:square}) vanishes if and only if
\begin{equation}
  \sin \phi = \mathrm{cotan} \, \vartheta  \: .
\end{equation}
Since $\sin \phi$ requires $\mathrm{cotan} \, \vartheta$ to be less than unity the above condition can only be satisfied if $\vartheta \in (\pi/4,\pi/2)$. There are then two solutions for $\phi$ between $0$ and $\pi$ (since $\sin \phi > 0$ for $\vartheta \in (\pi/4, \pi/2)$), which are symmetric with respect to $\pi/2$.
In fact, it can be shown that the polarization of the radiation of the rotating dipole for $\vartheta \in (\pi/4,\pi/2)$ corresponds to that of a rotating dipole for which $\vartheta' = \pi / 2 - \vartheta$ (as in Fig.~\ref{fig:dipole}(f)) but rotated by $\ang{90}$ with respect to the $x_2$-axis.
These different cases where Eq.~\eqref{eq:rot:dipole:angles:square} vanishes for a given circular polarization have a very simple geometrical interpretation. For $\vartheta < \pi/4$ the rotating dipole describes an ellipse whose long axis is oriented along the $x_3$-axis. The two directions of zero $\sigma^+$-polarized radiation correspond to the two directions from which the ellipse is observed as a circle, with the orientation of the dipole rotation opposite to that of the $\sigma^+$ polarization.
For these two directions the radiation is therefore purely $\sigma^-$-polarized, explaining why the zeros of radiation are found on the meridian where $\phi =  \pi /2$. For $\vartheta > \pi/4$ the long axis of the ellipse is along the $x_1$-direction, which explains the fact that the directions where one circular polarization is zero are found at the equator. By symmetry the directions where the $\sigma^-$-polarized radiation vanishes are symmetric to those of the $\sigma^+$-polarized radiation with respect to the $(\Vie{\hat{e}}{1}{},\Vie{\hat{e}}{3}{})$-plane.

These results, obtained for the polarization of the dipole radiation in free space, will prove to be useful for the qualitative understanding of the full angular distribution of the incoherent component of the MDRC and MDTC in the case of the scattering by a randomly rough dielectric surface.\\

\begin{figure*}[ht] 
\includegraphics[width=.49\linewidth , trim= 0.6cm .8cm 2.6cm 1.cm,clip]{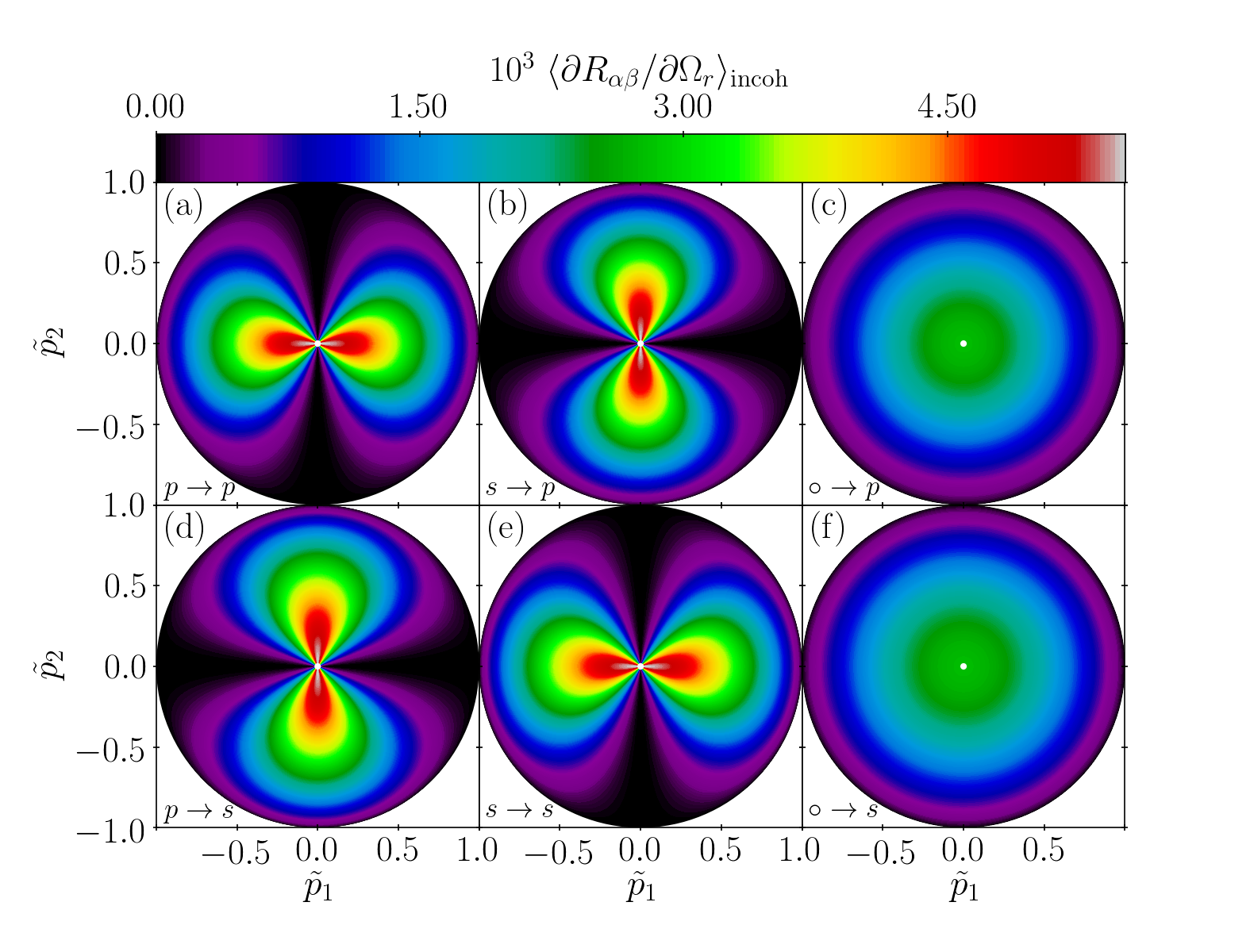}
\includegraphics[width=.49\linewidth , trim= 0.6cm .8cm 2.6cm 1.cm,clip]{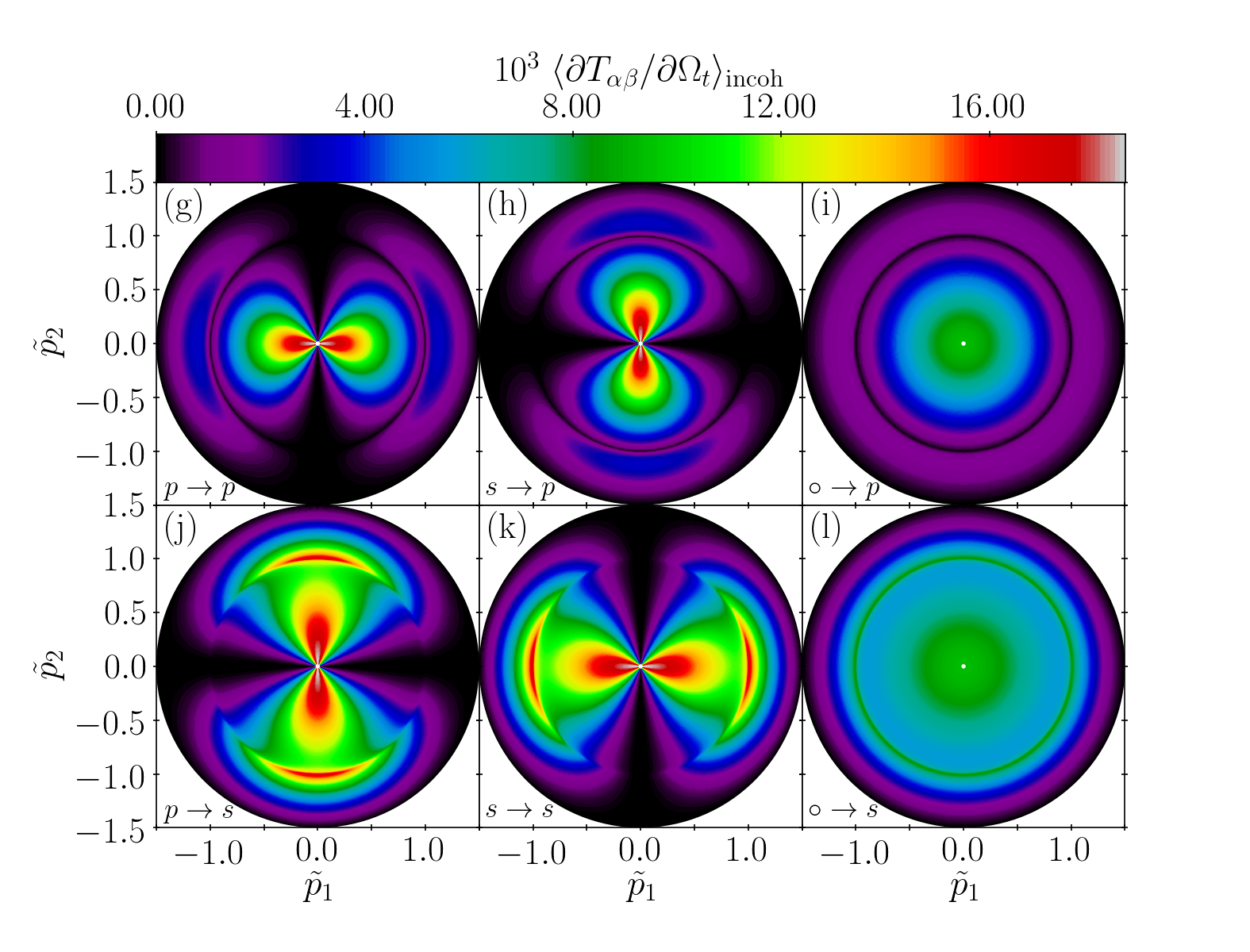}
\caption{
  The full angular distribution of the incoherent component of the MDRC and MDTC,
  $\left\langle \partial X_{\alpha\beta} / \partial \Omega_r \right\rangle_\mathrm{incoh}$ for $X=R$ or $T$, as function of the lateral wave vector $\mathbf{\tilde{p}}$ of the light that is scattered from a rough interface where the angle of incidence $\theta_0=\ang{0}$.
  The positions of the specular directions in reflection and transmission are indicated by white dots.
  The parameters assumed for the scattering geometry and used in performing the numerical calculation had values that are identical to those assumed in obtaining the results of Fig.~\protect\ref{fig:sapt_mdxc_incvacuum}.
  The sub-figures in Figs.~\ref{fig:2Dmdxc_incvacuum_0deg}(a)--(f) and  \ref{fig:2Dmdxc_incvacuum_0deg}(g)--(l) are both organized in the same manner and show how incident  $\beta$-polarized light is scattered by the one-rough-interface film geometry into $\alpha$-polarized light [with $\alpha=p,s$ and $\beta=p,s$] and denoted $\beta\rightarrow\alpha$.
  Moreover, the notation $\circ\rightarrow$ is taken to mean that the incident light was unpolarized. For instance, this means that the data shown in Fig.~\ref{fig:2Dmdxc_incvacuum_0deg}(c) result from the addition and division by a factor two of the the data sets presented in Figs.~\ref{fig:2Dmdxc_incvacuum_0deg}(a) and \ref{fig:2Dmdxc_incvacuum_0deg}(b); \emph{etc}.
  Finally, the in-plane intensity variations from Figs.~\protect\ref{fig:2Dmdxc_incvacuum_0deg}(a,\,e) and \ref{fig:2Dmdxc_incvacuum_0deg}(g,\,k) are the curves depicted in Figs.~\protect\ref{fig:sapt_mdxc_incvacuum}(a) and Figs.~\protect\ref{fig:sapt_mdxc_incvacuum}(d), respectively.}
\label{fig:2Dmdxc_incvacuum_0deg}
\end{figure*}

\begin{figure*}[ht] 
\includegraphics[width=.49\linewidth , trim= .6cm .8cm 2.6cm 1.cm,clip]{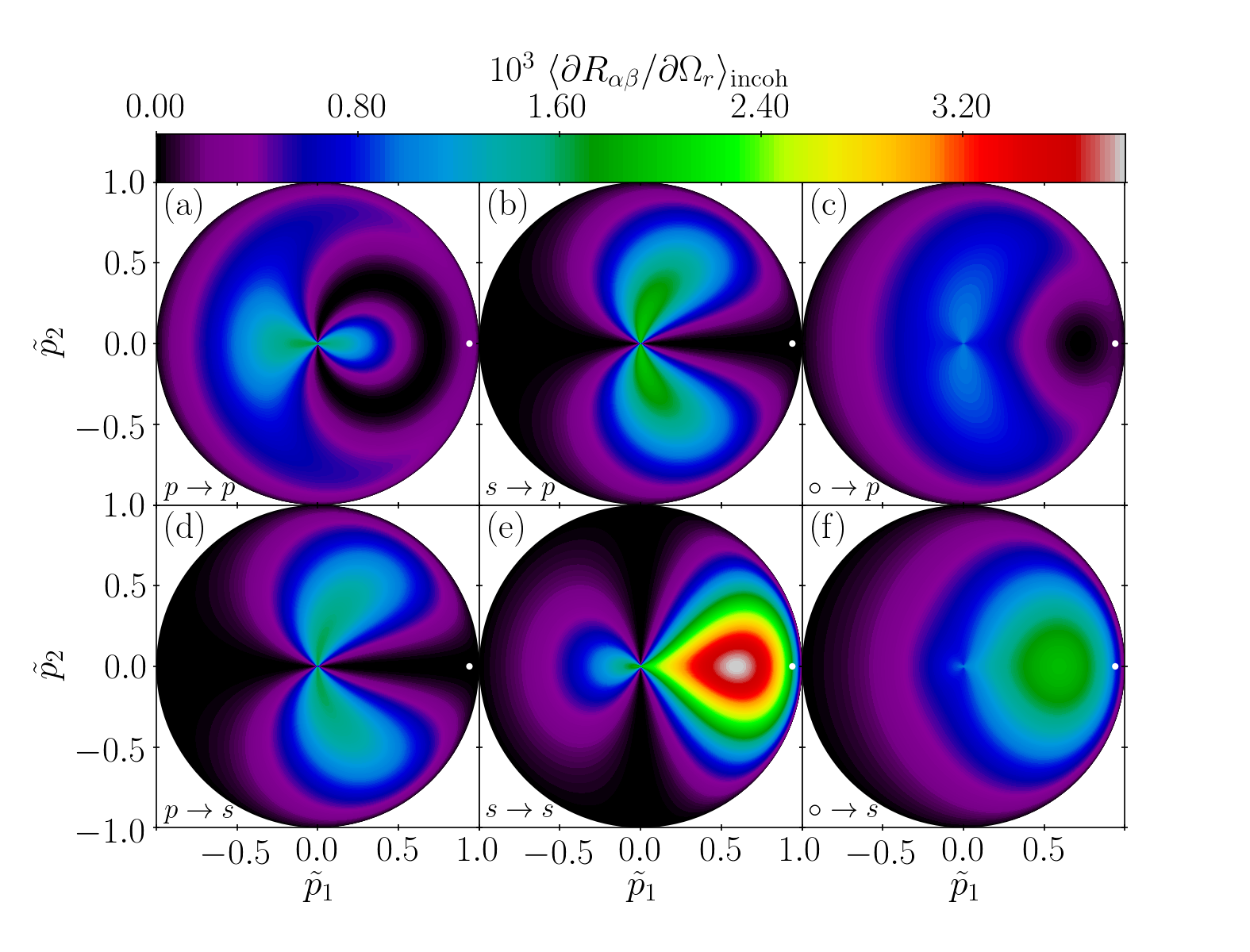}
\includegraphics[width=.49\linewidth , trim= .6cm .8cm 2.6cm 1.cm,clip]{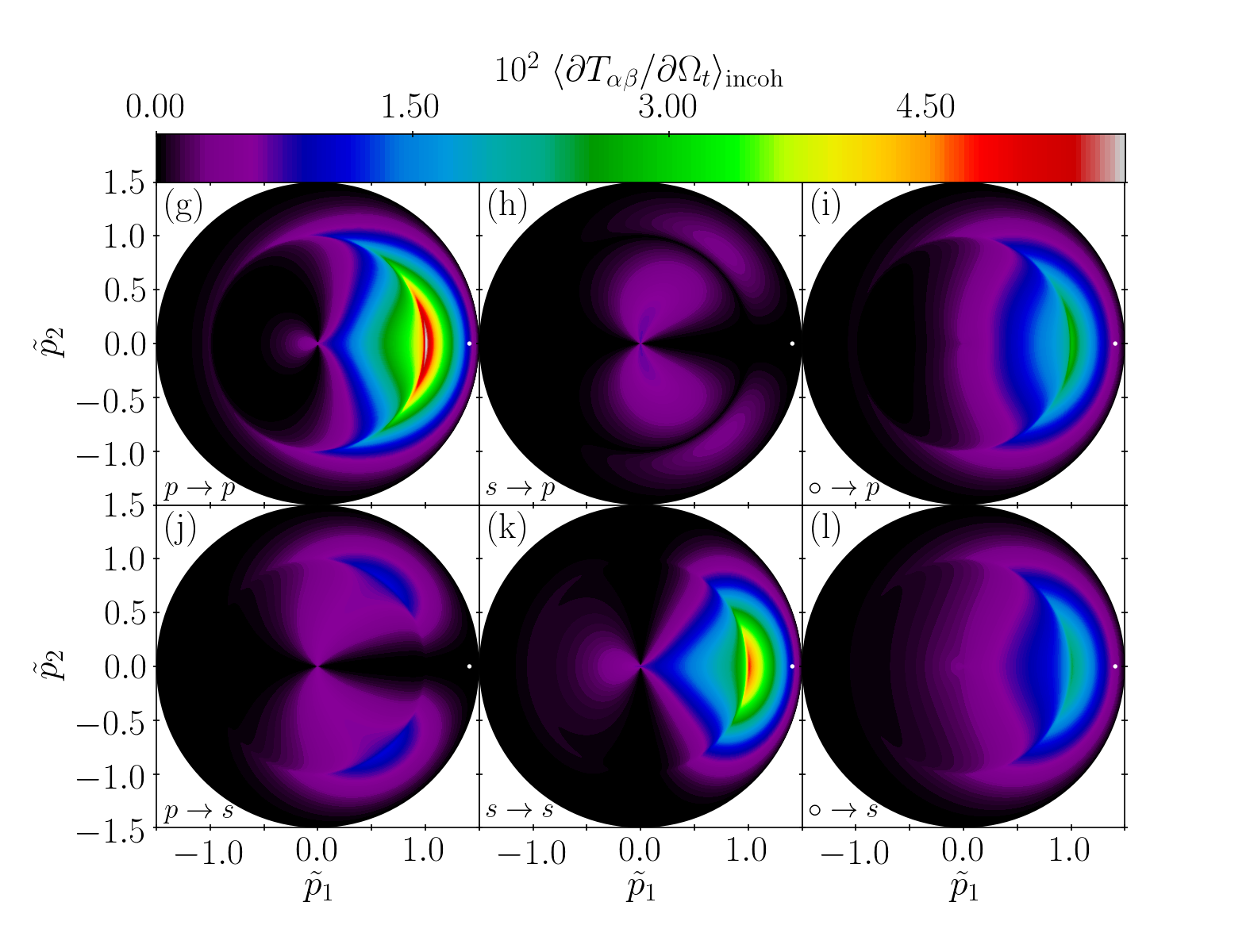}
\caption{
  Same as Fig.~\ref{fig:2Dmdxc_incvacuum_0deg}, but now for the angle of incidence $\theta_0=\ang{70}$.
}
\label{fig:2Dmdxc_incvacuum_70deg}
\end{figure*}

\begin{figure*}[t]
\centering
	\subfloat[]{%
		\includegraphics[width=.32\linewidth]{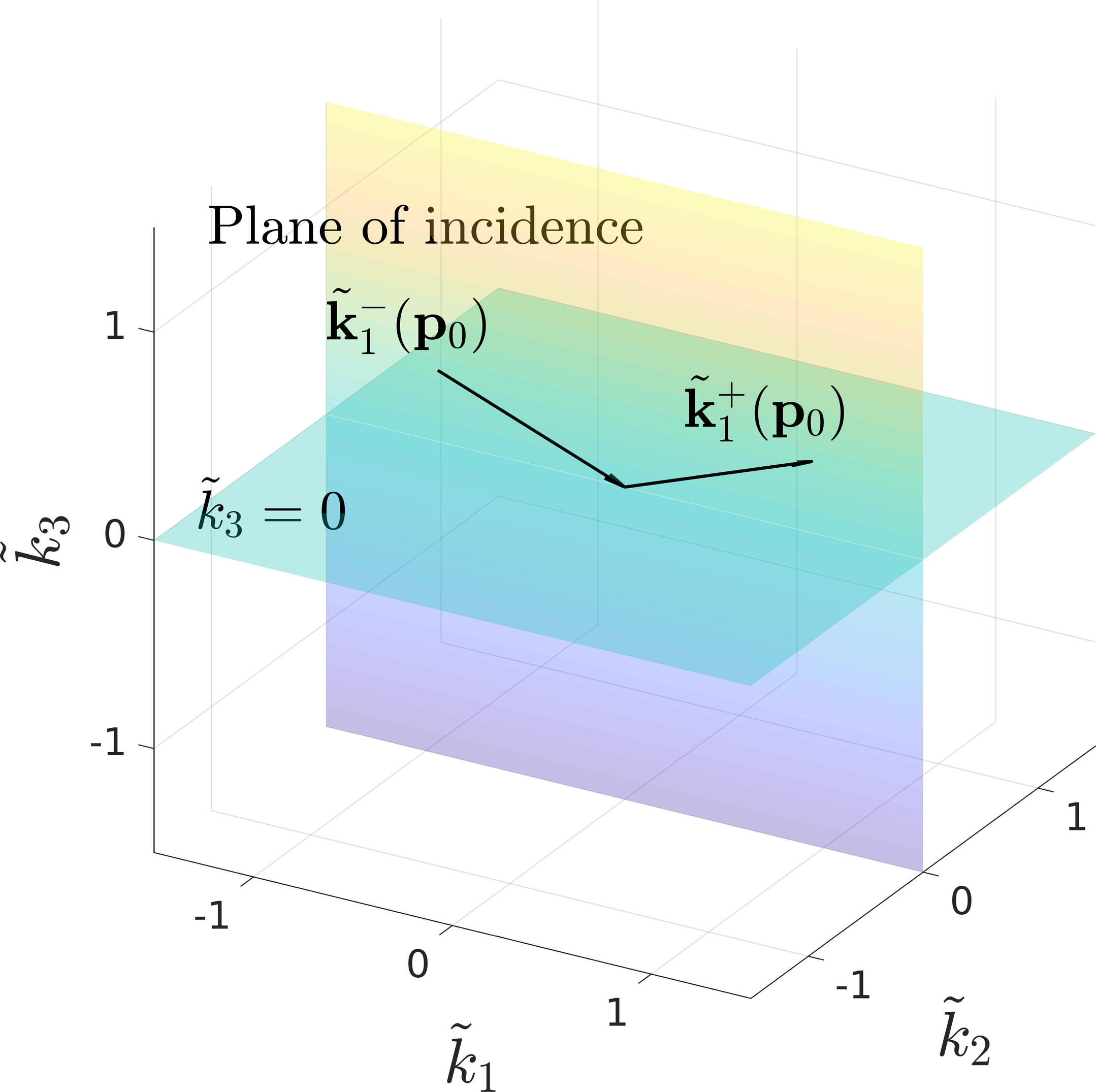}
  }
	\subfloat[]{%
		\includegraphics[width=.32\linewidth]{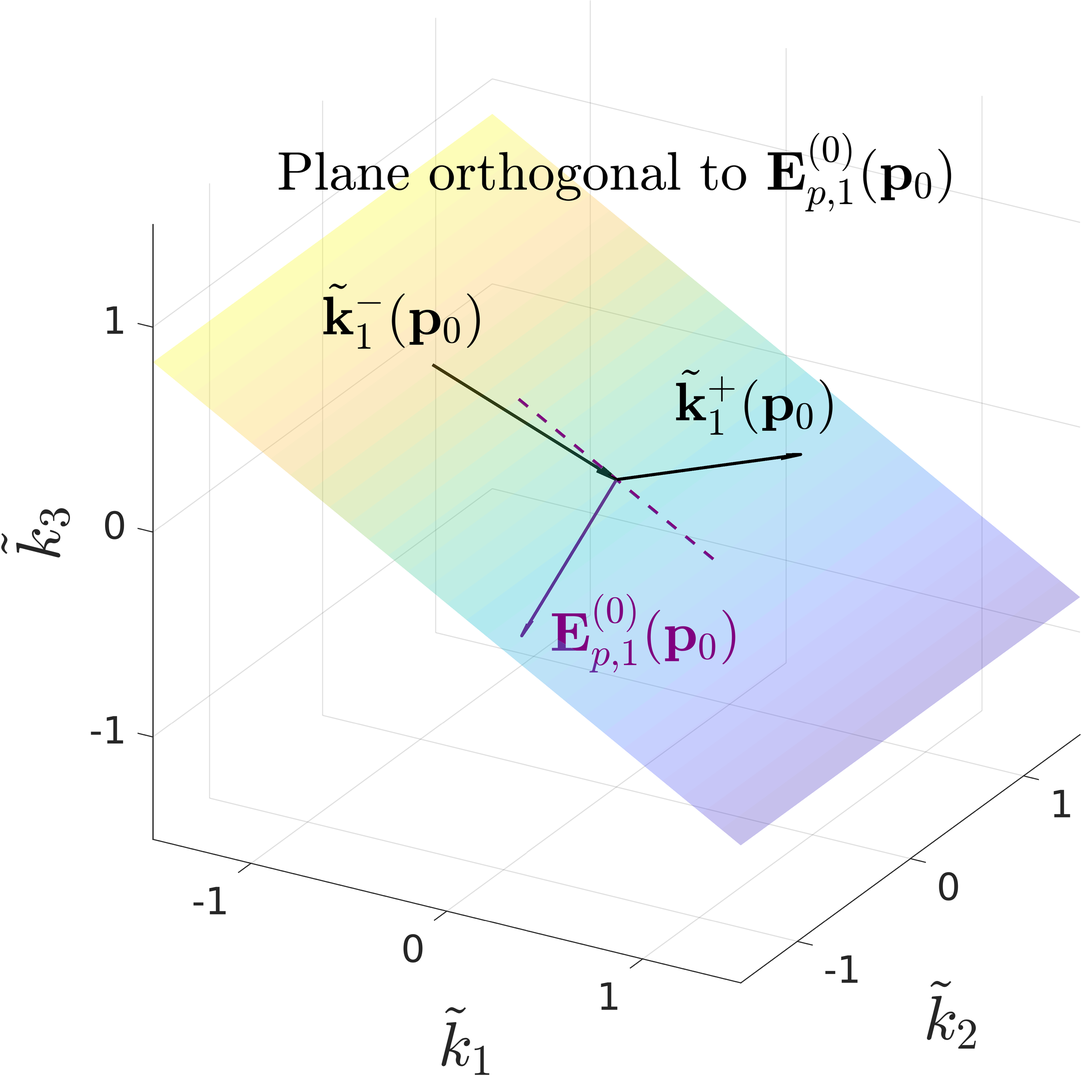}
  }
	\subfloat[]{%
		\includegraphics[width=.32\linewidth]{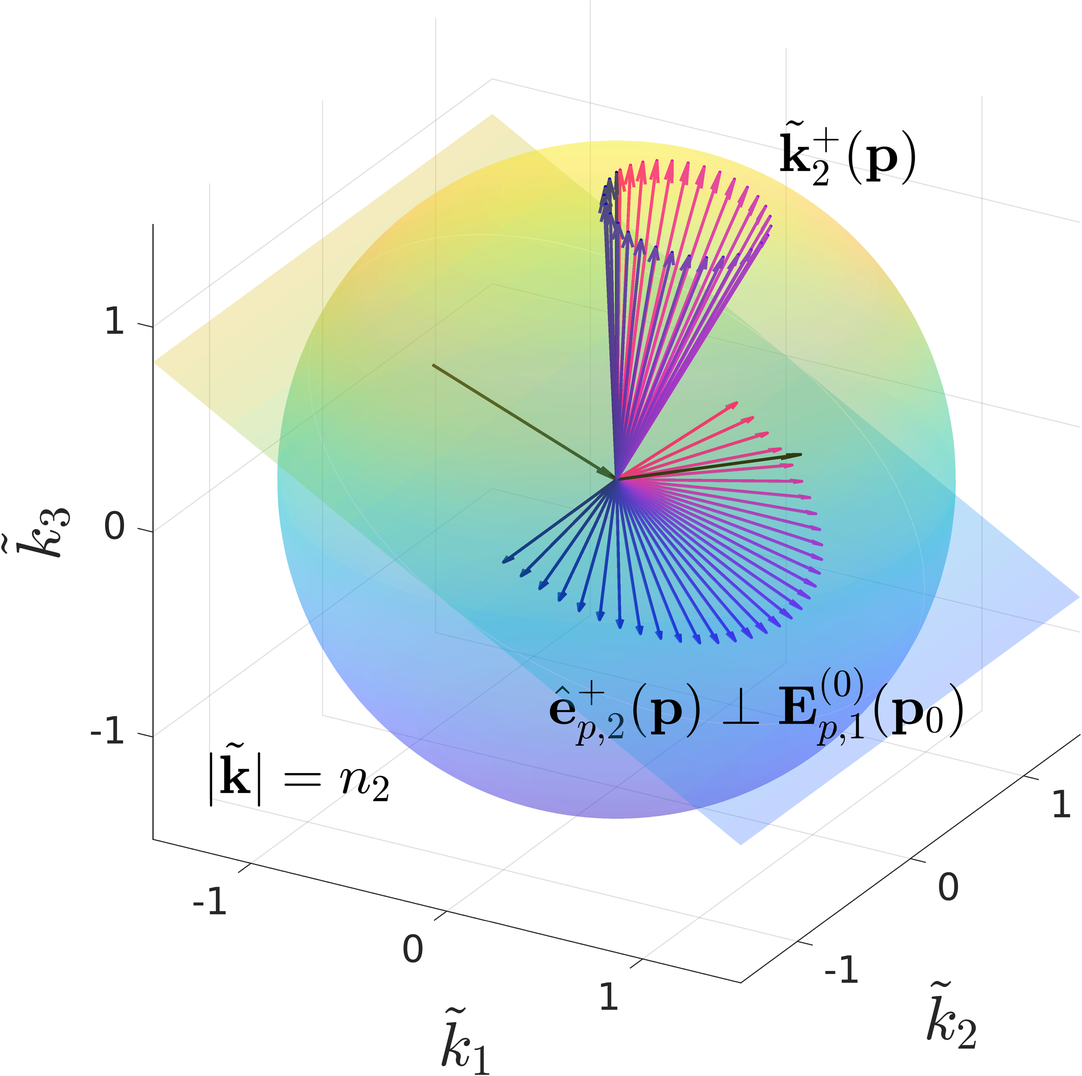}
  }

	\subfloat[]{%
		\includegraphics[width=.32\linewidth]{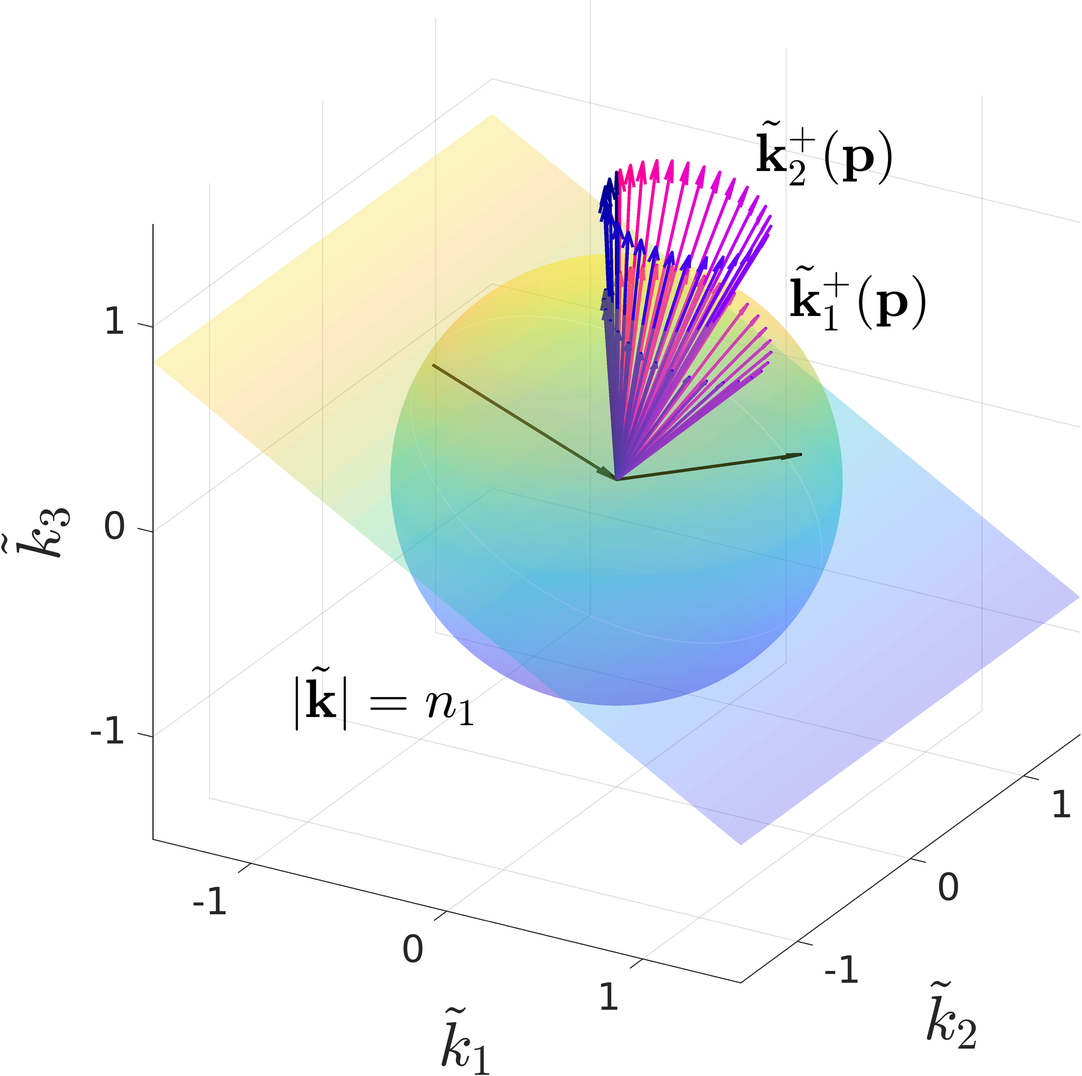}
  }
	\subfloat[]{%
		\includegraphics[width=.32\linewidth]{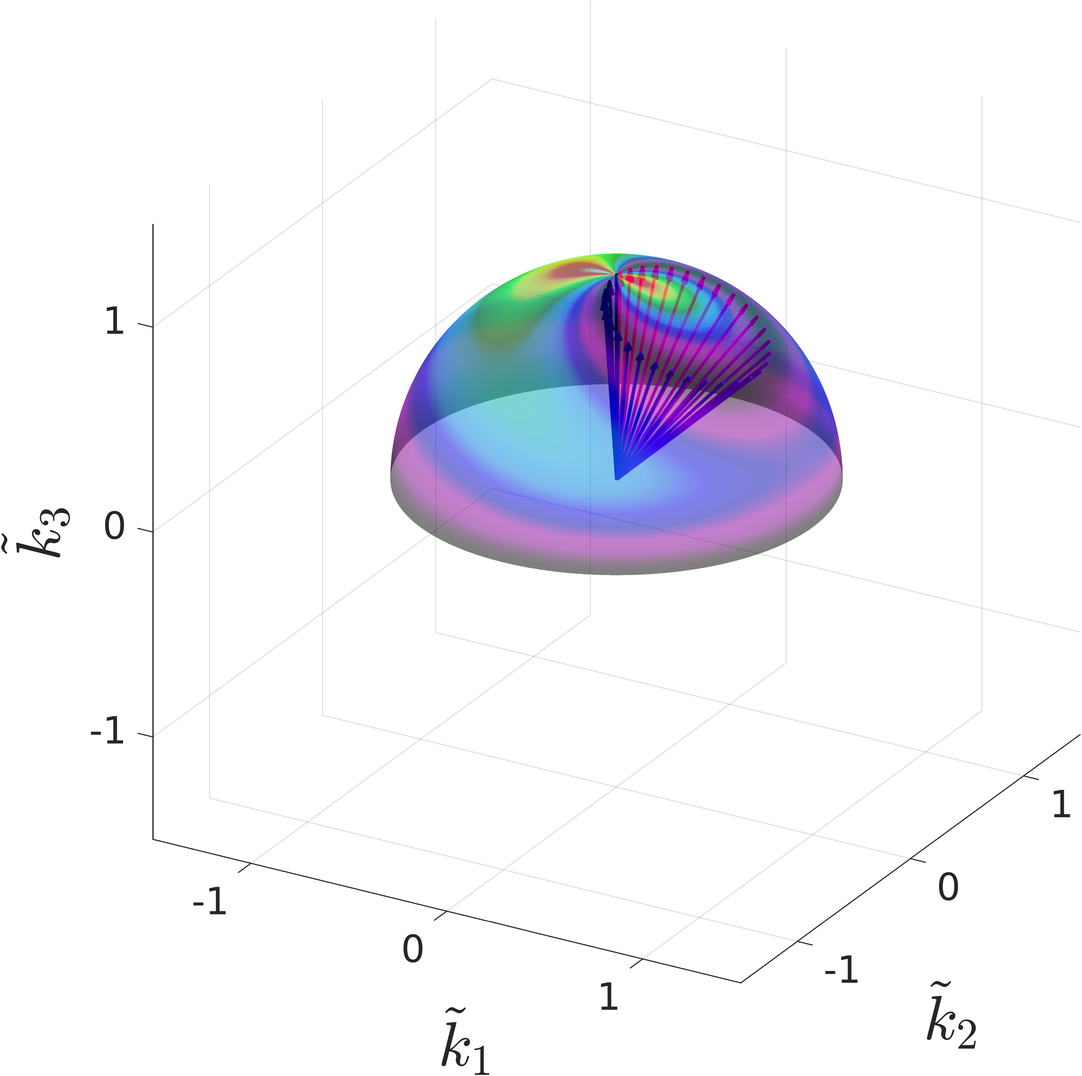}
  }
	\subfloat[]{%
		\includegraphics[width=.32\linewidth]{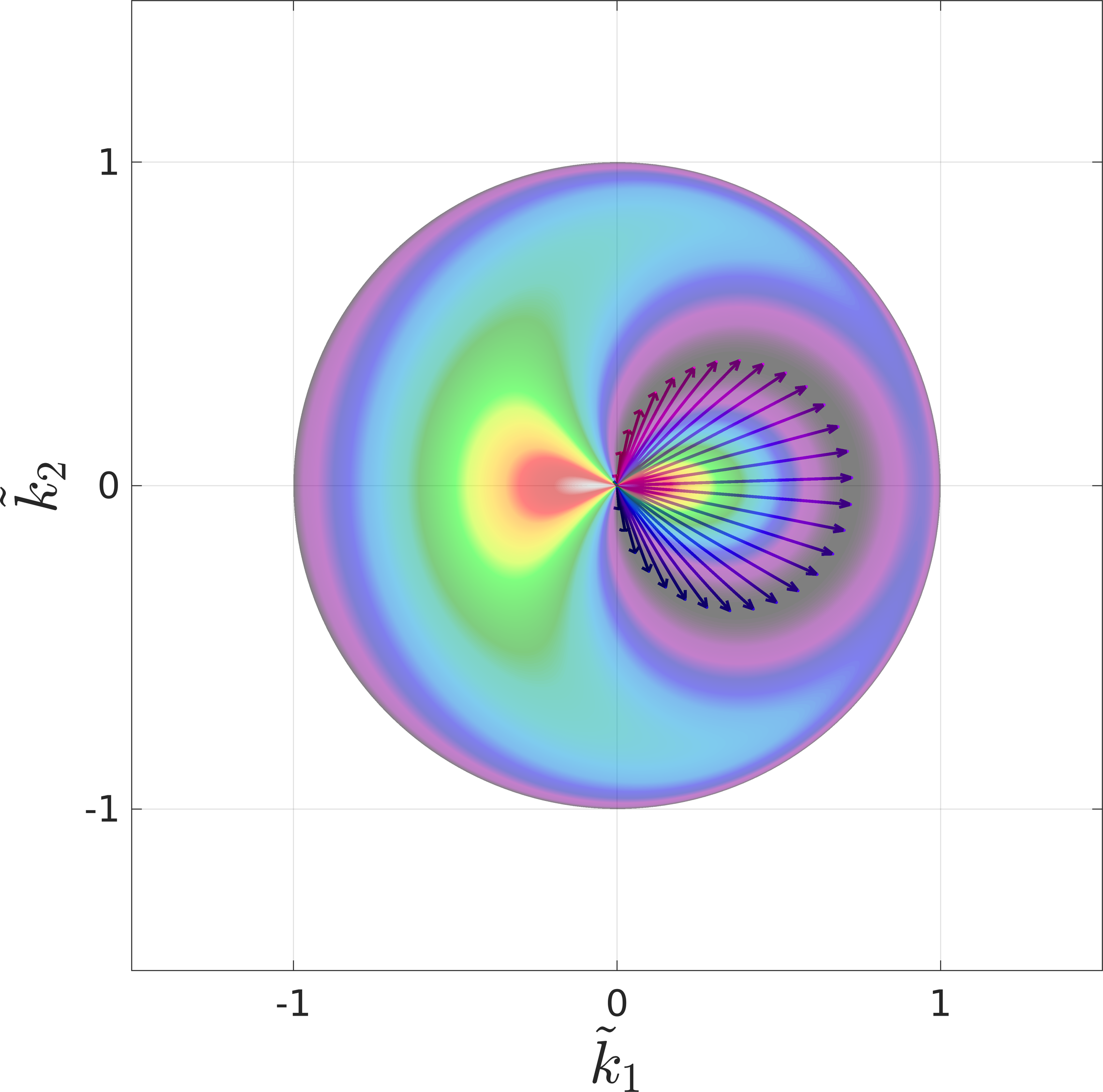}
  }
\caption{Illustration of the construction steps leading to the set of directions of zero $p \to p$ reflection for an angle of incidence of $\theta_0 = \ang{70}$. (a) Sketch of the average surface, the plane of incidence and the considered wave vectors of the incident and reflected zero order waves ($\Vie{k}{1}{-} (\Vie{p}{0}{})$ and $\Vie{k}{1}{-} (\Vie{p}{0}{})$).
(b) Construction of the total zero order field amplitude $\Vie{E}{p,1}{(0)} (\Vie{p}{0}{})$ and the plane orthogonal to it. Note that the incident wave vector does not in general belong in this plane as illustrated with the dashed indigo line indicating the intersection of the plane of incidence and the plane $\Vie{E}{p,1}{(0)} (\Vie{p}{0}{})^\perp$.
(c) Unit vectors belonging to the lower half $\Vie{E}{p,1}{(0)} (\Vie{p}{0}{})^\perp$-plane. They correspond to the possible polarization vectors $\Vie{\hat{e}}{p,2}{+}(\Vie{p}{}{})$ of Eq.~(\ref{eq:brewster_criterion}). The wave vectors $\Vie{k}{2}{+} (\Vie{p}{}{})$ associated to the polarization vectors $\Vie{\hat{e}}{p,2}{+}(\Vie{p}{}{})$ are then constructed according to Eqs.~(\ref{basis}).
Note that they lie on a sphere of radius $|\Vie{k}{}{}| = n_2 \,\omega / c$. The color associated to the vectors $\Vie{\hat{e}}{p,2}{+}(\Vie{p}{}{})$ and $\Vie{k}{2}{+} (\Vie{p}{}{})$ helps us to identify the $\Vie{k}{2}{+} (\Vie{p}{}{})$ associated to each $\Vie{\hat{e}}{p,2}{+}(\Vie{p}{}{})$ (they share the same color).
(d) The wave vectors $\Vie{k}{2}{+} (\Vie{p}{}{})$ are projected on the sphere of radius $|\Vie{k}{}{}| = n_1 \,\omega / c$ following the $x_3$-direction, hence giving the wave vectors $\Vie{k}{1}{+} (\Vie{p}{}{})$ of zero $p \to p$ reflection. (e) The incoherent component of the MDRC is shown on the scattering sphere together with the set of wave vectors $\Vie{k}{1}{+} (\Vie{p}{}{})$ obtained in (d).
(f) Projection of (e) in the $(\Vie{\hat{e}}{1}{},\Vie{\hat{e}}{2}{})$-plane. We verify in (e) and (f) that the constructed wave vectors indeed follow the curve of zero scattering of the incoherent component of the MDRC. Note that unit-less wave vectors are used in the illustration, i.e. $\tilde{k} = k c / \omega$.}
\label{fig:3D_geom_brewster}
\end{figure*}


\begin{figure*}[ht]
\includegraphics[width=.49\linewidth , trim= 0.cm 0.cm 1.cm 0.cm,clip]{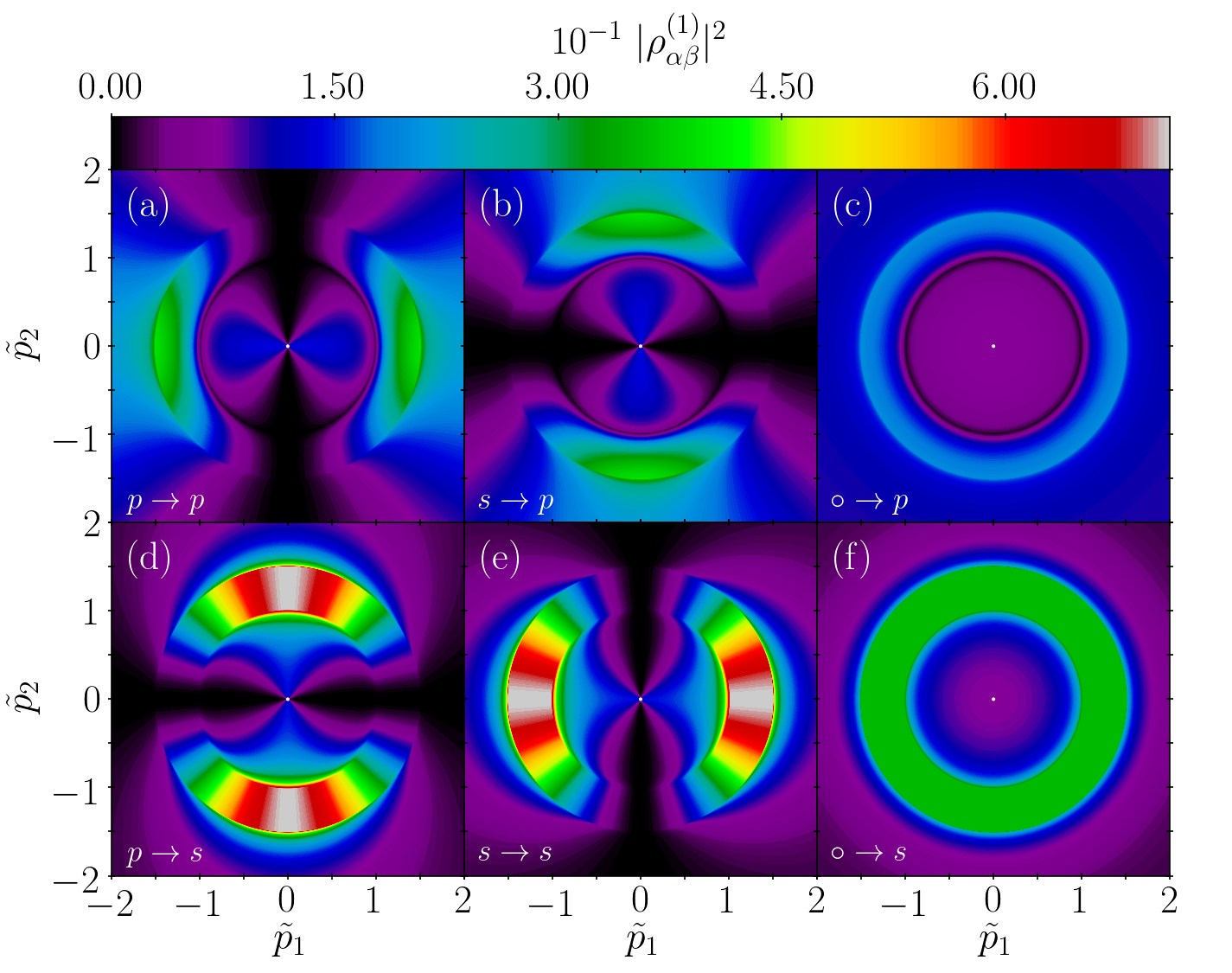}
\includegraphics[width=.49\linewidth , trim= 0.cm 0.cm 1.cm 0.cm,clip]{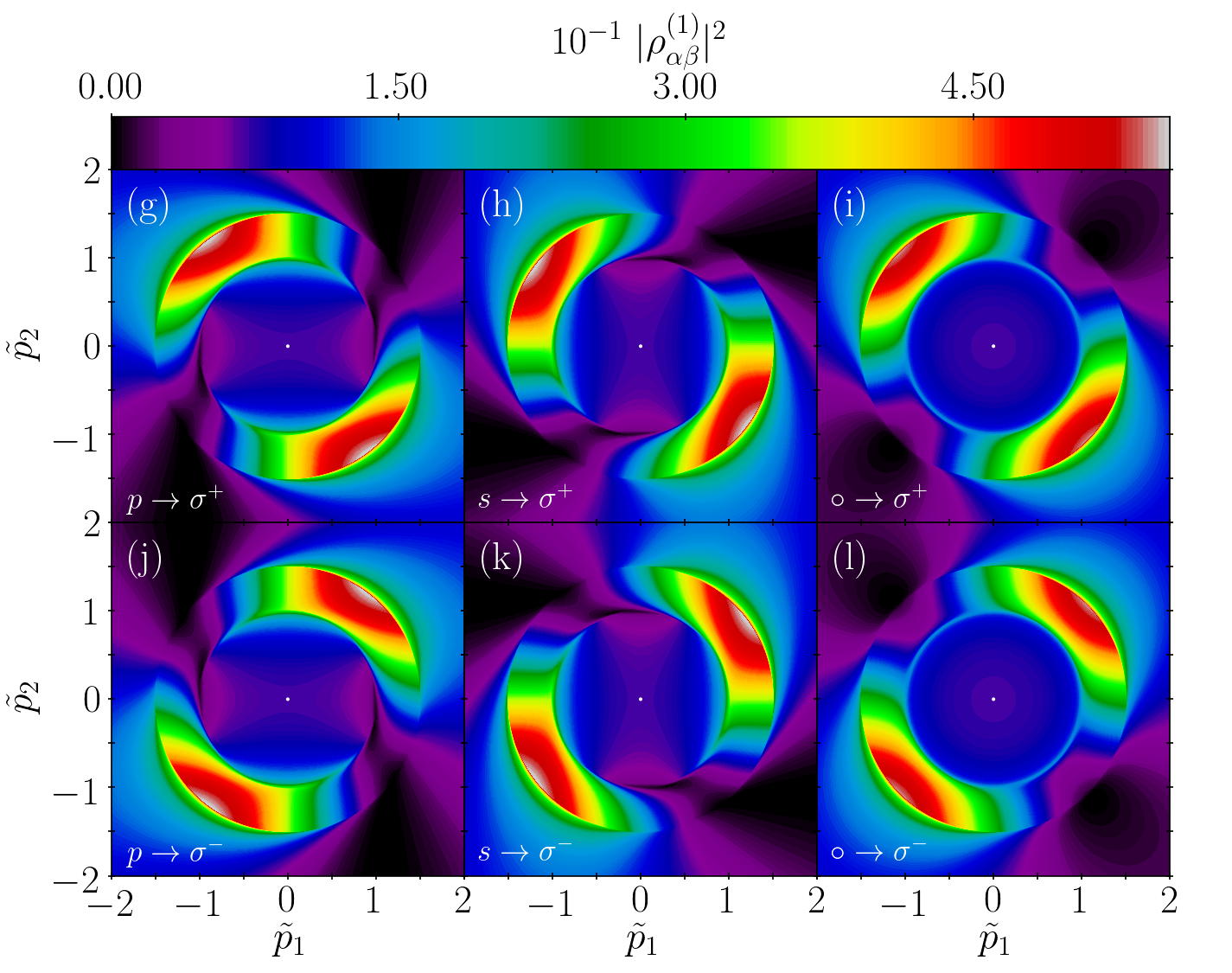}
\caption{The full angular distribution of $|\rho_{\alpha \beta}^{(1)}|^2$ for normal incidence, $\theta_0 = \ang{0}$, $\ve_1 = 2.25$, $\ve_2 = 1$, for incident polarization $\beta \in \{p, s\}$ or unpolarized ($\circ$) and outgoing polarization $\alpha \in \{p, s, \sigma^+, \sigma^- \}$.
}
\label{fig:2Dmdxc_incglass_0deg}
\end{figure*}

\begin{figure*}[ht]
\includegraphics[width=.49\linewidth , trim= 0.cm 0.cm 1.cm 0.cm,clip]{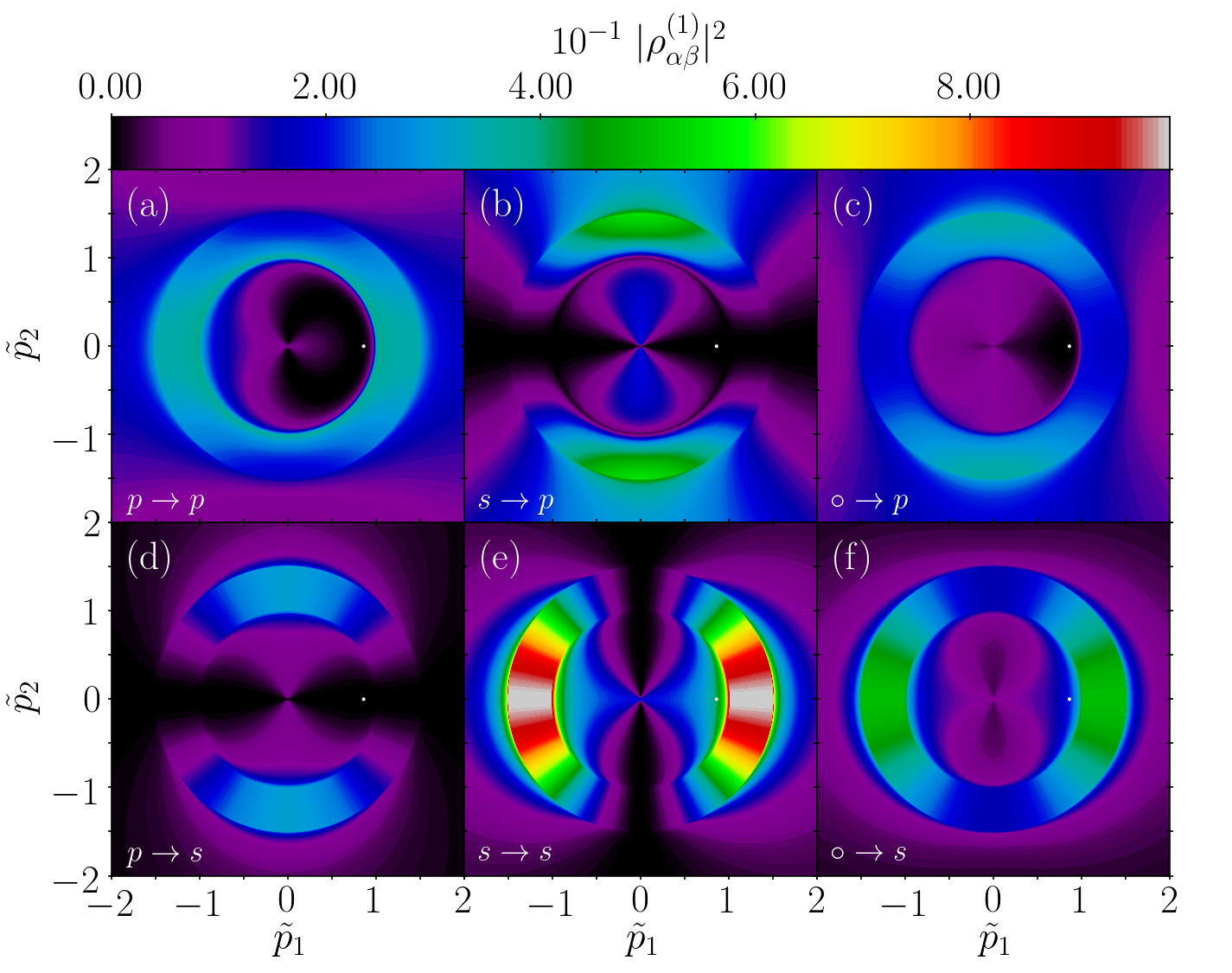}
\includegraphics[width=.49\linewidth , trim= 0.cm 0.cm 1.cm 0.cm,clip]{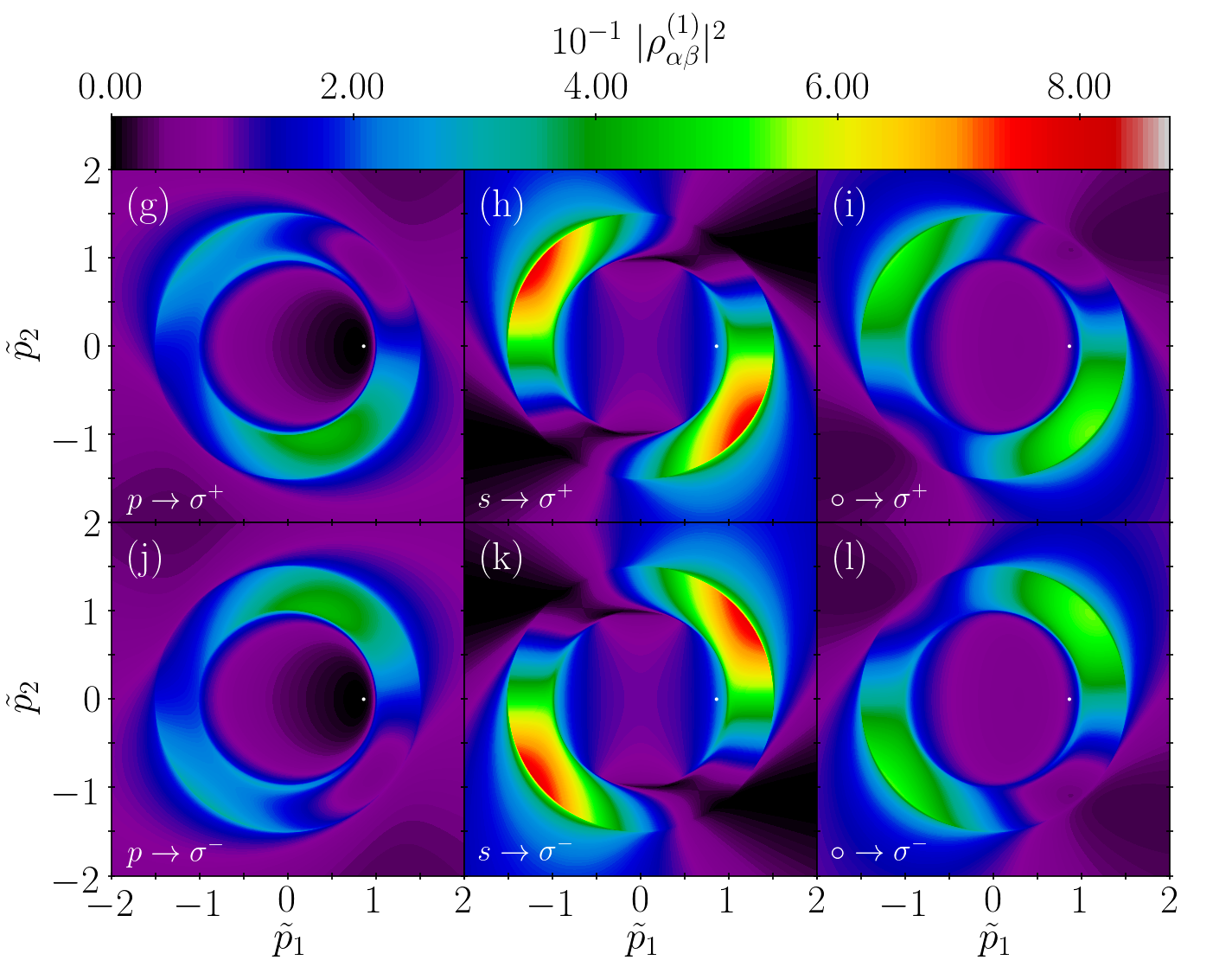}
\caption{Same as Fig.~\ref{fig:2Dmdxc_incglass_0deg} but for the angle of incidence $\theta_0 = \ang{35}$.
}
\label{fig:2Dmdxc_incglass_35deg}
\end{figure*}

\begin{figure*}[ht]
\includegraphics[width=.49\linewidth , trim= 0.cm 0.cm 1.cm 0.cm,clip]{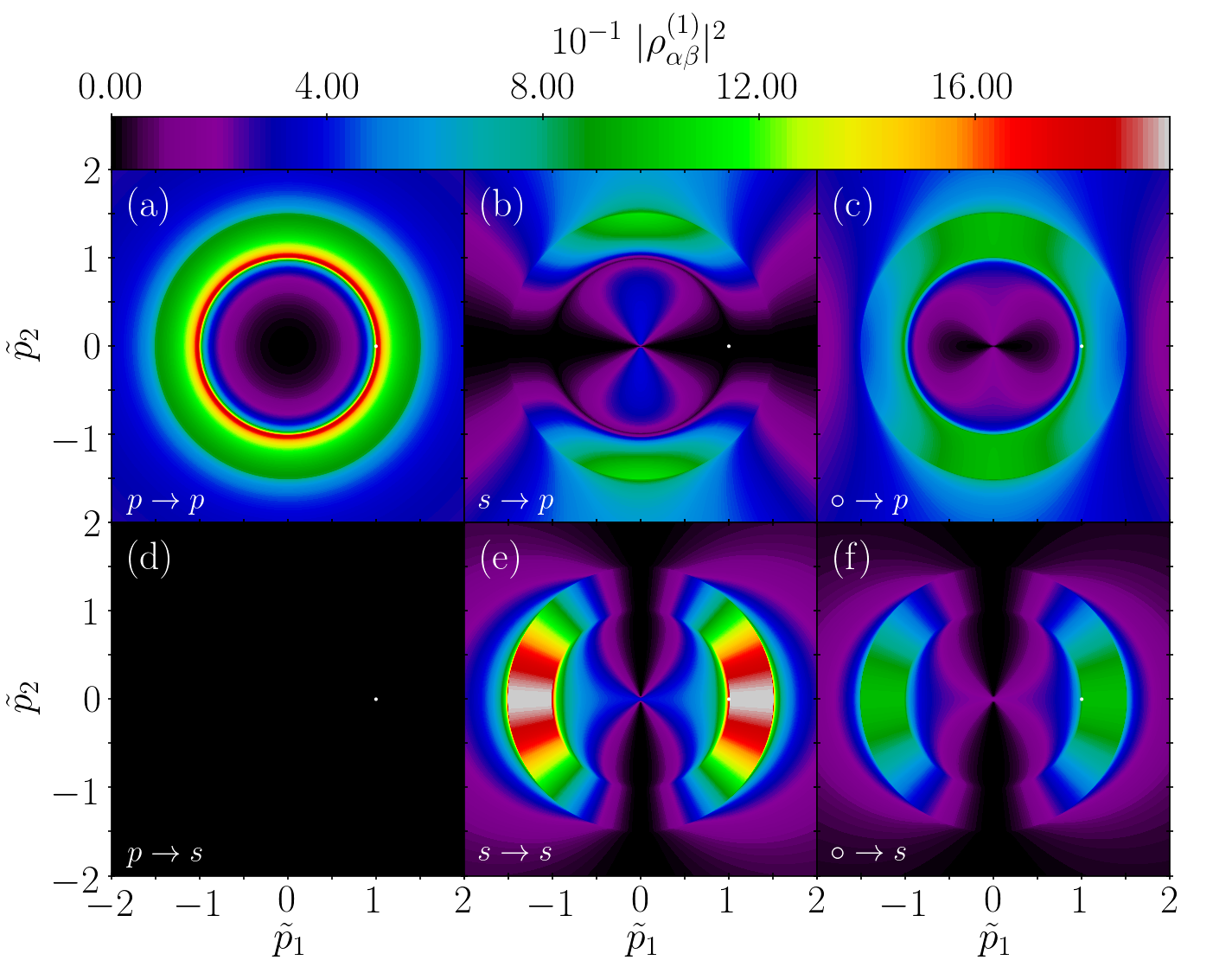}
\includegraphics[width=.49\linewidth , trim= 0.cm 0.cm 1.cm 0.cm,clip]{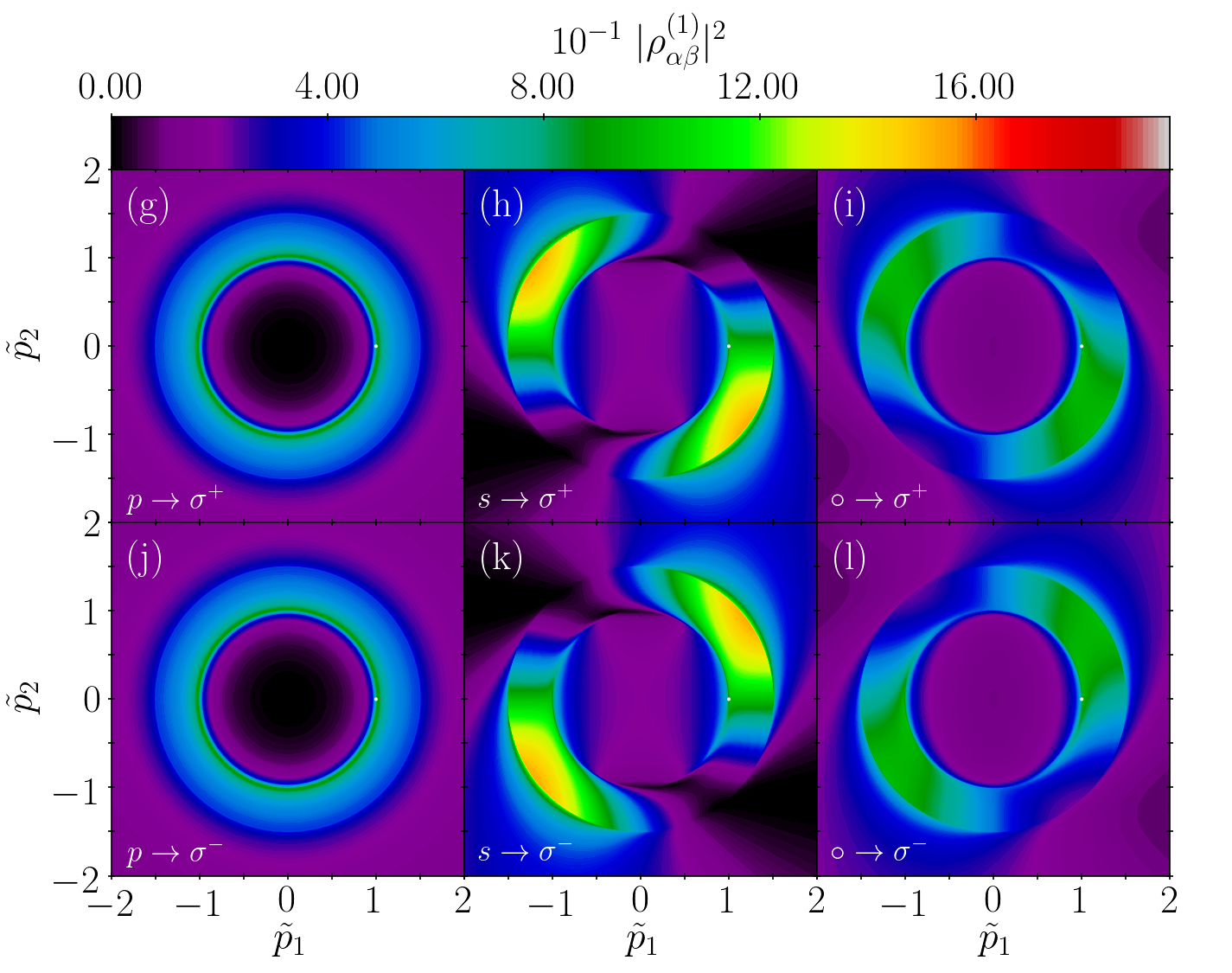}
\caption{Same as Fig.~\ref{fig:2Dmdxc_incglass_0deg} but for the angle of incidence equal to the critical angle for total internal reflection, $\theta_0 = \theta_c = \ang{41.81}$.
}
\label{fig:2Dmdxc_incglass_41deg}
\end{figure*}

\begin{figure*}[ht]
\includegraphics[width=.48\linewidth , trim= 0.cm 0.cm 1.cm 0.cm,clip]{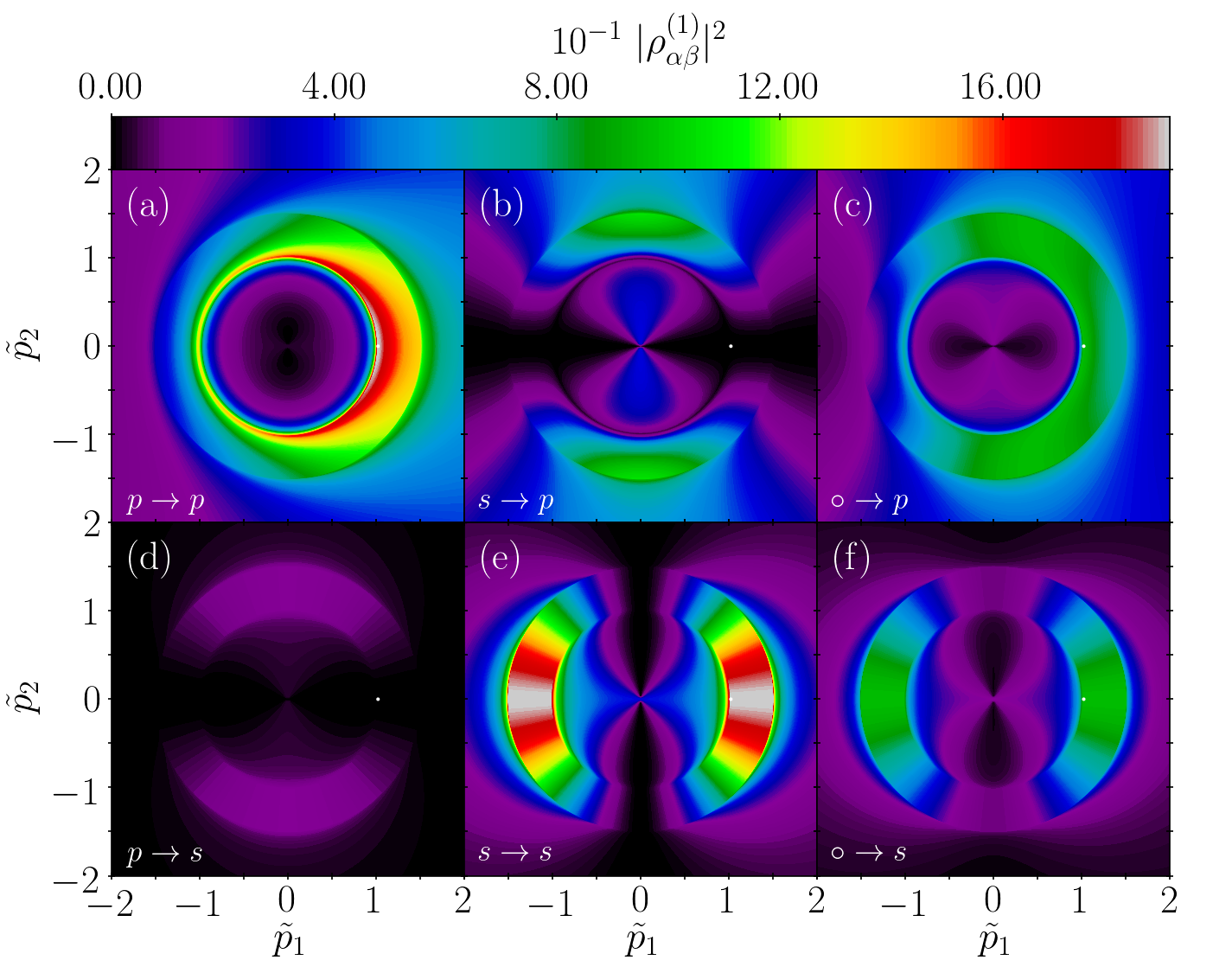}
\includegraphics[width=.48\linewidth , trim= 0.cm 0.cm 1.cm 0.cm,clip]{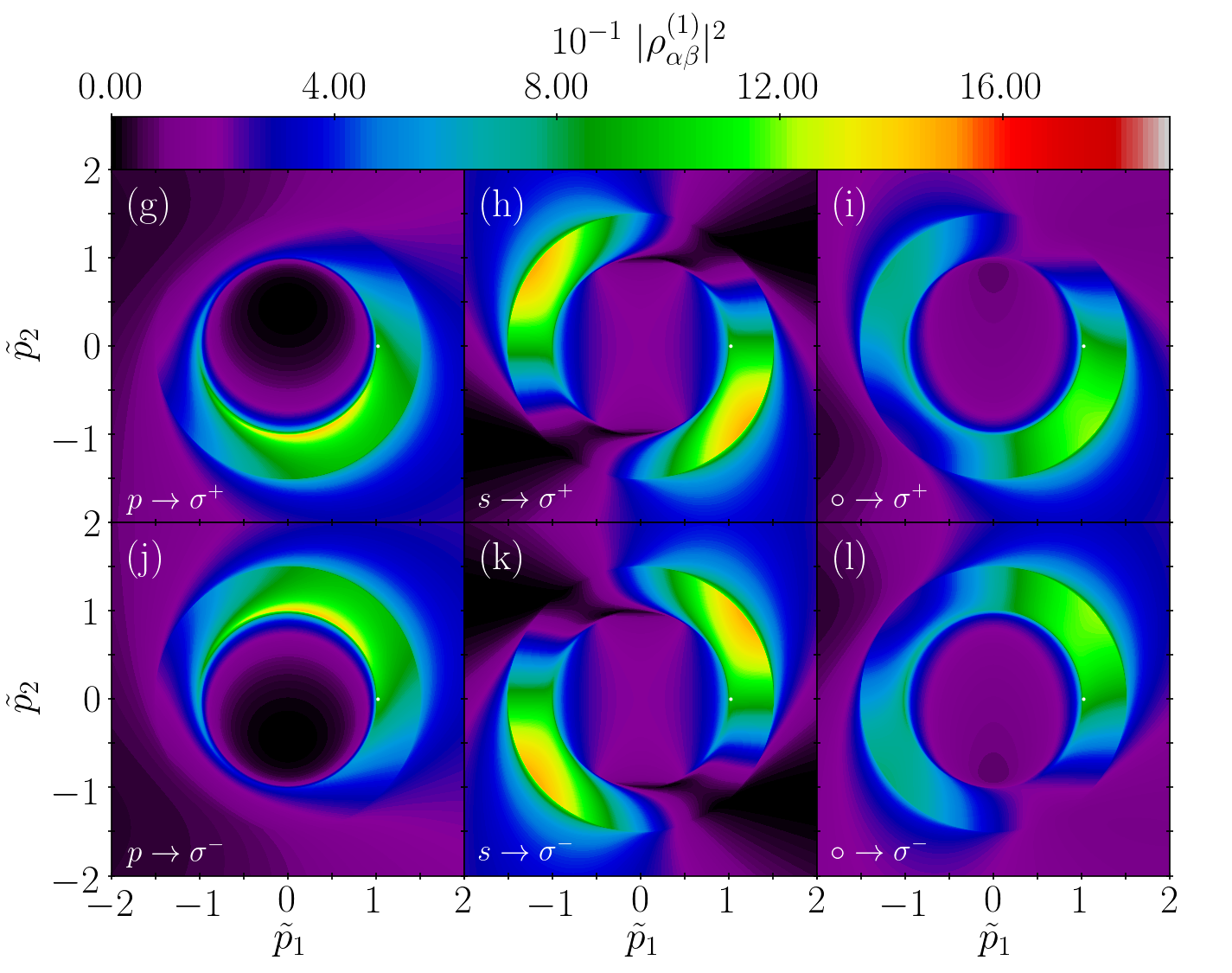}
\caption{Same as Fig.~\ref{fig:2Dmdxc_incglass_0deg} but for the angle of incidence $\theta_0 = \ang{43}$.
}
\label{fig:2Dmdxc_incglass_42deg}
\end{figure*}


\subsection{Full angular distributions of the MDRC and MDTC}\label{sec:full_distribution}

\emph{p-polarized Brewster scattering} --- Figure~\ref{fig:2Dmdxc_incvacuum_0deg} presents the full angular distributions of the diffuse contribution to the MDRC and MDTC for $\theta_0=\ang{0}$ and parameters equivalent to those assumed in Figs.~\ref{fig:sapt_mdxc_incvacuum}(a) and \ref{fig:sapt_mdxc_incvacuum}(d), respectively.
The overall dipole-like appearance of the lower left $2\times 2$ panels in each collection of panels in Fig.~\ref{fig:2Dmdxc_incvacuum_0deg} is reminiscent of the polarization pattern of the dipole radiation in free space discussed above in the case when the dipole oscillates in the $(\Vie{\hat{e}}{1}{}, \Vie{\hat{e}}{2}{})$-plane. For normal incidence all the zero order waves and the incident wave have an oscillating electric field either along $\Vie{\hat{e}}{1}{}$ for \textit{p} polarization or along $\Vie{\hat{e}}{2}{}$ for \textit{s} polarization. Thus the dipoles in the media\footnote{Here limited to one medium, but in general both media sharing the interface could be dielectrics.} oscillate along the direction of the incident field.
For an \textit{s}-polarized wave (field along $\Vie{\hat{e}}{2}{}$) we have seen that the dipole radiation in free space yields zero \textit{s}-polarized emission in the $(\Vie{\hat{e}}{2}{},\Vie{\hat{e}}{3}{})$-plane and an overall $|\sin (\phi_r - \pi/2) |$ intensity, which is consistent with what is observed in Fig.~\ref{fig:2Dmdxc_incvacuum_0deg}(f). Note that for a given polar angle of reflection $\theta_r$, the variation along $\phi_r$ of the incoherent component of the MDRC to lowest non-zero order in the surface roughness for $s \to s$ polarized scattering is \emph{exactly} proportional to $|\sin (\phi_r - \pi/2) |$ since $\rho_{ss}^{(1)}$ is proportional to $\Vie{\hat{e}}{s}{}(\Vie{p}{}{}) \cdot \Vie{\hat{e}}{s}{} (\Vie{p}{0}{})$, as can be seen from Eq.~\eqref{eq:rho1ss}, and this is the only $\phi_r$ dependence for normal incidence.
This observation holds for all the polarization couplings up to a rotation by $\pi / 2$ for cross-polarization. For example, for an s-polarized incident field and \textit{p}-polarized reflected light the $\phi_r$ dependence is proportional to $|\sin \phi_r|$.
For the transmitted light [Fig.~\ref{fig:2Dmdxc_incvacuum_0deg}(j-r)] the behavior is similar, but in addition we now observe the Yoneda phenomenon. This is the enhancement of the diffuse contribution to the MDTC intensity above the critical lateral wave vector for the scattered light $|\Vie{p}{}{}| > p_c$, as discussed extensively in Sec.~\ref{sec:yoneda}, which for normal incidence is directly observable for outgoing s-polarized light, especially in Fig.~\ref{fig:2Dmdxc_incvacuum_0deg}(r).
For transmitted \textit{p}-polarized light we observe a black ring of zero scattering intensity along the circle $|\Vie{p}{}{}| = p_c$. This is the two-dimensional extension of our discussion in Sec.~\ref{sec:brewster_1storder} for in-plane scattering, where we found that at normal incidence two Brewster waves with $\Vie{p}{B}{} = \pm p_c \, \Vie{\hat{e}}{1}{}$ could be found. Now we see that in two-dimensional $\Vie{p}{}{}$-space the solution to Eq.~\eqref{eq:brewster_criterion} is in fact given by $|\Vie{p}{}{}| = p_c$.
In terms of dipole radiation in free space this corresponds to the vanishing radiation of p-polarized light in the equatorial plane for the case $\vartheta = \ang{90}$ as illustrated in Fig.~\ref{fig:dipole}(c). A ring of zero intensity for the p-polarized \emph{reflected} waves can also be found, but then in the evanescent regime as a two-dimensional extension of the corresponding discussion for in-plane scattering.\\

The similitude with the polarization of the radiation emitted by an oscillating dipole in free space is clear for normal incidence. Let us now consider a larger angle of incidence, $\theta_0 = \ang{70}$, for which the diffuse contributions to the MDRC and MDTC for incidence in vacuum are shown in Fig.~\ref{fig:2Dmdxc_incvacuum_70deg}. First, we observe that for $p \to p$ reflection [Fig.~\ref{fig:2Dmdxc_incvacuum_70deg}(b)], there exists a closed curve of zero intensity in the forward scattering direction. Similarly, we observe a closed curve of zero intensity for $p \to p$ transmission [Fig.~\ref{fig:2Dmdxc_incvacuum_70deg}(k)] but in the backscattering region.
These features are analogous to those observed in the case of the \textit{p} polarization component of the dipole radiation in free space in the case where the dipole tilting angle is such that $\ang{0} < \vartheta < \ang{90}$, e.g. as is illustrated in Fig.~\ref{fig:dipole}(b) for $\vartheta = \ang{45}$. We can interpret the curves of zero intensity for $p \to p$ scattering in Fig.~\ref{fig:2Dmdxc_incvacuum_70deg} as the signature of an overall dipole radiation whose dipole moment is tilted from the $x_3$-axis by some angle $\vartheta$, where the polarization of the reflected light is derived from the northern hemisphere of the radiation polarization pattern while the polarization of the transmitted light is derived from the southern hemisphere of the radiation polarization pattern.

Let us now interpret Eq.~\eqref{eq:brewster_criterion} geometrically for $p \to p$ scattering for the case of reflection and $\theta_0 = \ang{70}$. This construction is a generalization of the one made for scattering in the plane of incidence presented in Sec.~\ref{sec:brewster_1storder}. Figure~\ref{fig:3D_geom_brewster} provides illustrations of the main steps of the geometrical construction of the set of directions of zero $p \to p$ reflection in three dimensions.
First, the wave vectors of incidence $\Vie{k}{1}{-}(\Vie{p}{0}{})$ and of the reflected zero order wave $\Vie{k}{1}{+}(\Vie{p}{0}{})$ are drawn [Fig.~\ref{fig:3D_geom_brewster}(a)].
Second, one determines the direction of the total zero order field amplitude $\Vie{E}{p,1}{(0)} (\Vie{p}{0}{})$ which is contained in the plane of incidence. The steps to geometrically construct the total zero order field have been treated in detail for \textit{s}- and \textit{p}- polarizations in Ref.~\citenum{Doyle1980}, and thus we do not show these here for clarity. Once $\Vie{E}{p,1}{(0)} (\Vie{p}{0}{})$ is determined, we can construct the plane orthogonal to it: $\Vie{E}{p,1}{(0)} (\Vie{p}{0}{})^\perp$ [Fig.~\ref{fig:3D_geom_brewster}(b)].
Note that in general this plane does \emph{not} contain the incident wave vector as made clear by the dashed line, showing the intersection of the plane of incidence with the plane $\Vie{E}{p,1}{(0)} (\Vie{p}{0}{})^\perp$. According to Eq.~\eqref{eq:brewster_criterion}, all the polarization vectors $\Vie{\hat{e}}{p,2}{+}(\Vie{p}{}{})$  must be contained in the plane $\Vie{E}{p,1}{(0)} (\Vie{p}{0}{})^\perp$.
Moreover, since the $\Vie{\hat{e}}{p,2}{+}(\Vie{p}{}{})$ vectors are normalized they are distributed on a circle of unit radius. The set of all $\Vie{\hat{e}}{p,2}{+}(\Vie{p}{}{})$ vectors satisfying Eq.~\eqref{eq:brewster_criterion} therefore spans a half circle in the plane $\Vie{E}{p,1}{(0)} (\Vie{p}{0}{})^\perp$ as shown on Fig.~\ref{fig:3D_geom_brewster}(c), where a sample of polarization vectors are represented. The fact that only the lower half circle is needed comes from the definition\footnote{One may extend the construction to all vectors on the circle defined as the intersection of the unit sphere and the plane $\Vie{E}{p,1}{(0)} (\Vie{p}{0}{})^\perp$, but it would result in constructing twice the same set of wave vectors of zero scattering.} of a polarization vector $\Vie{\hat{e}}{p,2}{+}(\Vie{p}{}{})$.
For each polarization vector satisfying Eq.~\eqref{eq:brewster_criterion} we can construct its corresponding wave vector $\Vie{k}{2}{+}(\Vie{p}{}{})$, using for example that the direction is given by $\Vie{\hat{e}}{p,2}{+}(\Vie{p}{}{}) \times [\Vie{\hat{e}}{p,2}{+}(\Vie{p}{}{}) \times \Vie{\hat{e}}{3}{}]$ and that $\Vie{k}{2}{+}(\Vie{p}{}{})$ lies on the northern hemisphere of radius $k_2$.
We thus obtain the set of all wave vectors $\Vie{k}{2}{+}(\Vie{p}{}{})$ whose corresponding p polarization vector satisfies Eq.~\eqref{eq:brewster_criterion}. A sample of such vectors are represented for $\Vie{\hat{e}}{p,2}{+}(\Vie{p}{}{})$ and $\Vie{k}{2}{+}(\Vie{p}{}{})$ in Fig.~\ref{fig:3D_geom_brewster}(c).
The last step consists in projecting the vectors $\Vie{k}{2}{+}(\Vie{p}{}{})$ along $\Vie{\hat{e}}{3}{}$ onto the sphere of radius $|\Vie{k}{}{}| = k_1$ to obtain the wave vectors $\Vie{k}{1}{+}(\Vie{p}{}{})$ of zero $p \to p$ reflection [Fig.~\ref{fig:3D_geom_brewster}(d)].
Figures~\ref{fig:3D_geom_brewster}(e) and \ref{fig:3D_geom_brewster}(f) show the resulting sampled wave vectors $\Vie{k}{1}{+}(\Vie{p}{}{})$ together with the diffuse contribution to the the MDRC, mapped to the hemisphere and its projection in the $(\Vie{\hat{e}}{1}{},\Vie{\hat{e}}{2}{})$-plane respectively. We verify that the set of constructed wave vectors correspond to the observed curve of zero intensity for $p \to p$ reflection.

Figure~\ref{fig:2Dmdxc_incvacuum_70deg}(n) shows that the $s \to p$ transmitted light exhibits a circle of zero intensity, for $|\Vie{p}{}{}| = k_1$ similar to what was observed for normal incidence [Fig.~\ref{fig:2Dmdxc_incvacuum_0deg}(n)]. This feature is also present in reflection but in the evanescent region, and is observed by considering the complex amplitude instead of the MDRC. The reason for the invariance of the circle of zero intensity with the angle of incidence for the $s \to p$ scattering is simple to understand in terms of the dipole radiation in free space. For \textit{s}-polarized incident light the dipoles in the media are all oriented along $\Vie{\hat{e}}{2}{}$, independent of the angle of incidence. Thus when measuring the p-polarized component of the radiated light we expect to obtain an underlying pattern of zero intensity consistent with that obtained in the case of the oscillating dipole in free space as illustrated in Fig.~\ref{fig:dipole}(c).\\

\emph{Circularly-polarized Brewster scattering} --- It is instructive to study the modulus square of the amplitudes rather than the MDRC and MDTC in order to appreciate the behavior of the amplitudes of the waves scattered in the evanescent region as well as the ones scattered in the propagating region. Furthermore, in order to illustrate, to our knowledge, a new effect which can be considered as a generalization of the Brewster scattering effect for light scattered from \textit{p}-polarized to circularly-polarized light, we show in Figs.~\ref{fig:2Dmdxc_incglass_0deg}-\ref{fig:2Dmdxc_incglass_42deg} $|\rho_{\alpha \beta}^{(1)}|^2$ in the $\Vie{p}{}{}$-plane for different polar angles of incidence.
We let $\beta \in \{p, s, \circ\}$ represent the polarization of the incident light, where $\circ$ indicates unpolarized light, and we let $\alpha \in \{ p, s, \sigma^+, \sigma^-\}$ represent the polarization of the light scattered from the surface. The subscripts $\sigma^\pm$ denote respectively left and right circular polarization states and the corresponding reflection amplitudes are derived from the p and s polarization states by
\begin{equation}
  \rho_{\sigma^\pm \beta}^{(1)} = \frac{1}{\sqrt{2}} \, \left[ \rho_{p \beta}^{(1)} \pm i \rho_{s \beta}^{(1)} \right]  \: ,
\label{eq:sigma_def}
\end{equation}
and similarly for the transmission amplitudes. We consider here only the case for which the medium of incidence is the denser one, as it exhibits a richer variety of Brewster effects when the reflected zero order wave undergoes total internal reflection. Note, however, that the effect can be observed both in the reflected \emph{and} the transmitted scattered light. In Sec.~\ref{sec:brewster_1storder} we have seen that the Brewster scattering effect exhibits a sudden transition when the reflected zero order wave undergoes total internal reflection.
We have seen that, when restricted to scattering in the plane of incidence, the direction of zero p-polarized reflected intensity goes towards the $x_3$-direction as the polar angle of incidence approaches the critical angle for total internal reflection. Then the zero direction suddenly disappears from the propagating region as the polar angle of incidence goes beyond the critical angle for total internal reflection. This sudden transition was argued to be attributed to a transition of the dipolar response of the media, going from an oscillating behavior to a rotating behavior due to the phase shift between the incident excitation and the reflected zero order wave. We are now studying this transition in the full $\Vie{p}{}{}$-plane with particular attention on the scattered \emph{circularly polarized} light, as it was shown in Sec.~\ref{sec:dipole} that the radiation emitted by a rotating dipole in free space exhibits characteristic signatures in the emitted circularly-polarized light out of the plane of incidence.

First, for polar angles of incidence smaller than the critical angle, $\theta_0 < \theta_c$, we have seen that both $\rho_{p \beta}^{(1)}$ and $\rho_{s \beta}^{(1)}$ are real for scattering angles smaller than the Yoneda threshold. In that case the right-hand side in Eq.~(\ref{eq:sigma_def}) vanishes if and only if both $\rho_{p \beta}^{(1)}$ and $\rho_{s \beta}^{(1)}$ are zero simultaneously. For an incident p-polarized wave, $\beta = p$, this occurs only where the curve of zero $p \to p$ scattering (cf. previous subsection) intersects with the plane of incidence in which $p \to s$ scattering is identically zero.
This is illustrated for normal incidence, $\theta_0 = \ang{0}$, in Figs.~\ref{fig:2Dmdxc_incglass_0deg}(a) and (d) showing $|\rho_{p p}^{(1)}|^2$ and $|\rho_{s p}^{(1)}|^2$ in the $\Vie{p}{}{}$-plane, where we recognize the curves of zero scattering for the p- and s-polarized light discussed in previous sections. It is also illustrated in Figs.~\ref{fig:2Dmdxc_incglass_0deg}(g) and (j) showing $|\rho_{\sigma^\pm p}^{(1)}|^2$ where two directions of zero $p \to \sigma^\pm$ scattering are present at $\Vie{p}{}{} = \pm p_c \Vie{\hat{e}}{1}{}$, although they are hard to spot on this figure.
The effect is clearer for oblique incidence, as in Figs.~\ref{fig:2Dmdxc_incglass_35deg}(g) and (j), for which $\theta_0 = \ang{35}$. Figures~\ref{fig:2Dmdxc_incglass_35deg}(g) and (j) show a clear unique direction of zero intensity in $p \to \sigma^\pm$ scattering in the plane of incidence.

As the angle of incidence reaches the critical angle of incidence, $\theta_0 = \theta_c = \ang{41.81}$, the direction of zero intensity in $p \to \sigma^\pm$ scattering reaches the $x_3$-direction, as illustrated in Figs.~\ref{fig:2Dmdxc_incglass_41deg}(g) and (j). Note that the $x_3$-direction also implies zero $p \to p$ scattering intensity as already explained earlier, and that the distribution of $|\rho_{p p}^{(1)}|^2$ and $|\rho_{\sigma^\pm p}^{(1)}|^2$ are cylindrically symmetric as shown in Fig.~\ref{fig:2Dmdxc_incglass_41deg}(a), (g) and (j).
The cylindrical symmetry can be understood based on the radiation of an oscillating dipole aligned with the $x_3$-axis. Indeed, we have seen in Section~\ref{sec:dipole} that the p-polarized radiation from such a dipole is cylindrically symmetric with zero radiation at the poles of the unit sphere. The radiation from such a dipole is also purely p-polarized, which has two consequences: (i) the s-polarized scattered light vanishes identically for \emph{all} $\Vie{p}{}{}$ [Fig.~\ref{fig:2Dmdxc_incglass_41deg}(d)]; (ii) the radiation can be decomposed into $\sigma^+$ and $\sigma^-$ components of \emph{equal} intensity, as can be observed in Figs.~\ref{fig:2Dmdxc_incglass_41deg}(g) and (j).
Even though we have now based our interpretation on the radiation of an oscillating dipole in free space for the sake of simplicity, it is straightforward to verify these assertions based on the expressions of the amplitudes given in Eq.~(\ref{eq:rho1:all}). For example, it is clear that for $\theta_0 = \theta_c$, the total zero order field $\Vie{E}{1,p}{(0)}(\Vie{p}{0}{})$ is along $\Vie{\hat{e}}{3}{}$ and the dot product in Eq.~(\ref{eq:rho1sp}) vanishes for all $\Vie{p}{}{}$.

For $\theta_0 > \theta_c$, it is convenient to expand the right-hand side in Eq.~(\ref{eq:sigma_def}). By inserting Eq.~(\ref{eq:rho1:all}) into Eq.~(\ref{eq:sigma_def}), the reduced first order reflection amplitude, $\hat{\rho}_{\sigma^\pm p}^{(1)}$, for $\sigma^\pm$-polarized light scattered from incident p-polarized light is then given by
\begin{equation}
  \hat{\rho}_{\sigma^\pm p}^{(1)} (\Vie{p}{}{}|\Vie{p}{0}{}) = \frac{1}{\sqrt{2}} \, [\gamma(\Vie{p}{}{}) \Vie{\hat{e}}{p,2}{+}(\Vie{p}{}{}) \pm i \Vie{\hat{e}}{s}{}(\Vie{p}{}{})] \cdot \Vie{E}{1,p}{(0)} (\Vie{p}{0}{}) \: .
\label{eq:sig+_p}
\end{equation}
\sloppy Here we have used the short-hand notation $\gamma(\Vie{p}{}{}) = (\Vie{\hat{e}}{p,2}{+} (\Vie{p}{}{}) \cdot \Vie{\hat{e}}{p,1}{+} (\Vie{p}{}{}))^{-1}$. For $\theta_0 > \theta_c$, the total zero order field amplitude $\Vie{E}{1,p}{(0)} (\Vie{p}{0}{})$ is complex. Therefore neither $\rho_{pp}^{(1)}$ nor $\rho_{sp}^{(1)}$ can be zero for propagating waves. We have seen in Section~\ref{sec:brewster_1storder} that a zero intensity $p \to p$ scattering point can be found in the evanescent region since $\Vie{\hat{e}}{p,2}{+}(\Vie{p}{}{})$ becomes complex. However, a zero point in $p \to \sigma^\pm$ scattering may be found in the propagating region.
Indeed, the fact that the square bracket in Eq.~(\ref{eq:sig+_p}) is complex even for purely real values of $\Vie{\hat{e}}{p,2}{+}(\Vie{p}{}{})$ and $\Vie{\hat{e}}{s}{}(\Vie{p}{}{})$ may compensate for the fact that $\Vie{E}{1,p}{(0)}(\Vie{p}{0}{})$ is complex and make the dot product in Eq.~(\ref{eq:sig+_p}) vanish. Note the similarity with the right-hand side in Eq.~(\ref{eq:rot:dipole}) for the case of the radiation emitted by a rotating dipole, with the important difference that the p polarization vector is that of the Snell-conjugate wave.
Since $\Vie{E}{1,p}{(0)}(\Vie{p}{0}{})$ represents a state of polarization of the media in which the dipole rotates in the plane of incidence (cf. discussion in Section~\ref{sec:brewster_1storder}), we expect to find a zero in the $\sigma^\pm$ scattering intensity on each side ($\phi_r = \pm \pi/2$) of the plane of incidence. This is indeed what we observe in Figs.~\ref{fig:2Dmdxc_incglass_42deg}(a) and (d) in the $|\rho_{\sigma^\pm p}^{(1)}|^2$ distribution of $p \to \sigma^\pm$ scattering.


Finally, let us comment on $s \to \sigma^\pm$ scattering. In Figs.~\ref{fig:2Dmdxc_incglass_0deg}-\ref{fig:2Dmdxc_incglass_42deg} it can be observed that the distribution for $|\rho_{\sigma^\pm s}^{(1)}|^2$ stays identical, up to an overall factor, as the angle of incidence varies. This can be understood from the dipole picture. For s-polarized incident light, the incident and zero order waves are s-polarized, so the dipoles oscillate along the $x_2$-direction independently of the angle of incidence.
For scattering in the plane of incidence the first order waves are purely s-polarized and the two $\sigma^\pm$ components have equal intensity. For scattering at $\phi_r = \pm \pi/2$, the first order waves are purely p-polarized and the two $\sigma^\pm$ components have again have equal intensity. We obtain the largest contrast between $\sigma^+$ and $\sigma^-$ for $\phi_r$ being a multiple of $\ang{45}$ since then the p- and s-polarized components are of similar amplitudes.

\section{Conclusion}\label{sec:conclusion}
Based on a perturbative solution of the reduced Rayleigh equations to first order in the surface profile function, we have achieved a detailed mathematical and physical analysis of the scattering of polarized light by a weakly rough interface between two dielectric media. The first order amplitudes are factorized as a product of a scalar component, mainly representing the relative phases of the different scattering paths, and a polarization component. The polarization component can be interpreted as the signature of the polarization state of the dipoles in the media induced by the incident and zero order fields.

We have seen that the Yoneda phenomenon can be explained simply based on a scalar wave, single scattering picture as an intensity enhancement induced by the evanescence of the component of a scattered couple mode existing in the lesser dense medium, while all the energy allocated to the couple mode is radiated away by the component existing in the denser medium. This mechanism clearly answers previous questions put forward in the literature: we conclude that the phenomenon results from a so-called single scattering mechanism, and is not associated with surface (eigen) modes. In particular, the Yoneda phenomenon is nothing else but the continuous analog of a Rayleigh anomaly for periodic dielectric grating, in the sense that the diffuse light here plays the role of probing what the efficiency of a diffracted order would be if it were tracked as the period of the grating would vary. This claim is easily verified with straightforward numerical calculation and the exact same perturbation analysis we have exposed here but adapted to gratings.

By factorizing the scalar behavior from that specific to a polarized wave, we have identified the geometrical criterion for the Brewster scattering phenomenon for p-polarized excitation, and more generally, for predicting the zeros of scattered intensity and amplitude for any polarization state. Simply put, these zeros are not different from those found for the radiation from a tilted oscillating dipole in free space, when the polarization of the emitted radiation is adequately measured in a fixed frame of reference. To be more accurate one may say that the physical essence is that of oscillating dipoles, but one must include the fact that arrays of dipoles yield conjugate waves as was described by e.g. Ewald and Doyle \cite{Ewald1916,Doyle1985}.
The directions of zero scattering (also for evanescent waves) can then be easily interpreted geometrically in terms of Snell-conjugate waves. Moreover, we have discovered an interesting phenomenon of circularly-polarized Brewster scattering in the reflected and transmitted light scattered out of the plane of incidence when the light is incident in the dense medium and the zero order wave undergoes total internal reflection. The physical mechanism responsible for this effect was explained based on the emission of dipoles rotating in the plane of incidence (and by Snell-conjugate waves), which are induced by the fact that the reflected and transmitted zero order waves are out of phase with the incident wave.

In the present work, particular attention is given to the average phase of the scattered waves compared to previously published works on the Yoneda and Brewster scattering phenomena. We have seen that the Brewster scattering phenomenon is associated with a phase jump, while the region of polar scattering angles beyond the Yoneda threshold is associated with a gradually changing phase. These considerations on the phase of the scattered waves can be of particular interest for testing the theory against experiment, e.g. the phase behavior could be tested by the use of interferometry techniques. A simple way to measure the phase behavior associated with the Yoneda and the Brewster scattering effects is to study the scattering of light by a thin dielectric film deposited on a dielectric substrate, as was recently suggested and observed numerically in Ref.~\citenum{Banon2018a}. For such a system, Sel\'{e}nyi rings, which are interference rings in the intensity of the diffusely scattered light, are expected to exhibit: (i) a reversal of angular positions of the maxima and minima of intensity of the rings for p-polarized light as the Brewster scattering angle is surpassed;
(ii) a gradual shift of the angular positions of the rings with respect to those predicted by the simple path difference argument for light scattered at angles beyond the Yoneda threshold due to the additional gradual phase change associated with the Yoneda phenomenon.
In addition, a scattering experiment such as the one achieved in Ref.~\citenum{Gonzalez-Alcalde2016}, but where the outgoing circularly-polarized light is measured instead of the linearly polarized light, would be of particular interest to verify the existence of a circularly-polarized Brewster scattering phenomenon out of the plane of incidence as it would strengthen the rotating dipole interpretation from which it originates.

Finally, we emphasize that the results presented in this work are approximate and are expected to be valid only for weakly rough surfaces. Additional experimental and theoretical investigations are therefore welcome to assess the range of validity of the presented hypotheses.

\begin{acknowledgments}
This research was supported in part by The Research Council of Norway Contract No. 216699. The research of I.S. was supported in part by the French National Research Agency (ANR-15-CHIN-0003).
This research was supported in part by NTNU and the Norwegian metacenter for High Performance Computing (NOTUR) by the allocation of computer time.
\end{acknowledgments}


\begin{widetext}
\appendix
\section{Perturbative solution}\label{AppendixA}

This appendix is devoted to the derivation of the method known as small amplitude perturbation theory (SAPT) for obtaining approximate solutions of the reduced Rayleigh equations. The basic principle of the method is to expand the kernel factor $\Cie{J}{l,m}{b,a}$ in a series of Fourier transforms of the power of the surface profile function $\zeta$, and to expand the unknown reflection and transmission amplitudes in a similar series and matching terms of the same order. The expansions can be expressed as follows
\begin{subequations}
\begin{align}
	\Cie{J}{l,m}{b,a} \ofuipi{p}{}{q}{} &= [b \alpha_l\of{p} - a \alpha_m\of{q}]^{-1} \int \exp[-i(\Vie{p}{}{} - \Vie{q}{}{})\cdot \bxp] \: \exp[-i (b \alpha_l\of{p} - a \alpha_m\of{q}) \: \zxp] \: \dtwox \nonumber\\
		&= \sum_{n=0}^\infty \frac{(-i)^n }{n!} \:  [b \alpha_l\of{p} - a \alpha_m\of{q}]^{n-1} \:\hat{\zeta}^{(n)}(\Vie{p}{}{} - \Vie{q}{}{}) \label{kernelExpansion}\\
	\Vie{R}{}{} \ofuipi{q}{}{p}{0} &= \sum_{j=0}^\infty \frac{(-i)^j }{j!} \: \Vie{R}{}{(j)} \ofuipi{q}{}{p}{0} \label{reflExpansion}\\
	\Vie{T}{}{} \ofuipi{q}{}{p}{0} &= \sum_{j=0}^\infty \frac{(-i)^j }{j!} \: \Vie{T}{}{(j)} \ofuipi{q}{}{p}{0} \: .\label{tranExpansion}
\end{align}\label{expansions}
\end{subequations}
In equation Eq.~(\ref{kernelExpansion}), we have defined the Fourier transform of the $n^\mathrm{th}$ power of $\zeta$, which we will refer to as the $n^\mathrm{th}$ \emph{Fourier moment of the surface profile}, as
\begin{equation}
\hat{\zeta}^{(n)}(\Vie{q}{}{}) = \int \zeta^n (\bxp) \: \exp[-i \Vie{q}{}{}\cdot\bxp ] \:\dtwox \: .
\end{equation}
We are now ready to proceed with the perturbative method.\\

\textbf{Reflection}:
We start by inserting Eqs.~(\ref{kernelExpansion},\ref{reflExpansion}) into the reduced Rayleigh equation Eq.~(\ref{RREint}) in the case of reflection [see Eq.~(\ref{matrices})]. We obtain
%
\begin{align}
&\sum_{n=0}^\infty \: \sum_{j=0}^\infty \: \frac{(-i)^{n+j}}{n! \: j!} \: \int  [ \alpha_2\of{p} - \alpha_1\of{q}]^{n-1} \: \hat{\zeta}^{(n)}(\Vie{p}{}{} - \Vie{q}{}{}) \Vie{M}{2,1}{+,+}\ofuipi{p}{}{q}{} \: \Vie{R}{}{(j)} \ofuipi{q}{}{p}{0} \:\dtwoq  \nonumber\\
&= - \sum_{m=0}^\infty \: \frac{(-i)^{m}}{m!} \: [\alpha_2\of{p} + \alpha_1(\Vie{p}{0}{})]^{m-1} \: \hat{\zeta}^{(m)}(\Vie{p}{}{} - \Vie{p}{0}{}) \: \Vie{M}{2,1}{+,-} \ofuipi{p}{}{p}{0} \: .
\end{align}
%
A summation over all $(n,j) \in \mathbb{N}^2$ is equivalent to a summation over subsets $\Cie{S}{m}{} = \{ (n,j) \in \mathbb{N}^2 \st n+j = m \}$ of pairs of constant sum $m = n+j$, i.e. that we have $\sum_{n,j = 0}^{\infty} \equiv \sum_{m = 0}^{\infty} \sum_{(n,j) \in \Cie{S}{m}{}}$, therefore the previous equation can be recast as
%
\begin{align}
\sum_{m=0}^\infty \: \frac{(-i)^{m}}{m!} \sum_{n=0}^m \:  \binom{m}{n} \: &\int  [ \alpha_2\of{p} - \alpha_1\of{q}]^{n-1} \: \hat{\zeta}^{(n)}(\Vie{p}{}{} - \Vie{q}{}{}) \: \Vie{M}{2,1}{+,+} \ofuipi{p}{}{q}{} \: \Vie{R}{}{(m-n)} \ofuipi{q}{}{p}{0} \:\dtwoq \nonumber\\&= - \sum_{m=0}^\infty \: \frac{(-i)^{m}}{m!} \: [\alpha_2\of{p} + \alpha_1(\Vie{p}{0}{})]^{m-1} \: \hat{\zeta}^{(m)}(\Vie{p}{}{} - \Vie{p}{0}{}) \: \Vie{M}{2,1}{+,-}\ofuipi{p}{}{p}{0} \: .
\end{align}
%
Note that here we have used that $\frac{1}{n! \: (m-n)!} = \frac{1}{m!} \binom{m}{n}$ by definition of the binomial coefficients. The perturbation procedure consists in matching orders in both side of the equation. The order zero only consists of one term $n=m=0$ and gives
\begin{align}
&\int  [ \alpha_2\of{p} - \alpha_1\of{q}]^{-1} \: \hat{\zeta}^{(0)}(\Vie{p}{}{} - \Vie{q}{}{}) \: \Vie{M}{2,1}{+,+} \ofuipi{p}{}{q}{} \: \Vie{R}{}{(0)} \ofuipi{q}{}{p}{0} \:\dtwoq \nonumber\\
&= - [\alpha_2\of{p} + \alpha_1(\Vie{p}{0}{})]^{-1} \: \hat{\zeta}^{(0)}(\Vie{p}{}{} - \Vie{p}{0}{}) \: \Vie{M}{2,1}{+,-}\ofuipi{p}{}{p}{0} \: .
\end{align}
By using that $\hat{\zeta}^{(0)}(\Vie{p}{}{} - \Vie{q}{}{}) = (2\pi)^2 \, \delta(\Vie{p}{}{} - \Vie{q}{}{})$, we finally obtain the zero order reflection amplitude
\begin{align}
\Vie{R}{}{(0)} \ofuipi{p}{}{p}{0} &= (2\pi)^2 \, \delta(\Vie{p}{}{} - \Vie{p}{0}{}) \: \frac{\alpha_1(\Vie{p}{0}{}) - \alpha_2(\Vie{p}{0}{})}{\alpha_2(\Vie{p}{0}{}) + \alpha_1(\Vie{p}{0}{})} \: \left[\Vie{M}{2,1}{+,+}\ofuipi{p}{0}{p}{0}\right]^{-1}  \: \Vie{M}{2,1}{+,-}\ofuipi{p}{0}{p}{0} = (2\pi)^2 \, \delta(\Vie{p}{}{} - \Vie{p}{0}{}) \: \boldsymbol{\rho}^{(0)} (\Vie{p}{0}{})
\label{Aeq:sapt0:ref}
\end{align}
We have just obtained that the zero order of the reflection amplitude corresponds exactly to the reflection amplitude for a planar surface and it is straightforward to show that $\boldsymbol{\rho}^{(0)} (\Vie{p}{0}{})$ is a diagonal matrix containing the Fresnel amplitudes. This was to be expected in the sense that the zero order of the surface profile corresponds to its average plane. For orders $m \geq 1$, we have
\begin{align}
\sum_{n=0}^m \: \binom{m}{n} \: \int  &[ \alpha_2\of{p} - \alpha_1\of{q}]^{n-1} \: \hat{\zeta}^{(n)}(\Vie{p}{}{} - \Vie{q}{}{}) \: \Vie{M}{2,1}{+,+}\ofuipi{p}{}{q}{} \: \Vie{R}{}{(m-n)} \ofuipi{q}{}{p}{0} \:\dtwoq \nonumber\\= - &[\alpha_2\of{p} + \alpha_1(\Vie{p}{0}{})]^{m-1} \: \hat{\zeta}^{(m)}(\Vie{p}{}{} - \Vie{p}{0}{}) \: \Vie{M}{2,1}{+,-}\ofuipi{p}{}{p}{0}\: ,
\end{align}
which by isolating the term of interest, $n=0$, gives $\Vie{R}{}{(m)}$ as a function of $\Vie{R}{}{(m-1)} \cdots \Vie{R}{}{(0)}$, in other words we have a recursive relation for determining all orders,
\begin{align}
\Vie{R}{}{(m)} \ofuipi{p}{}{p}{0} =& [\alpha_1\of{p} - \alpha_2\of{p}]\left[ \Vie{M}{2,1}{+,+}\ofuipi{p}{}{p}{}\right]^{-1} \: \bigg[ (\alpha_2\of{p} + \alpha_1(\Vie{p}{0}{}))^{m-1} \: \hat{\zeta}^{(m)}(\Vie{p}{}{} - \Vie{p}{0}{}) \: \Vie{M}{2,1}{+,-} \ofuipi{p}{}{p}{0} \nonumber\\& + \sum_{n=1}^m \: \binom{m}{n} \: \int  [\alpha_2\of{p} - \alpha_1\of{q}]^{n-1} \: \hat{\zeta}^{(n)}(\Vie{p}{}{} - \Vie{q}{}{}) \: \Vie{M}{2,1}{+,+}\ofuipi{p}{}{q}{} \: \Vie{R}{}{(m-n)} \ofuipi{q}{}{p}{0} \: \dtwoq \bigg] \: .
\label{recursive:singlePerturbation}
\end{align}
In general, the evaluation of high orders would require the evaluation of as many integrals as the order and can become costly. For the first order, only one such integral is to be evaluated and is straightforward thanks to the fact that $\Vie{R}{}{(0)} (\Vie{q}{}{}|\Vie{p}{0}{}) \propto \delta (\Vie{q}{}{} - \Vie{p}{0}{})$. Applying the above equation for $m=1$ gives
\begin{align}
\Vie{R}{}{(1)} \ofuipi{p}{}{p}{0} =& \: [\alpha_1\of{p} - \alpha_2\of{p}] \left[ \Vie{M}{2,1}{+,+} \ofuipi{p}{}{p}{}\right]^{-1} \: \bigg[   \hat{\zeta}^{(1)}(\Vie{p}{}{} - \Vie{p}{0}{}) \: \Vie{M}{2,1}{+,-}\ofuipi{p}{}{p}{0} \nonumber\\&  + \: \int  \hat{\zeta}^{(1)}(\Vie{p}{}{} - \Vie{q}{}{}) \: \Vie{M}{2,1}{+,+}\ofuipi{p}{}{q}{} \: \Vie{R}{}{(0)} \ofuipi{q}{}{p}{0} \: \dtwoq  \bigg] \nonumber\\
=& \: [\alpha_1\of{p} - \alpha_2\of{p}] \: \hat{\zeta}^{(1)}(\Vie{p}{}{} - \Vie{p}{0}{}) \:\left[ \Vie{M}{2,1}{+,+}\ofuipi{p}{}{p}{}\right]^{-1} \: \Big[  \Vie{M}{2,1}{+,-}\ofuipi{p}{}{p}{0}  + \Vie{M}{2,1}{+,+}\ofuipi{p}{}{p}{0}  \:  \boldsymbol{\rho}^{(0)} (\Vie{p}{0}{}) \Big]\nn\\
=& \: [\alpha_1\of{p} - \alpha_2\of{p}] \: \hat{\zeta}^{(1)}(\Vie{p}{}{} - \Vie{p}{0}{}) \: \boldsymbol{\hat{\rho}}^{(1)}\ofuipi{p}{}{p}{0} = \hat{\zeta}^{(1)}(\Vie{p}{}{} - \Vie{p}{0}{}) \: \boldsymbol{\rho}^{(1)}\ofuipi{p}{}{p}{0} \: . \label{Aeq:sapt1:ref}
\end{align}
In Eq.~(\ref{eq:sapt1:ref}), we define the amplitude $\boldsymbol{\hat{\rho}}^{(1)}\ofuipi{p}{}{p}{0}$ and $\boldsymbol{\rho}^{(1)}\ofuipi{p}{}{p}{0} = (\alpha_1\of{p} - \alpha_2\of{p}) \: \boldsymbol{\hat{\rho}}^{(1)}\ofuipi{p}{}{p}{0}$. The reason for these two alternative expressions is that the first one gives a factorization which is more easily interpreted from a physical point of view while the latter factorization aims at separating what depends on the realization of the surface profile, which is just $\hat{\zeta}$ here, and the amplitude factor $\boldsymbol{\rho}^{(1)}\ofuipi{p}{}{p}{0}$ which remains independent of the specific realization of the surface profile (see Section~\ref{sec:theory}).

\textbf{Transmission}: Repeating the reasoning for the transmission amplitudes, we start by inserting Eqs.~(\ref{kernelExpansion},\ref{tranExpansion}) into Eq.~(\ref{RREint}) for transmission [see Eq.~(\ref{matrices})] and get
\begin{align}
\sum_{n=0}^\infty \: \sum_{j=0}^\infty \: \frac{(-i)^{n+j}}{n! \: j!} \: &\int  [- \alpha_1\of{p} + \alpha_2\of{q}]^{n-1} \: \hat{\zeta}^{(n)}(\Vie{p}{}{} - \Vie{q}{}{}) \: \Vie{M}{1,2}{-,-} \ofuipi{p}{}{q}{}\: \Vie{T}{}{(j)} \ofuipi{q}{}{p}{0} \:\dtwoq  \nonumber\\&= \frac{2 \, \sqrt{\epsilon_1 \epsilon_2} \, \alpha_1(\Vie{p}{0}{})}{\epsilon_2 - \epsilon_1} \, (2\pi)^2 \, \delta(\mathbf{p} - \Vie{p}{0}{}) \, \mathbf{I}_2 \: .
\end{align}
By using the same re-summation argument as for reflection, the previous equation thus becomes
\begin{align}
\sum_{m=0}^\infty \: \sum_{n=0}^m \: \frac{(-i)^{m}}{m!} \binom{m}{n} \: &\int  [- \alpha_1\of{p} + \alpha_2\of{q}]^{n-1} \: \hat{\zeta}^{(n)}(\Vie{p}{}{} - \Vie{q}{}{}) \: \Vie{M}{1,2}{-,-}\ofuipi{p}{}{q}{} \: \Vie{T}{}{(m-n)} \ofuipi{q}{}{p}{0} \: \dtwoq \nonumber\\&= \frac{2 \, \sqrt{\epsilon_1 \epsilon_2} \, \alpha_1(\Vie{p}{0}{})}{\epsilon_2 - \epsilon_1} \, (2\pi)^2 \, \delta(\mathbf{p} - \Vie{p}{0}{}) \, \mathbf{I}_2 \: .
\end{align}
Next we match the zero order to the right hand side and the other orders to zero. The zero order only consists of one term $n=m=0$ and gives
\begin{align}
\Vie{T}{}{(0)} \ofuipi{p}{}{p}{0} &= \frac{2 \, \sqrt{\epsilon_1 \epsilon_2} \alpha_1(\Vie{p}{0}{})}{\epsilon_2 - \epsilon_1} \, (2\pi)^2 \, \delta(\mathbf{p} - \Vie{p}{0}{}) \: [\alpha_2(\Vie{p}{0}{}) - \alpha_1(\Vie{p}{0}{})] \: \left[ \Vie{M}{1,2}{-,-} \ofuipi{p}{0}{p}{0}\right]^{-1} \\
& = (2\pi)^2 \, \delta(\mathbf{p} - \Vie{p}{0}{}) \: \boldsymbol{\tau}^{(0)} (\Vie{p}{0}{})
\end{align}
Here we have used that $\hat{\zeta}^{(0)}(\Vie{p}{}{} - \Vie{q}{}{}) = (2\pi)^2 \, \delta(\Vie{p}{}{} - \Vie{q}{}{})$. As observed for the reflection amplitudes, we have just obtained that the zero order of the transmission amplitudes corresponds exactly to the transmission amplitudes for a planar surface, i.e. that $\boldsymbol{\tau}^{(0)} (\Vie{p}{0}{})$ is a diagonal matrix containing the Fresnel transmission amplitudes. For orders $m \geq 1$, we have
\begin{equation}
\sum_{n=0}^m \: \binom{m}{n} \: \int  [\alpha_2\of{q} - \alpha_1\of{p}]^{n-1} \: \hat{\zeta}^{(n)}(\Vie{p}{}{} - \Vie{q}{}{}) \: \Vie{M}{1,2}{-,-}\ofuipi{p}{}{q}{} \: \Vie{T}{}{(m-n)} \ofuipi{q}{}{p}{0} \: \dtwoq = \Vie{0}{}{} \: ,
\end{equation}
which by isolating the term of interest, $n=0$ gives $\Vie{T}{}{(m)}$ as a function of $\Vie{T}{}{(m-1)} \cdots \Vie{T}{}{(0)}$, in other words we have a recursive relation for determining all orders,
\begin{align}
\Vie{T}{}{(m)} \ofuipi{p}{}{p}{0}= &[\alpha_1\of{p} - \alpha_2\of{p}] \: \left[\Vie{M}{1,2}{-,-}\ofuipi{p}{}{p}{}\right]^{-1} \nonumber\\ &\sum_{n=1}^m \: \binom{m}{n} \: \int  [\alpha_2\of{q} - \alpha_1\of{p}]^{n-1} \: \hat{\zeta}^{(n)}(\Vie{p}{}{} - \Vie{q}{}{}) \: \Vie{M}{1,2}{-,-}\ofuipi{p}{}{q}{}\: \Vie{T}{}{(m-n)} \ofuipi{q}{}{p}{0} \: \dtwoq  \: .
\end{align}
Applying the above equation for $m=1$ and using that $\Vie{T}{}{(0)} \ofuipi{q}{}{p}{0} \propto \delta (\Vie{q}{}{} - \Vie{p}{0}{})$ gives
\begin{align}
\Vie{T}{}{(1)} \ofuipi{p}{}{p}{0} =& [\alpha_1\of{p} - \alpha_2\of{p}] \: \hat{\zeta}^{(1)}(\Vie{p}{}{} - \Vie{p}{0}{}) \: \left[\Vie{M}{1,2}{-,-}\ofuipi{p}{}{p}{} \right]^{-1} \: \Vie{M}{1,2}{-,-}\ofuipi{p}{}{p}{0} \: \boldsymbol{\tau}^{(0)} (\Vie{p}{0}{}) \nn\\
=&[\alpha_1\of{p} - \alpha_2\of{p}] \: \hat{\zeta}^{(1)}(\Vie{p}{}{} - \Vie{p}{0}{}) \: \boldsymbol{\hat{\tau}}^{(1)}\ofuipi{p}{}{p}{0} = \hat{\zeta}^{(1)}(\Vie{p}{}{} - \Vie{p}{0}{}) \: \boldsymbol{\tau}^{(1)}\ofuipi{p}{}{p}{0}\: . \label{Aeq:sapt1:tra}
\end{align}

\section{Differential reflection coefficient}\label{AppendixB}

Assuming we have obtained the reflection amplitudes $R_{\alpha \beta} (\Vie{p}{}{} \st \Vie{p}{0}{} )$ either by using the perturbative approach or by the purely numerical simulation, we can now proceed to express the differential reflection coefficient~(DRC) defined as the time-averaged flux radiated around a given scattering direction $(\theta_r,\phi)$ per unit solid angle and per unit incident flux and denoted $\partial R / \partial \Omega_r (\Vie{p}{}{} \st \Vie{p}{0}{} )$. Let a virtual hemisphere of radius $r \gg c/\omega$ lie on the plane $x_3 = 0$ on top of the scattering system. The support of this hemisphere is a disk of area $S = \pi r^2$. We consider the scattering from a \emph{truncated} version of the scattering system in which the surface profiles are set to be flat outside the disk support. Consequently, the field amplitudes we will manipulate are not strictly speaking those of the full system of interest but will converge to them as $r \to \infty$. We will nevertheless keep the same notations as that from the full system introduced in Section~\ref{sec:theory} for simplicity. The time-averaged flux incident on this disk is given by
\begin{align}
P_{\mathrm{inc}/S} &= - \mathrm{Re} \frac{c}{8 \pi} \int_S  \left[ \Vie{E}{0}{*} (\Vie{p}{0}{}) \times \left( \frac{c}{\omega}  \, \Vie{k}{1}{-} (\Vie{p}{0}{}) \times \Vie{E}{0}{} (\Vie{p}{0}{}) \right) \right] \cdot \hat{\mathbf{e}}_3 \: \exp \left[ -i (\Vie{k}{1}{-*} (\Vie{p}{0}{}) - \Vie{k}{1}{-} (\Vie{p}{0}{})) \cdot \Vie{x}{}{} \right] \dtwox \nonumber \\
&= - \frac{c^2}{8 \pi \omega} \, \mathrm{Re}  \int_S  \left[ | \Vie{E}{0}{} (\Vie{p}{0}{}) |^2 \, \Vie{k}{1}{-} (\Vie{p}{0}{})   -
\left( \Vie{E}{0}{*} (\Vie{p}{0}{})  \cdot \Vie{k}{1}{-} (\Vie{p}{0}{}) \right) \cdot \Vie{E}{0}{} (\Vie{p}{0}{})  \right] \cdot \hat{\mathbf{e}}_3 \dtwox \nonumber \\
                   &= S \, \frac{c^2}{8 \pi \omega}  \, \alpha_1 (\Vie{p}{0}{}) \, | \Vie{E}{0}{} (\Vie{p}{0}{}) |^2
\nonumber \\
&= S \, \frac{c^2}{8 \pi \omega}  \, \alpha_1 (\Vie{p}{0}{}) \, \left[ \left|\Cie{E}{0,p}{}\right|^2  + \left|\Cie{E}{0,s}{}\right|^2 \right].
\end{align}
Here, the $^*$ denotes the complex conjugate, and incident field amplitude $\Vie{E}{0}{} (\Vie{p}{0}{}) = \Cie{E}{0,p}{} \, \hat{\mathbf{e}}_p^{\mathnormal{-}} (\Vie{p}{0}{})  + \Cie{E}{0,s}{} \, \hat{\mathbf{e}}_s  (\Vie{p}{0}{})$ as defined in Eq.~(\ref{incField}), the vector identity
$\mathbf{a}\times(\mathbf{b}\times\mathbf{c}) = (\mathbf{a}\cdot\mathbf{c})\mathbf{b} - (\mathbf{a}\cdot\mathbf{b})\mathbf{c}$ and the orthogonality between the field and the wave vector $\Vie{E}{0}{*} (\Vie{p}{0}{})  \cdot \Vie{k}{1}{-} (\Vie{p}{0}{}) = 0$ have been used.
Note that the flux incident on the disk is proportional to the disk area. Let us now consider the outgoing flux crossing an elementary surface $\mathrm{d}\sigma = r^2 \sin \theta_r \mathrm{d}\theta_r \mathrm{d}\phi = r^2 \mathrm{d}\Omega_r$ around a point $\Vie{r}{}{} = r \, (\sin \theta_r \cos \phi \, \hat{\mathbf{e}}_1 + \sin \theta_r \sin \phi \, \hat{\mathbf{e}}_2 + \cos \theta_r \, \hat{\mathbf{e}}_3 ) = r \, \hat{\mathbf{n}}$. The flux crossing this elementary surface is given by
\begin{equation}
P_{\mathrm{d}\sigma} = \frac{c}{8 \pi} \, \mathrm{Re}  \left[ \Vie{E}{1}{+*} (\Vie{r}{}{}) \times \Vie{H}{1}{+} (\Vie{r}{}{}) \right] \cdot \hat{\mathbf{n}} \, \mathrm{d}\sigma .
\label{outpower}
\end{equation}
We then use the well-known asymptotic expansion of the field in the far-field given by (see Refs.~\cite{Agarwal1977} and \cite{Miyamoto1962})
\begin{subequations}
\begin{align}
\Vie{E}{1}{+} (\Vie{r}{}{}) &\sim - i \, \epsilon_1^{1/2} \frac{\omega}{2 \pi \, c} \cos \theta_r \, \frac{\exp (i \epsilon_1^{1/2} \frac{\omega}{c} r)}{r} \, \Vie{E}{1}{+} (\Vie{p}{}{})
\\
\Vie{H}{1}{+} (\Vie{r}{}{}) &\sim - i \, \epsilon_1 \frac{\omega}{2 \pi \, c} \cos \theta_r \, \frac{\exp (i \epsilon_1^{1/2} \frac{\omega}{c} r)}{r} \, \hat{\mathbf{n}} \times \Vie{E}{1}{+} (\Vie{p}{}{})
\end{align}
\label{asymptotic:field}%
\end{subequations}
where $\Vie{p}{}{} = \sqrt{\epsilon_1} \frac{\omega}{c} (\sin \theta_r \cos \phi \, \hat{\mathbf{e}}_1 + \sin \theta_r \sin \phi \, \hat{\mathbf{e}}_2 )$. This asymptotic approximation will become more and more accurate as we let $r \to \infty$. Plugging Eq.~(\ref{asymptotic:field}) into Eq.~(\ref{outpower}) we obtain
\begin{equation}
P_{\mathrm{d}\sigma} = \epsilon_1^{3/2}  \left(\frac{\omega}{2 \pi \, c}\right)^2 \, \cos^2 \theta_r \, \frac{c}{8 \pi} \, | \Vie{E}{1}{+} (\Vie{p}{}{})|^2 \, \mathrm{d}\Omega_r = \epsilon_1^{3/2}  \left(\frac{\omega}{2 \pi \, c}\right)^2 \, \cos^2 \theta_r \, \frac{c}{8 \pi} \, \left( | \Cie{E}{1,p}{+} (\Vie{p}{}{})|^2  + | \Cie{E}{1,s}{+} (\Vie{p}{}{})|^2\right) \, \mathrm{d}\Omega_r .
\end{equation}
The total differential reflection coefficient is then given by
\begin{equation}
  \frac{\partial R}{\partial \Omega_r} (\Vie{p}{}{} \st \Vie{p}{0}{} )
  =
  \lim_{r\to\infty} \, \frac{P_{\mathrm{d}\sigma}}{P_{\mathrm{inc}/S} \, \mathrm{d}\Omega_r}
  =
  \lim_{r\to\infty} \,
  \frac{\epsilon_1}{S}
  \left(\frac{\omega}{2 \pi \, c}\right)^2
  \frac{ \cos^2 \theta_r }{ \cos \theta_0 }
  %
  \, \frac{| \Cie{E}{1,p}{+} (\Vie{p}{}{})|^2  + | \Cie{E}{1,s}{+} (\Vie{p}{}{})|^2}{| \Cie{E}{0,p}{} |^2  + | \Cie{E}{0,s}{} |^2} .
\label{total:DRC}
\end{equation}
From the total differential reflection coefficient given by Eq.~(\ref{total:DRC}), we deduce the differential reflection coefficient when an incident plane wave of polarization $\beta$, with lateral wave vector $\Vie{p}{0}{}$ is reflected into a plane wave of polarization $\alpha$ with lateral wave vector $\Vie{p}{}{}$ given as
\begin{equation}
  \frac{\partial R_{\alpha \beta}}{\partial \Omega_r} (\Vie{p}{}{} \st \Vie{p}{0}{} )
  =
  \lim_{r\to\infty} \,
  \frac{\epsilon_1}{S}
  \left(\frac{\omega}{2 \pi \, c}\right)^2
  \frac{ \cos^2 \theta_r }{ \cos \theta_0 }
  %
  \, \left|R_{\alpha \beta} (\Vie{p}{}{} \st \Vie{p}{0}{})\right|^2
  =
  \lim_{r\to\infty} \, \frac{\partial R_{\alpha \beta}^{(S)}}{\partial \Omega_r} (\Vie{p}{}{} \st \Vie{p}{0}{} ) .
\label{ab:DRC}
\end{equation}
As we are interested in averaging the optical response over realizations of the surface profiles, we consider the following ensemble average
\begin{equation}
  \bigg\langle \frac{\partial R_{\alpha \beta}^{(S)}}{\partial \Omega_r} (\Vie{p}{}{} \st \Vie{p}{0}{} ) \bigg\rangle
  =
  \frac{\epsilon_1}{S}
  \left(\frac{\omega}{2 \pi \, c}\right)^2
  \frac{ \cos^2 \theta_r }{ \cos \theta_0 }
  %
  \, \left\langle |R_{\alpha \beta} (\Vie{p}{}{} \st \Vie{p}{0}{})|^2 \right\rangle  .
\end{equation}
A similar derivation for the differential transmitted coefficient yields
\begin{equation}
  \bigg\langle \frac{\partial T_{\alpha \beta}^{(S)}}{\partial \Omega_t} (\Vie{p}{}{} \st \Vie{p}{0}{} ) \bigg\rangle
  =
  \frac{\epsilon_2^{3/2}}{\epsilon_1^{1/2} S}
  \left(\frac{\omega}{2 \pi \, c}\right)^2
  \frac{ \cos^2 \theta_t }{ \cos \theta_0 }
  %
  \, \left\langle |T_{\alpha \beta} (\Vie{p}{}{} \st \Vie{p}{0}{})|^2 \right\rangle  .
\end{equation}

By decomposing the reflection amplitudes as the sum of the mean and fluctuation (deviation from the mean)
\begin{equation}
R_{\alpha \beta} (\Vie{p}{}{} \st \Vie{p}{0}{}) = \left\langle R_{\alpha \beta} (\Vie{p}{}{} \st \Vie{p}{0}{}) \right\rangle  + \left[ R_{\alpha \beta} (\Vie{p}{}{} \st \Vie{p}{0}{}) - \left\langle R_{\alpha \beta} (\Vie{p}{}{} \st \Vie{p}{0}{}) \right\rangle \right] \: ,
\end{equation}
we can decompose the MDRC as the sum of a coherent term and an incoherent term
\begin{equation}
  \bigg\langle \frac{\partial R_{\alpha \beta}^{(S)}}{\partial \Omega_r} (\Vie{p}{}{} \st \Vie{p}{0}{} ) \bigg\rangle
  = \bigg\langle \frac{\partial R_{\alpha \beta}^{(S)}}{\partial \Omega_r} (\Vie{p}{}{} \st \Vie{p}{0}{} ) \bigg\rangle_\mathrm{coh}
  +
  \bigg\langle \frac{\partial R_{\alpha \beta}^{(S)}}{\partial \Omega_r} (\Vie{p}{}{} \st \Vie{p}{0}{} ) \bigg\rangle _\mathrm{incoh}\: ,
\end{equation}
where
\begin{subequations}
\begin{align}
  \bigg\langle \frac{\partial R_{\alpha \beta}^{(S)}}{\partial \Omega_r} (\Vie{p}{}{} \st \Vie{p}{0}{} ) \bigg\rangle_\mathrm{coh}
  &=
  \frac{\epsilon_1}{S}
  \left(\frac{\omega}{2 \pi \, c}\right)^2
  \frac{ \cos^2 \theta_r }{ \cos \theta_0 }
  %
    \, \left| \left\langle R_{\alpha \beta} (\Vie{p}{}{} \st \Vie{p}{0}{}) \right\rangle \right|^2
  \\
  \bigg\langle \frac{\partial R_{\alpha \beta}^{(S)}}{\partial \Omega_r} (\Vie{p}{}{} \st \Vie{p}{0}{} ) \bigg\rangle_\mathrm{incoh}
  &=
  \frac{\epsilon_1}{S}
  \left(\frac{\omega}{2 \pi \, c}\right)^2
  \frac{ \cos^2 \theta_r }{ \cos \theta_0 }
  %
    \, \left[ \left\langle |R_{\alpha \beta} (\Vie{p}{}{} \st \Vie{p}{0}{})|^2 \right\rangle - \left| \left\langle R_{\alpha \beta} (\Vie{p}{}{} \st \Vie{p}{0}{}) \right\rangle \right|^2 \right].
\label{mdrc:inco}%
\end{align}
\end{subequations}
If we now use the expression found in Appendix \ref{AppendixA} for the reflection amplitudes to first order in the surface profile,
\begin{equation}
\mathbf{R} \ofuipi{p}{}{p}{0} \approx  \Vie{R}{}{(0)} \ofuipi{p}{}{p}{0} - i \Vie{R}{}{(1)} \ofuipi{p}{}{p}{0} \: ,
\end{equation}
where $\Vie{R}{}{(0)} \ofuipi{p}{}{p}{0}$ is the response from the corresponding system with planar interface, Eq.~(\ref{eq:sapt0:ref}), and $\Vie{R}{}{(1)} \ofuipi{p}{}{p}{0}$ is given in Eq.~(\ref{eq:sapt1:ref}), we obtain that the factor in the square bracket in Eq.~(\ref{mdrc:inco}) reads
\begin{equation}
  \left\langle |R_{\alpha \beta} (\Vie{p}{}{} \st \Vie{p}{0}{})|^2 \right\rangle - \left| \left\langle R_{\alpha \beta} (\Vie{p}{}{} \st \Vie{p}{0}{}) \right\rangle \right|^2
= \left\langle \left|R_{\alpha \beta}^{(1)} (\Vie{p}{}{} \st \Vie{p}{0}{})\right|^2 \right\rangle
= \left\langle | \hat{\zeta}_{S} (\Vie{p}{}{} - \Vie{p}{0}{}) |^2 \right\rangle \: |\rho_{\alpha \beta}^{(1)} \ofuipi{p}{}{p}{0} |^2  .
\end{equation}
Note here that we are still dealing with a scattering system whose surface profiles are flat outside the disk of radius $r$, hence the subscript $S$.  For the statistical properties attributed to the surface profiles in Sec.~\ref{sec:scatt:sys}, we have
\begin{align}
\left\langle \hat{\zeta}_{S} (\Vie{q}{}{}) \hat{\zeta}_{S}^* (\Vie{q}{}{}) \right\rangle &= \left\langle \int_S \int_S \zeta (\Vie{x}{}{}) \zeta (\mathbf{x}') \exp \left[i \Vie{q}{}{} \cdot (\Vie{x}{}{} - \mathbf{x}') \right] \: \mathrm{d}^2x \:  \mathrm{d}^2x' \right\rangle \nonumber \\
&=  \int_S \int_S \left\langle \zeta (\Vie{x}{}{}) \zeta (\mathbf{x}') \right\rangle \: \exp \left[i \Vie{q}{}{} \cdot (\Vie{x}{}{} - \mathbf{x}') \right]  \: \mathrm{d}^2x \:  \mathrm{d}^2x'  \nonumber \\
&=  \int_S \int_S \sigma^2 \: W(\Vie{x}{}{} - \mathbf{x}') \: \exp \left[i \Vie{q}{}{} \cdot (\Vie{x}{}{} - \mathbf{x}') \right] \: \mathrm{d}^2x \:  \mathrm{d}^2x'  .
\end{align}
Here we have used the definition of the Fourier transform, and the fact that ensemble average commutes with the integration of the surfaces and the definition of the correlation function. Via the change of variable $\Vie{u}{}{} = \Vie{x}{}{} - \Vie{x}{}{}'$ we obtain
\begin{equation}
\left\langle \hat{\zeta}_{S} (\Vie{q}{}{}) \hat{\zeta}_{S}^* (\Vie{q}{}{}) \right\rangle = S \: \sigma^2 \: \int_S W(\Vie{u}{}{}) \: \exp (i \Vie{q}{}{} \cdot \Vie{u}{}{} ) \: \mathrm{d}^2u  = S \: \sigma^2 \: g_S (\Vie{q}{}{})  .
\label{pow:spec:trunc}
\end{equation}
Thus
\begin{equation}
  \left\langle |R_{\alpha \beta} (\Vie{p}{}{} \st \Vie{p}{0}{})|^2 \right\rangle - \left| \left\langle R_{\alpha \beta} (\Vie{p}{}{} \st \Vie{p}{0}{}) \right\rangle \right|^2
  =  S \: \sigma^2 \: g_S (\Vie{p}{}{} -\Vie{p}{0}{}) \, \left|\rho_{\alpha \beta}^{(1)} \ofuipi{p}{}{p}{0} \right|^2  .
\end{equation}
Finally, by plugging the above equation into Eq.~(\ref{mdrc:inco}), the surface area $S$ cancels and letting $r \to \infty$, $g_S \to g$ (where we remind the reader that $g$ is the power spectrum of the surface profiles) and we finally obtain the expression for the incoherent component of the MDRC for the entire (infinite) system under the first order approximation of the reflected amplitudes in product of the surface profiles
\begin{equation}
\left\langle \frac{\partial R_{\alpha \beta} }{\partial \Omega_r} (\Vie{p}{}{} | \Vie{p}{0}{}) \right\rangle_{\mathrm{incoh}} = \: \epsilon_1 \left(\frac{\omega}{2 \pi c} \right)^2 \: \frac{\cos^2 \theta_r}{\cos \theta_0} \: g (\Vie{p}{}{} -\Vie{p}{0}{}) \: \sigma^2 \: \left|\rho_{\alpha \beta}^{(1)} \ofuipi{p}{}{p}{0} \right|^2  .
\label{eq:incoMDRC:final:app}
\end{equation}
Similarly, for the transmitted light we obtain
\begin{equation}
\left\langle \frac{\partial T_{\alpha \beta} }{\partial \Omega_t} (\Vie{p}{}{} | \Vie{p}{0}{}) \right\rangle_{\mathrm{incoh}} = \: \frac{\epsilon_2^{3/2}}{\epsilon_1^{1/2}} \left(\frac{\omega}{2 \pi c} \right)^2 \: \frac{\cos^2 \theta_t}{\cos \theta_0} \: g (\Vie{p}{}{} -\Vie{p}{0}{}) \: \sigma^2 \: \left|\tau_{\alpha \beta}^{(1)} \ofuipi{p}{}{p}{0} \right|^2  .
\label{eq:incoMDTC:final:app}
\end{equation}

\end{widetext}

%
\bibliography{scatterbib_hetland}

\begin{thebibliography}{46}%
\makeatletter
\providecommand \@ifxundefined [1]{%
 \@ifx{#1\undefined}
}%
\providecommand \@ifnum [1]{%
 \ifnum #1\expandafter \@firstoftwo
 \else \expandafter \@secondoftwo
 \fi
}%
\providecommand \@ifx [1]{%
 \ifx #1\expandafter \@firstoftwo
 \else \expandafter \@secondoftwo
 \fi
}%
\providecommand \natexlab [1]{#1}%
\providecommand \enquote  [1]{``#1''}%
\providecommand \bibnamefont  [1]{#1}%
\providecommand \bibfnamefont [1]{#1}%
\providecommand \citenamefont [1]{#1}%
\providecommand \href@noop [0]{\@secondoftwo}%
\providecommand \href [0]{\begingroup \@sanitize@url \@href}%
\providecommand \@href[1]{\@@startlink{#1}\@@href}%
\providecommand \@@href[1]{\endgroup#1\@@endlink}%
\providecommand \@sanitize@url [0]{\catcode `\\12\catcode `\$12\catcode
  `\&12\catcode `\#12\catcode `\^12\catcode `\_12\catcode `\%12\relax}%
\providecommand \@@startlink[1]{}%
\providecommand \@@endlink[0]{}%
\providecommand \url  [0]{\begingroup\@sanitize@url \@url }%
\providecommand \@url [1]{\endgroup\@href {#1}{\urlprefix }}%
\providecommand \urlprefix  [0]{URL }%
\providecommand \Eprint [0]{\href }%
\providecommand \doibase [0]{http://dx.doi.org/}%
\providecommand \selectlanguage [0]{\@gobble}%
\providecommand \bibinfo  [0]{\@secondoftwo}%
\providecommand \bibfield  [0]{\@secondoftwo}%
\providecommand \translation [1]{[#1]}%
\providecommand \BibitemOpen [0]{}%
\providecommand \bibitemStop [0]{}%
\providecommand \bibitemNoStop [0]{.\EOS\space}%
\providecommand \EOS [0]{\spacefactor3000\relax}%
\providecommand \BibitemShut  [1]{\csname bibitem#1\endcsname}%
\let\auto@bib@innerbib\@empty
\bibitem [{\citenamefont {Hetland}\ \emph {et~al.}(2016)\citenamefont
  {Hetland}, \citenamefont {Maradudin}, \citenamefont {Nordam},\ and\
  \citenamefont {Simonsen}}]{Hetland2016a}%
  \BibitemOpen
  \bibfield  {author} {\bibinfo {author} {\bibfnamefont {{\O}.~S.}\
  \bibnamefont {Hetland}}, \bibinfo {author} {\bibfnamefont {A.~A.}\
  \bibnamefont {Maradudin}}, \bibinfo {author} {\bibfnamefont {T.}~\bibnamefont
  {Nordam}}, \ and\ \bibinfo {author} {\bibfnamefont {I.}~\bibnamefont
  {Simonsen}},\ }\href {\doibase 10.1103/PhysRevA.93.053819} {\bibfield
  {journal} {\bibinfo  {journal} {Phys. Rev. A}\ }\textbf {\bibinfo {volume}
  {93}},\ \bibinfo {pages} {053819} (\bibinfo {year} {2016})}\BibitemShut
  {NoStop}%
\bibitem [{\citenamefont {Hetland}\ \emph {et~al.}(2017)\citenamefont
  {Hetland}, \citenamefont {Maradudin}, \citenamefont {Nordam}, \citenamefont
  {Letnes},\ and\ \citenamefont {Simonsen}}]{Hetland2017}%
  \BibitemOpen
  \bibfield  {author} {\bibinfo {author} {\bibfnamefont {{\O}.~S.}\
  \bibnamefont {Hetland}}, \bibinfo {author} {\bibfnamefont {A.~A.}\
  \bibnamefont {Maradudin}}, \bibinfo {author} {\bibfnamefont {T.}~\bibnamefont
  {Nordam}}, \bibinfo {author} {\bibfnamefont {P.~A.}\ \bibnamefont {Letnes}},
  \ and\ \bibinfo {author} {\bibfnamefont {I.}~\bibnamefont {Simonsen}},\
  }\href {\doibase 10.1103/PhysRevA.95.043808} {\bibfield  {journal} {\bibinfo
  {journal} {Phys. Rev. A}\ }\textbf {\bibinfo {volume} {95}},\ \bibinfo
  {pages} {043808} (\bibinfo {year} {2017})}\BibitemShut {NoStop}%
\bibitem [{\citenamefont {Yoneda}(1963)}]{Yoneda1963}%
  \BibitemOpen
  \bibfield  {author} {\bibinfo {author} {\bibfnamefont {Y.}~\bibnamefont
  {Yoneda}},\ }\href {\doibase 10.1103/PhysRev.131.2010} {\bibfield  {journal}
  {\bibinfo  {journal} {Phys. Rev.}\ }\textbf {\bibinfo {volume} {131}},\
  \bibinfo {pages} {2010} (\bibinfo {year} {1963})}\BibitemShut {NoStop}%
\bibitem [{\citenamefont {Vineyard}(1982)}]{Vineyard1982}%
  \BibitemOpen
  \bibfield  {author} {\bibinfo {author} {\bibfnamefont {G.~H.}\ \bibnamefont
  {Vineyard}},\ }\href {\doibase 10.1103/PhysRevB.26.4146} {\bibfield
  {journal} {\bibinfo  {journal} {Physical Review B}\ }\textbf {\bibinfo
  {volume} {26}},\ \bibinfo {pages} {4146} (\bibinfo {year}
  {1982})}\BibitemShut {NoStop}%
\bibitem [{\citenamefont {Sinha}\ \emph {et~al.}(1988)\citenamefont {Sinha},
  \citenamefont {Sirota}, \citenamefont {Garoff},\ and\ \citenamefont
  {Stanley}}]{Sinha1988}%
  \BibitemOpen
  \bibfield  {author} {\bibinfo {author} {\bibfnamefont {S.~K.}\ \bibnamefont
  {Sinha}}, \bibinfo {author} {\bibfnamefont {E.~B.}\ \bibnamefont {Sirota}},
  \bibinfo {author} {\bibfnamefont {S.}~\bibnamefont {Garoff}}, \ and\ \bibinfo
  {author} {\bibfnamefont {H.~B.}\ \bibnamefont {Stanley}},\ }\href {\doibase
  10.1103/PhysRevB.38.2297} {\bibfield  {journal} {\bibinfo  {journal} {Phys.
  Rev. B}\ }\textbf {\bibinfo {volume} {38}},\ \bibinfo {pages} {2297}
  (\bibinfo {year} {1988})}\BibitemShut {NoStop}%
\bibitem [{\citenamefont {Gorodnichev}\ \emph {et~al.}(1988)\citenamefont
  {Gorodnichev}, \citenamefont {Dudarev}, \citenamefont {Rogozkin},\ and\
  \citenamefont {Ryazanov}}]{Gorodnichev1988}%
  \BibitemOpen
  \bibfield  {author} {\bibinfo {author} {\bibfnamefont {E.~E.}\ \bibnamefont
  {Gorodnichev}}, \bibinfo {author} {\bibfnamefont {S.~L.}\ \bibnamefont
  {Dudarev}}, \bibinfo {author} {\bibfnamefont {D.~B.}\ \bibnamefont
  {Rogozkin}}, \ and\ \bibinfo {author} {\bibfnamefont {M.~I.}\ \bibnamefont
  {Ryazanov}},\ }\href@noop {} {\bibfield  {journal} {\bibinfo  {journal} {Sov.
  Phys. JETP Lett.}\ }\textbf {\bibinfo {volume} {48}},\ \bibinfo {pages} {147}
  (\bibinfo {year} {1988})}\BibitemShut {NoStop}%
\bibitem [{\citenamefont {Leskova}\ and\ \citenamefont
  {Maradudin}(1997)}]{Leskova1997}%
  \BibitemOpen
  \bibfield  {author} {\bibinfo {author} {\bibfnamefont {T.~A.}\ \bibnamefont
  {Leskova}}\ and\ \bibinfo {author} {\bibfnamefont {A.~A.}\ \bibnamefont
  {Maradudin}},\ }\href {\doibase 10.1080/13616679709409807} {\bibfield
  {journal} {\bibinfo  {journal} {Waves in Random Media}\ }\textbf {\bibinfo
  {volume} {7}},\ \bibinfo {pages} {395} (\bibinfo {year} {1997})}\BibitemShut
  {NoStop}%
\bibitem [{\citenamefont {Renaud}\ \emph {et~al.}(2009)\citenamefont {Renaud},
  \citenamefont {Lazzari},\ and\ \citenamefont {Leroy}}]{Renaud2009}%
  \BibitemOpen
  \bibfield  {author} {\bibinfo {author} {\bibfnamefont {G.}~\bibnamefont
  {Renaud}}, \bibinfo {author} {\bibfnamefont {R.}~\bibnamefont {Lazzari}}, \
  and\ \bibinfo {author} {\bibfnamefont {F.}~\bibnamefont {Leroy}},\ }\href
  {\doibase http://dx.doi.org/10.1016/j.surfrep.2009.07.002} {\bibfield
  {journal} {\bibinfo  {journal} {Surf. Sci. Rep.}\ }\textbf {\bibinfo {volume}
  {64}},\ \bibinfo {pages} {255 } (\bibinfo {year} {2009})}\BibitemShut
  {NoStop}%
\bibitem [{\citenamefont {Dosch}(1987)}]{Dosch1987}%
  \BibitemOpen
  \bibfield  {author} {\bibinfo {author} {\bibfnamefont {H.}~\bibnamefont
  {Dosch}},\ }\href {\doibase 10.1103/PhysRevB.35.2137} {\bibfield  {journal}
  {\bibinfo  {journal} {Phys. Rev. B}\ }\textbf {\bibinfo {volume} {35}},\
  \bibinfo {pages} {2137} (\bibinfo {year} {1987})}\BibitemShut {NoStop}%
\bibitem [{\citenamefont {Stepanov}(2000)}]{Stepanov2000}%
  \BibitemOpen
  \bibfield  {author} {\bibinfo {author} {\bibfnamefont {S.}~\bibnamefont
  {Stepanov}},\ }in\ \href@noop {} {\emph {\bibinfo {booktitle} {Exploration of
  Subsurface Phenomena by Particle Scattering}}},\ \bibinfo {editor} {edited
  by\ \bibinfo {editor} {\bibfnamefont {N.}~\bibnamefont {Lam}}, \bibinfo
  {editor} {\bibfnamefont {C.}~\bibnamefont {Melendres}}, \ and\ \bibinfo
  {editor} {\bibfnamefont {S.}~\bibnamefont {Sinha}}}\ (\bibinfo  {publisher}
  {International Advanced Studies Institute Press, North East, Maryland},\
  \bibinfo {year} {2000})\ pp.\ \bibinfo {pages} {119--137}\BibitemShut
  {NoStop}%
\bibitem [{\citenamefont {Kitahara}\ \emph {et~al.}(2002)\citenamefont
  {Kitahara}, \citenamefont {Inoue}, \citenamefont {Kikkawa}, \citenamefont
  {Matsushita},\ and\ \citenamefont {Takahashi}}]{Kitahara2002}%
  \BibitemOpen
  \bibfield  {author} {\bibinfo {author} {\bibfnamefont {A.}~\bibnamefont
  {Kitahara}}, \bibinfo {author} {\bibfnamefont {K.}~\bibnamefont {Inoue}},
  \bibinfo {author} {\bibfnamefont {H.}~\bibnamefont {Kikkawa}}, \bibinfo
  {author} {\bibfnamefont {K.}~\bibnamefont {Matsushita}}, \ and\ \bibinfo
  {author} {\bibfnamefont {I.}~\bibnamefont {Takahashi}},\ }\href
  {http://pfwww.kek.jp/acr2002pdf/part_b/pf02b083.pdf} {\emph {\bibinfo {title}
  {{X-ray Reflectivity study on polymeric surfaces near glass transition
  temperature}}}},\ \bibinfo {type} {Photon Factory Activity Report No. 20,
  Part B (unpublished)}\ (\bibinfo {year} {2002})\ \bibinfo {note} {p.
  83}\BibitemShut {NoStop}%
\bibitem [{\citenamefont {Gasse}\ \emph {et~al.}(2016)\citenamefont {Gasse},
  \citenamefont {L{\"u}tzenkirchen-Hecht}, \citenamefont {Wagner},\ and\
  \citenamefont {Frahm}}]{Gasse2016}%
  \BibitemOpen
  \bibfield  {author} {\bibinfo {author} {\bibfnamefont {J.-C.}\ \bibnamefont
  {Gasse}}, \bibinfo {author} {\bibfnamefont {D.}~\bibnamefont
  {L{\"u}tzenkirchen-Hecht}}, \bibinfo {author} {\bibfnamefont
  {R.}~\bibnamefont {Wagner}}, \ and\ \bibinfo {author} {\bibfnamefont
  {R.}~\bibnamefont {Frahm}},\ }\href
  {http://stacks.iop.org/1742-6596/712/i=1/a=012028} {\bibfield  {journal}
  {\bibinfo  {journal} {J. Phys.: Conf. Ser.}\ }\textbf {\bibinfo {volume}
  {712}},\ \bibinfo {pages} {012028} (\bibinfo {year} {2016})}\BibitemShut
  {NoStop}%
\bibitem [{\citenamefont {Kawanishi}\ \emph {et~al.}(1997)\citenamefont
  {Kawanishi}, \citenamefont {Ogura},\ and\ \citenamefont
  {Wang}}]{Kawanishi1997}%
  \BibitemOpen
  \bibfield  {author} {\bibinfo {author} {\bibfnamefont {T.}~\bibnamefont
  {Kawanishi}}, \bibinfo {author} {\bibfnamefont {H.}~\bibnamefont {Ogura}}, \
  and\ \bibinfo {author} {\bibfnamefont {Z.~L.}\ \bibnamefont {Wang}},\ }\href
  {\doibase 10.1080/13616679709409805} {\bibfield  {journal} {\bibinfo
  {journal} {Wave. Random Media}\ }\textbf {\bibinfo {volume} {7}},\ \bibinfo
  {pages} {351} (\bibinfo {year} {1997})}\BibitemShut {NoStop}%
\bibitem [{Note1()}]{Note1}%
  \BibitemOpen
  \bibinfo {note} {In an earlier numerical investigation of light scattering
  from one-dimensional dielectric rough surfaces, Nieto-Vesperinas and
  S\'{a}nchez-Gil~\cite {Nieto-Vesperinas1992} observed ``sidelobes'' in the
  angular intensity distributions. They did, however, not associate this
  observation with the Yoneda phenomenon.}\BibitemShut {Stop}%
\bibitem [{\citenamefont {Tamir}(1982)}]{Book:Tamir1982}%
  \BibitemOpen
  \bibfield  {author} {\bibinfo {author} {\bibfnamefont {T.}~\bibnamefont
  {Tamir}},\ }in\ \href@noop {} {\emph {\bibinfo {booktitle} {Electromagnetic
  Surface Modes}}},\ \bibinfo {editor} {edited by\ \bibinfo {editor}
  {\bibfnamefont {A.~D.}\ \bibnamefont {Boardman}}}\ (\bibinfo  {publisher}
  {John Wiley \& Sons},\ \bibinfo {address} {New York},\ \bibinfo {year}
  {1982})\ Chap.~\bibinfo {chapter} {13}\BibitemShut {NoStop}%
\bibitem [{\citenamefont {Gonz\'{a}lez-Alcalde}\ \emph
  {et~al.}(2016)\citenamefont {Gonz\'{a}lez-Alcalde}, \citenamefont {Banon},
  \citenamefont {Hetland}, \citenamefont {Maradudin}, \citenamefont
  {M\'{e}ndez}, \citenamefont {Nordam},\ and\ \citenamefont
  {Simonsen}}]{Gonzalez-Alcalde2016}%
  \BibitemOpen
  \bibfield  {author} {\bibinfo {author} {\bibfnamefont {A.~K.}\ \bibnamefont
  {Gonz\'{a}lez-Alcalde}}, \bibinfo {author} {\bibfnamefont {J.-P.}\
  \bibnamefont {Banon}}, \bibinfo {author} {\bibfnamefont {{\O}.~S.}\
  \bibnamefont {Hetland}}, \bibinfo {author} {\bibfnamefont {A.~A.}\
  \bibnamefont {Maradudin}}, \bibinfo {author} {\bibfnamefont {E.~R.}\
  \bibnamefont {M\'{e}ndez}}, \bibinfo {author} {\bibfnamefont
  {T.}~\bibnamefont {Nordam}}, \ and\ \bibinfo {author} {\bibfnamefont
  {I.}~\bibnamefont {Simonsen}},\ }\href {\doibase 10.1364/OE.24.025995}
  {\bibfield  {journal} {\bibinfo  {journal} {Opt. Express}\ }\textbf {\bibinfo
  {volume} {24}},\ \bibinfo {pages} {25995} (\bibinfo {year}
  {2016})}\BibitemShut {NoStop}%
\bibitem [{\citenamefont {Brewster}(1815)}]{Brewster1815}%
  \BibitemOpen
  \bibfield  {author} {\bibinfo {author} {\bibfnamefont {D.}~\bibnamefont
  {Brewster}},\ }\href {\doibase 10.1098/rstl.1815.0010} {\bibfield  {journal}
  {\bibinfo  {journal} {Philos. Trans. R. Soc. London}\ }\textbf {\bibinfo
  {volume} {105}},\ \bibinfo {pages} {125} (\bibinfo {year}
  {1815})}\BibitemShut {NoStop}%
\bibitem [{\citenamefont {Lakhtakia}(1989)}]{Lakhtakia1989}%
  \BibitemOpen
  \bibfield  {author} {\bibinfo {author} {\bibfnamefont {A.}~\bibnamefont
  {Lakhtakia}},\ }\href {\doibase 10.1364/ON.15.6.000014} {\bibfield  {journal}
  {\bibinfo  {journal} {Opt. News}\ }\textbf {\bibinfo {volume} {15}},\
  \bibinfo {pages} {14} (\bibinfo {year} {1989})}\BibitemShut {NoStop}%
\bibitem [{\citenamefont {Jackson}(1999)}]{jackson}%
  \BibitemOpen
  \bibfield  {author} {\bibinfo {author} {\bibfnamefont {J.~D.}\ \bibnamefont
  {Jackson}},\ }\href@noop {} {\emph {\bibinfo {title} {{Classical
  Electrodynamics}}}},\ \bibinfo {edition} {3rd}\ ed.\ (\bibinfo  {publisher}
  {John Wiley \& Sons, New York},\ \bibinfo {year} {1999})\BibitemShut
  {NoStop}%
\bibitem [{\citenamefont {Hecht}(2002)}]{book:hecht2002}%
  \BibitemOpen
  \bibfield  {author} {\bibinfo {author} {\bibfnamefont {E.}~\bibnamefont
  {Hecht}},\ }\href@noop {} {\emph {\bibinfo {title} {Optics}}}\ (\bibinfo
  {publisher} {Addison-Wesley},\ \bibinfo {year} {2002})\BibitemShut {NoStop}%
\bibitem [{\citenamefont {Sommerfeld}(1923)}]{Sommerfeld1923}%
  \BibitemOpen
  \bibfield  {author} {\bibinfo {author} {\bibfnamefont {A.}~\bibnamefont
  {Sommerfeld}},\ }\href {\doibase 10.1364/JOSA.7.000501} {\bibfield  {journal}
  {\bibinfo  {journal} {J. Opt. Soc. Am.}\ }\textbf {\bibinfo {volume} {7}},\
  \bibinfo {pages} {501} (\bibinfo {year} {1923})}\BibitemShut {NoStop}%
\bibitem [{\citenamefont {Greffet}\ and\ \citenamefont
  {Sentenac}(1991)}]{Greffet1991}%
  \BibitemOpen
  \bibfield  {author} {\bibinfo {author} {\bibfnamefont {J.-J.}\ \bibnamefont
  {Greffet}}\ and\ \bibinfo {author} {\bibfnamefont {A.}~\bibnamefont
  {Sentenac}},\ }in\ \href {\doibase 10.1117/12.49634} {\emph {\bibinfo
  {booktitle} {Wave Propagation and Scattering in Varied Media II}}},\ Vol.\
  \bibinfo {volume} {1558},\ \bibinfo {editor} {edited by\ \bibinfo {editor}
  {\bibfnamefont {V.~K.}\ \bibnamefont {Varadan}}}\ (\bibinfo {year} {1991})\
  pp.\ \bibinfo {pages} {288--294}\BibitemShut {NoStop}%
\bibitem [{\citenamefont {Sentenac}\ and\ \citenamefont
  {Greffet}(1998)}]{Sentenac1998}%
  \BibitemOpen
  \bibfield  {author} {\bibinfo {author} {\bibfnamefont {A.}~\bibnamefont
  {Sentenac}}\ and\ \bibinfo {author} {\bibfnamefont {J.-J.}\ \bibnamefont
  {Greffet}},\ }\href {\doibase 10.1364/JOSAA.15.000528} {\bibfield  {journal}
  {\bibinfo  {journal} {J. Opt. Soc. Am. A}\ }\textbf {\bibinfo {volume}
  {15}},\ \bibinfo {pages} {528} (\bibinfo {year} {1998})}\BibitemShut
  {NoStop}%
\bibitem [{\citenamefont {Calvo-Perez}\ \emph {et~al.}(1999)\citenamefont
  {Calvo-Perez}, \citenamefont {Sentenac},\ and\ \citenamefont
  {Greffet}}]{Calvo-Perez1999}%
  \BibitemOpen
  \bibfield  {author} {\bibinfo {author} {\bibfnamefont {O.}~\bibnamefont
  {Calvo-Perez}}, \bibinfo {author} {\bibfnamefont {A.}~\bibnamefont
  {Sentenac}}, \ and\ \bibinfo {author} {\bibfnamefont {J.~J.}\ \bibnamefont
  {Greffet}},\ }\href {\doibase 10.1029/1998RS900027} {\bibfield  {journal}
  {\bibinfo  {journal} {Radio Sci.}\ }\textbf {\bibinfo {volume} {34}},\
  \bibinfo {pages} {311} (\bibinfo {year} {1999})}\BibitemShut {NoStop}%
\bibitem [{\citenamefont {Lekner}(1987)}]{Lekner1987}%
  \BibitemOpen
  \bibfield  {author} {\bibinfo {author} {\bibfnamefont {J.}~\bibnamefont
  {Lekner}},\ }\href@noop {} {\emph {\bibinfo {title} {Theory of Reflection of
  Electromagnetic and Particle Waves}}},\ Vol.~\bibinfo {volume} {3}\ (\bibinfo
   {publisher} {Springer Science \& Business Media},\ \bibinfo {year}
  {1987})\BibitemShut {NoStop}%
\bibitem [{\citenamefont {Doyle}(1985)}]{Doyle1985}%
  \BibitemOpen
  \bibfield  {author} {\bibinfo {author} {\bibfnamefont {W.~T.}\ \bibnamefont
  {Doyle}},\ }\href {\doibase 10.1119/1.14201} {\bibfield  {journal} {\bibinfo
  {journal} {Am. J. Phys.}\ }\textbf {\bibinfo {volume} {53}},\ \bibinfo
  {pages} {463} (\bibinfo {year} {1985})}\BibitemShut {NoStop}%
\bibitem [{\citenamefont {Sein}(1970)}]{Sein1970}%
  \BibitemOpen
  \bibfield  {author} {\bibinfo {author} {\bibfnamefont {J.}~\bibnamefont
  {Sein}},\ }\href {\doibase 10.1016/0030-4018(70)90008-8} {\bibfield
  {journal} {\bibinfo  {journal} {Opt. Commun.}\ }\textbf {\bibinfo {volume}
  {2}},\ \bibinfo {pages} {170} (\bibinfo {year} {1970})}\BibitemShut {NoStop}%
\bibitem [{\citenamefont {Pattanayak}\ and\ \citenamefont
  {Wolf}(1972)}]{Pattanayak1972}%
  \BibitemOpen
  \bibfield  {author} {\bibinfo {author} {\bibfnamefont {D.}~\bibnamefont
  {Pattanayak}}\ and\ \bibinfo {author} {\bibfnamefont {E.}~\bibnamefont
  {Wolf}},\ }\href {\doibase 10.1016/0030-4018(72)90178-2} {\bibfield
  {journal} {\bibinfo  {journal} {Opt. Commun.}\ }\textbf {\bibinfo {volume}
  {6}},\ \bibinfo {pages} {217} (\bibinfo {year} {1972})}\BibitemShut {NoStop}%
\bibitem [{\citenamefont {Ewald}(1916)}]{Ewald1916}%
  \BibitemOpen
  \bibfield  {author} {\bibinfo {author} {\bibfnamefont {P.~P.}\ \bibnamefont
  {Ewald}},\ }\href@noop {} {\bibfield  {journal} {\bibinfo  {journal} {Annalen
  der Physik}\ }\textbf {\bibinfo {volume} {354}},\ \bibinfo {pages} {1}
  (\bibinfo {year} {1916})}\BibitemShut {NoStop}%
\bibitem [{\citenamefont {Banon}\ \emph {et~al.}(2018)\citenamefont {Banon},
  \citenamefont {{S. Hetland}},\ and\ \citenamefont {Simonsen}}]{Banon2018a}%
  \BibitemOpen
  \bibfield  {author} {\bibinfo {author} {\bibfnamefont {J.-P.}\ \bibnamefont
  {Banon}}, \bibinfo {author} {\bibfnamefont {{\O}.}~\bibnamefont {{S.
  Hetland}}}, \ and\ \bibinfo {author} {\bibfnamefont {I.}~\bibnamefont
  {Simonsen}},\ }\href {\doibase https://doi.org/10.1016/j.aop.2017.12.003}
  {\bibfield  {journal} {\bibinfo  {journal} {Annals of Physics}\ }\textbf
  {\bibinfo {volume} {389}},\ \bibinfo {pages} {352 } (\bibinfo {year}
  {2018})}\BibitemShut {NoStop}%
\bibitem [{\citenamefont {Soubret}\ \emph
  {et~al.}(2001{\natexlab{a}})\citenamefont {Soubret}, \citenamefont
  {Berginc},\ and\ \citenamefont {Bourrely}}]{Soubret2001a}%
  \BibitemOpen
  \bibfield  {author} {\bibinfo {author} {\bibfnamefont {A.}~\bibnamefont
  {Soubret}}, \bibinfo {author} {\bibfnamefont {G.}~\bibnamefont {Berginc}}, \
  and\ \bibinfo {author} {\bibfnamefont {C.}~\bibnamefont {Bourrely}},\ }\href
  {\doibase 10.1103/PhysRevB.63.245411} {\bibfield  {journal} {\bibinfo
  {journal} {Phys. Rev. B}\ }\textbf {\bibinfo {volume} {63}},\ \bibinfo
  {pages} {245411} (\bibinfo {year} {2001}{\natexlab{a}})}\BibitemShut
  {NoStop}%
\bibitem [{\citenamefont {Brown}\ \emph {et~al.}(1984)\citenamefont {Brown},
  \citenamefont {Celli}, \citenamefont {Haller},\ and\ \citenamefont
  {Marvin}}]{Brown1984}%
  \BibitemOpen
  \bibfield  {author} {\bibinfo {author} {\bibfnamefont {G.}~\bibnamefont
  {Brown}}, \bibinfo {author} {\bibfnamefont {V.}~\bibnamefont {Celli}},
  \bibinfo {author} {\bibfnamefont {M.}~\bibnamefont {Haller}}, \ and\ \bibinfo
  {author} {\bibfnamefont {A.}~\bibnamefont {Marvin}},\ }\href {\doibase
  10.1016/0039-6028(84)90619-8} {\bibfield  {journal} {\bibinfo  {journal}
  {Surf. Sci.}\ }\textbf {\bibinfo {volume} {136}},\ \bibinfo {pages} {381}
  (\bibinfo {year} {1984})}\BibitemShut {NoStop}%
\bibitem [{\citenamefont {Soubret}\ \emph
  {et~al.}(2001{\natexlab{b}})\citenamefont {Soubret}, \citenamefont
  {Berginc},\ and\ \citenamefont {Bourrely}}]{Soubret2001}%
  \BibitemOpen
  \bibfield  {author} {\bibinfo {author} {\bibfnamefont {A.}~\bibnamefont
  {Soubret}}, \bibinfo {author} {\bibfnamefont {G.}~\bibnamefont {Berginc}}, \
  and\ \bibinfo {author} {\bibfnamefont {C.}~\bibnamefont {Bourrely}},\ }\href
  {\doibase 10.1364/JOSAA.18.002778} {\bibfield  {journal} {\bibinfo  {journal}
  {J. Opt. Soc. Am. A}\ }\textbf {\bibinfo {volume} {18}},\ \bibinfo {pages}
  {2778} (\bibinfo {year} {2001}{\natexlab{b}})}\BibitemShut {NoStop}%
\bibitem [{\citenamefont {Banon}(2018)}]{banon:thesis}%
  \BibitemOpen
  \bibfield  {author} {\bibinfo {author} {\bibfnamefont {J.-P.}\ \bibnamefont
  {Banon}},\ }\emph {\bibinfo {title} {On the simulation of electromagnetic
  wave scattering by periodic and randomly rough layered structures based on
  the reduced {R}ayleigh equations : theory, numerical analysis and
  applications}},\ \href {http://hdl.handle.net/11250/2570845} {Ph.D. thesis},\
  \bibinfo  {school} {Norwegian University of Science and Technology}, \bibinfo
  {address} {Trondheim, Norway} (\bibinfo {year} {2018})\BibitemShut {NoStop}%
\bibitem [{\citenamefont {Maradudin}\ and\ \citenamefont
  {M{\'{e}}ndez}(1993)}]{Maradudin1993}%
  \BibitemOpen
  \bibfield  {author} {\bibinfo {author} {\bibfnamefont {A.~A.}\ \bibnamefont
  {Maradudin}}\ and\ \bibinfo {author} {\bibfnamefont {E.~R.}\ \bibnamefont
  {M{\'{e}}ndez}},\ }\href {\doibase 10.1364/AO.32.003335} {\bibfield
  {journal} {\bibinfo  {journal} {Appl. Opt.}\ }\textbf {\bibinfo {volume}
  {32}},\ \bibinfo {pages} {3335} (\bibinfo {year} {1993})}\BibitemShut
  {NoStop}%
\bibitem [{\citenamefont {Maradudin}\ \emph {et~al.}(1995)\citenamefont
  {Maradudin}, \citenamefont {McGurn},\ and\ \citenamefont
  {M{\'{e}}ndez}}]{Maradudin1995}%
  \BibitemOpen
  \bibfield  {author} {\bibinfo {author} {\bibfnamefont {A.}~\bibnamefont
  {Maradudin}}, \bibinfo {author} {\bibfnamefont {A.~R.}\ \bibnamefont
  {McGurn}}, \ and\ \bibinfo {author} {\bibfnamefont {E.~R.}\ \bibnamefont
  {M{\'{e}}ndez}},\ }\href {\doibase 10.1364/JOSAA.12.002500} {\bibfield
  {journal} {\bibinfo  {journal} {J. Opt. Soc. Am. A}\ }\textbf {\bibinfo
  {volume} {12}},\ \bibinfo {pages} {2500} (\bibinfo {year}
  {1995})}\BibitemShut {NoStop}%
\bibitem [{\citenamefont {McGurn}\ and\ \citenamefont
  {Maradudin}(1996)}]{Mcgurn1996}%
  \BibitemOpen
  \bibfield  {author} {\bibinfo {author} {\bibfnamefont {A.~R.}\ \bibnamefont
  {McGurn}}\ and\ \bibinfo {author} {\bibfnamefont {A.~A.}\ \bibnamefont
  {Maradudin}},\ }\href {\doibase 10.1088/0959-7174/6/3/006} {\bibfield
  {journal} {\bibinfo  {journal} {Waves in Random Media}\ }\textbf {\bibinfo
  {volume} {6}},\ \bibinfo {pages} {251} (\bibinfo {year} {1996})}\BibitemShut
  {NoStop}%
\bibitem [{\citenamefont {Simonsen}(2010)}]{Simonsen2010}%
  \BibitemOpen
  \bibfield  {author} {\bibinfo {author} {\bibfnamefont {I.}~\bibnamefont
  {Simonsen}},\ }\href {\doibase 10.1140/epjst/e2010-01221-4} {\bibfield
  {journal} {\bibinfo  {journal} {Eur. Phys. J.-Spec. Top.}\ }\textbf {\bibinfo
  {volume} {181}},\ \bibinfo {pages} {1} (\bibinfo {year} {2010})}\BibitemShut
  {NoStop}%
\bibitem [{\citenamefont {{Kr{\"o}ger}}\ and\ \citenamefont
  {{Kretschmann}}(1970)}]{kroger}%
  \BibitemOpen
  \bibfield  {author} {\bibinfo {author} {\bibfnamefont {E.}~\bibnamefont
  {{Kr{\"o}ger}}}\ and\ \bibinfo {author} {\bibfnamefont {E.}~\bibnamefont
  {{Kretschmann}}},\ }\href@noop {} {\bibfield  {journal} {\bibinfo  {journal}
  {Zeitschrift fur Physik}\ }\textbf {\bibinfo {volume} {237}},\ \bibinfo
  {pages} {1} (\bibinfo {year} {1970})}\BibitemShut {NoStop}%
\bibitem [{Note2()}]{Note2}%
  \BibitemOpen
  \bibinfo {note} {The dot product here must be taken as the Hermitian inner
  product for complex vectors $\protect \mathbf {a} \cdot \protect \mathbf {b}
  = \DOTSB \sum@ \slimits@ _j a_j^* b_j$.}\BibitemShut {Stop}%
\bibitem [{Note3()}]{Note3}%
  \BibitemOpen
  \bibinfo {note} {Here limited to one medium, but in general both media
  sharing the interface could be dielectrics.}\BibitemShut {Stop}%
\bibitem [{\citenamefont {Doyle}(1980)}]{Doyle1980}%
  \BibitemOpen
  \bibfield  {author} {\bibinfo {author} {\bibfnamefont {W.~T.}\ \bibnamefont
  {Doyle}},\ }\href {\doibase 10.1119/1.12042} {\bibfield  {journal} {\bibinfo
  {journal} {Am. J. Phys.}\ }\textbf {\bibinfo {volume} {48}},\ \bibinfo
  {pages} {643} (\bibinfo {year} {1980})}\BibitemShut {NoStop}%
\bibitem [{Note4()}]{Note4}%
  \BibitemOpen
  \bibinfo {note} {One may extend the construction to all vectors on the circle
  defined as the intersection of the unit sphere and the plane $\mathop
  {\protect \mathbf {E}_{p,1}^{(0)}} (\mathop {\protect \mathbf
  {p}_{0}^{}})^\perp $, but it would result in constructing twice the same set
  of wave vectors of zero scattering.}\BibitemShut {Stop}%
\bibitem [{\citenamefont {Agarwal}(1977)}]{Agarwal1977}%
  \BibitemOpen
  \bibfield  {author} {\bibinfo {author} {\bibfnamefont {G.~S.}\ \bibnamefont
  {Agarwal}},\ }\href@noop {} {\bibfield  {journal} {\bibinfo  {journal} {Phys.
  Rev. B}\ }\textbf {\bibinfo {volume} {15}},\ \bibinfo {pages} {2371}
  (\bibinfo {year} {1977})}\BibitemShut {NoStop}%
\bibitem [{\citenamefont {Miyamoto}\ and\ \citenamefont
  {Wolf}(1962)}]{Miyamoto1962}%
  \BibitemOpen
  \bibfield  {author} {\bibinfo {author} {\bibfnamefont {K.}~\bibnamefont
  {Miyamoto}}\ and\ \bibinfo {author} {\bibfnamefont {E.}~\bibnamefont
  {Wolf}},\ }\href@noop {} {\bibfield  {journal} {\bibinfo  {journal} {J. Opt.
  Soc. Am.}\ }\textbf {\bibinfo {volume} {52}},\ \bibinfo {pages} {615}
  (\bibinfo {year} {1962})}\BibitemShut {NoStop}%
\bibitem [{\citenamefont {Nieto-Vesperinas}\ and\ \citenamefont
  {S\'{a}nchez-Gil}(1992)}]{Nieto-Vesperinas1992}%
  \BibitemOpen
  \bibfield  {author} {\bibinfo {author} {\bibfnamefont {M.}~\bibnamefont
  {Nieto-Vesperinas}}\ and\ \bibinfo {author} {\bibfnamefont {J.~A.}\
  \bibnamefont {S\'{a}nchez-Gil}},\ }\href {\doibase 10.1364/JOSAA.9.000424}
  {\bibfield  {journal} {\bibinfo  {journal} {J. Opt. Soc. Am. A}\ }\textbf
  {\bibinfo {volume} {9}},\ \bibinfo {pages} {424} (\bibinfo {year}
  {1992})}\BibitemShut {NoStop}%
\end{thebibliography}%

\end{document}